\documentstyle[preprint,prb,aps]{revtex}

\begin{document}

\draft

\title{Helicity Modulus and Fluctuating Type II Superconductors: 
Elastic Approximation and Numerical Simulations}
\author{Tao Chen and S. Teitel}
\address{Department of Physics and Astronomy, University of Rochester, 
Rochester, NY 14627\\}
\date{\today}
\maketitle
\begin{abstract}

We develop the helicity modulus as a criterion for superconducting
order in the mixed phase of a fluctuating type II superconductor.
We show that there is a duality relation between this helicity modulus 
and the superfluid density of a system of analog 2D bosons.
We show that the vortex line lattice exhibits a perfect
Meissner effect with respect to a shearing perturbation of the
applied magnetic field, and this becomes our criterion for
``longitudinal superconductivity'' parallel to the applied
field.  We present arguments based on the 2D boson analogy,
as well as the results of numerical simulations, that suggest that longitudinal
superconductivity can persist into the vortex line liquid state
for systems of finite thickness, comparable to those commonly 
found in experiments.

\end{abstract}
\newpage
\narrowtext

\section{Introduction}
\label{sintro}

The mixed state of a type II superconductor in an applied
magnetic field ${\bf H}$ is characterized, in 
{\it mean field} theory, by a spatially varying order parameter
$\psi({\bf r})$ whose amplitude vanishes continuously
as $T_{\rm c2}(H)$ is approached from below.\cite{R1}  
While this description is adequate for traditional superconductors,
the importance of thermal fluctuations in determining the behavior
of the high temperature superconductors\cite{R2} now requires one to find
a reasonable criterion for superconducting coherence in the mixed 
state, that is defined in terms of an average over all fluctuating
configurations $\psi({\bf r})$.  One possibility is the correlation function
$\langle\psi^*({\bf r})\psi(0)\rangle$.  However controversy has
arisen over the proper gauge invariant definition for this
correlation function;\cite{R3,R4,R5,R6,R9} the most straightforward
definition leads in three dimensions (3D) to correlations
which decay exponentially\cite{R3,R4} (albeit with a long decay length) even 
within the Abrikosov vortex line lattice state, once harmonic elastic
fluctuations of the vortex lines are included.  The flux flow
resistance of an unpinned vortex lattice in a completely clean
material also is contrary to the conventional idea of a superconductor
as a state with zero resistance.  In this paper we propose using the
helicity modulus as a clear equilibrium quantity that can distinguish
superconducting from normal behavior in the mixed state.
We will show that the helicity modulus, which for a $neutral$ superfluid
is proportional to the superfluid density,\cite{R7} is in a superconductor
(or $charged$ superfluid) related to the magnetic susceptibility
of the system to a small perturbation in applied magnetic field, about
the uniform applied ${\bf H}$.  Recall that it is the
magnetic response, rather than electrical resistivity, that gives the true
defining equilibrium signature of the Meissner transition in either a type I
or type II superconductor.  Here we will show that
for the mixed state of a type II superconductor, 
the vortex line lattice displays (in the absence of dislocations) a perfect
Meissner effect with respect to a certain type of $shear$ perturbation
of the applied field, for which the screening currents run parallel
to ${\bf H}$.  Such a shear Meissner effect, also referred to as
{\it longitudinal superconductivity},\cite{R8} we will take as the defining
equilibrium criterion for superconducting order within the mixed state.
We will then give a set of arguments, including the results of 
numerical simulations, that suggest that for system sizes of
experimental interest, longitudinal superconductivity can persist
above the vortex line lattice melting transition, into the vortex
line liquid state.  Some of our results have been briefly presented 
earlier,\cite{R9,R10} for the case of an isotropic system.
Here we provide much greater detail, and generalize our formalism
to the uniaxial anisotropic case.

The rest of this paper is organized as follows.
In Sec.\,II we define our London model for a continuous anisotropic
supercondutor, giving the mapping of the Hamiltonian from its
representation in terms of the wavefunction phase angle, to its
representation in terms of interacting vortex lines.  In Sec.\,III
we define the helicity modulus $\Upsilon_{\mu\nu}({\bf q})$, 
discuss its relation to the magnetic susceptibility, and describe
the important physical parameters that may be extracted from it.
We also discuss in some detail the mapping between the interacting
vortex lines and an analog system of interacting two dimensional
(2D) bosons.\cite{R11}  We show that an interesting duality exists between
the helicity modulus of the 3D superconductor model and the
helicity modulus of the 2D analog bosons, for both the
superconductor with a finite magnetic penetration length $\lambda$
(our main concern in this work),
and the superconductor in the $\lambda\to\infty$ approximation.
In Sec.\,IV we analyze the helicity modulus within the elastic
approximation\cite{R12,R12.1} for a vortex line lattice, and demonstrate the
existence of the shear Meissner effect.  We 
calculate how the penetration length for the shear
perturbation increases with temperature, due to second
order elastic fluctuations.  In Sec.\,V we consider the
vortex line liquid, and show how the hydrodynamic 
approximation\cite{R13}
yields the disappearance of the perfect shear Meissner effect.
We discuss how the Kosterlitz--Thouless transition\cite{R14} of the
analog 2D bosons can yield a cross-over 
from a normal vortex line liquid to
a line liquid with longitudinal superconductivity, and estimate
this cross-over temperature as a function of system thickness
and applied magnetic field.  
In Sec.\,VI we discuss our numerical Monte
Carlo (MC) simulations.  We define the Hamiltonian and helicity
modulus for a discretized {\it lattice superconductor},\cite{R15,R16}
discuss our MC algorithm, and present our numerical results
for an isotropic model.  We compute the helicity modulus
and other measures of vortex fluctuations, and find
evidence for longitudinal superconductivity within
the vortex line liquid state.  In Sec.\,VII we present
our conclusions and discussion.

\section{London Superconductor Model}
\label{slondon}

We will model our uniaxial superconductor as a three dimensional
continuum with the weak coupling direction parallel to the ${\bf\hat z}$
axis.  The bare magnetic penetration length along this weak direction
is $\lambda_z$, while $\lambda_\perp$ is the penetration length
within the more strongly coupled $xy$ planes.  $\eta\equiv\lambda_z/
\lambda_\perp$ is the anisotropy parameter.

The Ginzburg-Landau Hamiltonian for the Gibbs ensemble,
in the London approximation, can then be written as,

\begin{equation}
   {\cal H}[\theta,{\bf A}]={1\over 2}J_\perp\int d^3r
   \left[\sum_{\mu=x,y,z}\eta_\mu^{-2}(\nabla_\mu\theta
   -A_\mu)^2
   +\lambda^2_\perp \left| {\bf \nabla}\times ({\bf A}-{\bf 
   A}^{\rm ext})\right|^2\right]\enspace,
\label{eLG1}
\end{equation}
where 

\begin{equation}
   J_\perp\equiv\phi_0^2/(16\pi^3\lambda^2_\perp)
\label{eJperp}
\end{equation}
is the coupling within the $xy$ plane ($\phi_0=hc/2e$ is the flux quantum),

\begin{equation}
   \eta_x=\eta_y\equiv 1, \quad \eta_z\equiv\eta=\lambda_z/\lambda_\perp
\label{eeta}
\end{equation}
define the anisotropy, and $(\phi_0/2\pi){\bf A}$ and 
$(\phi_0/2\pi){\bf A}^{ext}$ are the
vector potentials for the internal and applied magnetic fields,

\begin{equation}
   {\bf\nabla}\times {\bf A}=2\pi {\bf b},\qquad
   {\bf\nabla}\times {\bf A}^{\rm ext}=2\pi {\bf h}\enspace,
\label{efh}
\end{equation}
where ${\bf b}={\bf B}/\phi_0$ and ${\bf h}={\bf H}/\phi_0$ are 
the densities of flux quanta of the magnetic field ${\bf B}({\bf r})$ inside 
the superconductor, and the externally applied field ${\bf H}({\bf r})$.
In Eq.(\ref{eLG1}), $\theta({\bf r})$ and ${\bf A}({\bf r})$ are thermally 
fluctuating
variables to be averaged over in a partition function sum, while
${\bf A}^{\rm ext}({\bf r})$ is a fixed (quenched) field.

It will be useful to introduce the induced magnetic vector potential,
\begin{equation}
{\bf A}^{\rm ind}\equiv{\bf A}-{\bf A}^{\rm ext},\qquad
{\bf\nabla}\times {\bf A}^{\rm ind}=2\pi({\bf b}-{\bf h})=2\pi{\bf 
b}^{\rm ind}\enspace,
\label{ebind}
\end{equation}
in terms of which the Hamiltonian becomes,
\begin{equation}
   {\cal H}[\theta,{\bf A}]={1\over 2}J_\perp\int d^3r
   \left[\sum_{\mu}\eta_\mu^{-2}(\nabla_\mu\theta
    -A^{\rm ext}_\mu-A^{\rm ind}_\mu)^2
   +\lambda^2_\perp \left| {\bf \nabla}\times{\bf A}^{\rm 
   ind}\right|^2\right]\enspace,
\label{eLG1.1}
\end{equation}
and the partition function is to be viewed as a sum over ${\bf A}^{\rm 
ind}$.

The Hamiltonian of Eq.(\ref{eLG1.1}) can now be mapped onto
to a model of interacting vortex lines.  If we define the superfluid
velocity,
\begin{equation}
{\bf v}={\bf\nabla}\theta-{\bf A}^{\rm ext}-{\bf A}^{\rm ind}
={\bf\nabla}\theta-{\bf A}\enspace,
\label{ev}
\end{equation}
then by Eq.(\ref{efh}) we have
\begin{equation}
{\bf\nabla}\times{\bf v}=2\pi({\bf n}-{\bf h}-{\bf b}^{\rm ind})
=2\pi({\bf n}-{\bf b})\enspace,
\label{ecurlv}
\end{equation}
where ${\bf n}\equiv{1\over 2\pi}{\bf\nabla}\times{\bf\nabla}\theta$ 
is the vortex line density, consisting of singular lines of integer
vorticity in the phase angle $\theta({\bf r})$.

Defining the Fourier transforms,
\begin{equation}
   {\bf v}_q=\int d^3r\,e^{i{\bf q}\cdot {\bf r}}{\bf v}({\bf 
   r})\enspace,
   \qquad\qquad {\bf v}({\bf r})={1\over{\cal V}}\sum_q e^{-i{\bf q}\cdot 
   {\bf r}}{\bf v}_q
\label{eFT}
\end{equation}
(where ${\cal V}$ is the system volume), we can then write
the Hamiltonian (\ref{eLG1.1}) as,
%
\begin{equation}
   {\cal H}={J_\perp\over 2{\cal V}}\sum_{q,\mu}
   \Big[ \eta_\mu^{-2} v_{q\mu} v_{-q\mu}
   +4\pi^2\lambda_\perp^2 b_{q\mu}^{\rm ind}
   b_{-q\mu}^{\rm ind}\Big]\enspace,
\label{eLG4}
\end{equation}
and solve Eq.(\ref{ecurlv}) as,
\begin{equation}
{\bf v}_q=-2\pi i\left[{\bf q}\chi_q +
{{\bf q}\times({\bf n}_q-{\bf h}_q-{\bf b}_q^{\rm ind})
\over q^2}\right]
\label{evsol}
\end{equation}
where $\chi({\bf r})$ is any smooth scalar function.
Substituting Eq.(\ref{evsol}) into Eq.(\ref{eLG4}), and completing
the square in $\chi_q$, results in,
\begin{eqnarray}
  {\cal H}&=&{4\pi^2J_\perp\over 2{\cal V}}\sum_q \Bigg[
  [{\bf n}_q-{\bf h}_q-{\bf b}_q^{\rm ind}]
  \cdot {\bf V}^0_q
  \cdot [{\bf n}_{-q}-{\bf h}_{-q}-{\bf b}_{-q}^{\rm ind}]
  \nonumber\\
   && \qquad\qquad\qquad\qquad\qquad+(q_\perp^2+\eta^{-2} q_z^2)
  \delta\chi_q\delta\chi_{-q} +\lambda_\perp^2{\bf b}_q^{\rm ind}
  \cdot{\bf b}_{-q}^{\rm ind}\Bigg]\enspace,
\label{eLG5}
\end{eqnarray}
where 
\begin{equation}
  {\bf V}^0_q={1\over q^2}\left[{\bf I}-
  {\lambda_z^2-\lambda_\perp^2\over \lambda_z^2q_\perp^2
  +\lambda_\perp^2 q_z^2}
  ({\bf\hat z}\times {\bf q})({\bf\hat z}\times {\bf q})\right]
\label{eVq0}
\end{equation}
is the ``bare'' vortex line interaction tensor, before screening by magnetic
field fluctuations, and
$\delta\chi_q\equiv\chi_q-\chi^0_q$ is the 
fluctuation of $\chi_q$ away from the value
\begin{equation}
  \chi^0_q={(\lambda_z^2-\lambda_\perp^2) q_z[{\bf q}\times ({\bf n}_q
  -{\bf h}_q-{\bf b}_q^{\rm ind})]_z\over
  q^2(\lambda_z^2q^2+\lambda_\perp^2 q_z^2)}\enspace.
\label{echi0}
\end{equation}
Substituting $\chi^0_q$ into Eq.(\ref{evsol}) gives the superfluid
velocity ${\bf v}_q^0$ that minimizes ${\cal H}$ for a given
configuration of ${\bf n}_q-{\bf h}_q-{\bf b}_q^{\rm ind}$.
$\delta\chi_q$ represents a smooth ``spin-wave'' fluctuation
about this ${\bf v}_q^0$.

We can now complete the square in ${\bf b}_q^{\rm ind}$ in 
Eq.(\ref{eLG5}), subject
to the constraint that ${\bf q}\cdot{\bf b}_q^{\rm ind}=0$,
to get,
\begin{equation}
   {\cal H}={4\pi^2J_\perp\over 2{\cal V}}\sum_q \Bigg[
   [{\bf n}_q-{\bf h}_q]\cdot {\bf V}_q\cdot[{\bf n}_{-q}
   -{\bf h}_{-q}]+
   (q_\perp^2+\eta^{-2} q_z^2)\delta\chi_q
   \delta\chi_{-q}+\delta {\bf b}_q\cdot {\bf U}_q
   \cdot\delta {\bf b}_{-q}
   \Bigg]\enspace,
\label{eLG6}
\end{equation}
where the tensor
\begin{eqnarray}
   {\bf V}_q & = & {\bf V}^0_q-{\bf V}^0_q
   \cdot{\bf U}^{-1}_q\cdot{\bf V}^0_q
    =\lambda_\perp^2{\bf U}^{-1}_q\cdot{\bf V}^0_q \nonumber\\
    & = & {\lambda_\perp^2\over 1+\lambda_\perp^2q^2}\left[{\bf I}-
    {\lambda_z^2-\lambda_\perp^2\over 1+\lambda_z^2q_\perp^2+
    \lambda_\perp^2q_z^2}({\bf\hat z}\times{\bf q})({\bf\hat z}\times{\bf q})
    \right]
\label{eVq}
\end{eqnarray}
is the uniaxial anisotropic generalization\cite{R12.1} 
of the familiar London vortex line interaction, and
\begin{equation}
   {\bf U}_q=\lambda_\perp^2{\bf I}+{\bf V}^0_q
\label{eUq}
\end{equation}
is the interaction tensor for fluctuations of magnetic field,
$\delta{\bf b}_q\equiv {\bf b}_q^{\rm ind}-{\bf b}_q^{\rm ind,\,0}$,
about the value
\begin{equation}
   {\bf b}^{\rm ind,\,0}_q={\bf U}^{-1}_q\cdot{\bf V}^0_q
   \cdot[{\bf n}_q-{\bf h}_q]=
   {1\over\lambda_\perp^2}{\bf V}_q\cdot[{\bf n}_q-{\bf h}_q]
\label{efq0}
\end{equation}
that minimizes ${\cal H}$ for a given configuration ${\bf n}_q-
{\bf h}_q$.

Eq.(\ref{eLG6}) represents the Ginzburg-Landau Hamiltonian written in terms
of decoupled spin wave, magnetic field, and vortex line fluctuations.  The
partition function is to be summed over all smooth
$\delta\chi_q$, all smooth 
$\delta {\bf b}_q$ subject to the constraint ${\bf q}\cdot
\delta {\bf b}_q=0$, and all singular vortex line 
distributions ${\bf n}_q$ with conserved vorticity ${\bf q}\cdot{\bf 
n}_q=0$.

The interaction ${\bf V}_q$ of Eq.(\ref{eVq}) 
is given as a tensor, with non-vanishing off diagonal components. 
However, as shown by Carneiro {\it et al}.,\cite{R18} 
one can exploit the conservation
of vorticity to rewrite ${\bf V}_q$ in a purely diagonal way.  Using

\begin{eqnarray}
   {\bf n}_q \cdot ({\bf\hat z}\times{\bf q})({\bf\hat z} \times{\bf q})
   \cdot{\bf n}_{-q} & = & q_\perp^2{\bf n}_{q\perp}
   \cdot{\bf n}_{-q\perp}-
   [{\bf q}_\perp\cdot{\bf n}_{q\perp}][{\bf q}_\perp\cdot{\bf n}_{-q\perp}]
   \nonumber\\
   & = & q_\perp^2{\bf n}_{q\perp}\cdot{\bf n}_{-q\perp}
   -q_z^2n_{qz}n_{-qz}\enspace,
\label{eVtrans}
\end{eqnarray}
where ${\bf n}_\perp\equiv (n_x,n_y)$ is the transverse
part of the vorticity, and a similar result for ${\bf h}_q$,
we can rewrite the vortex line interaction part of the Hamiltonian as,

\begin{equation}
   {\cal H}_{\rm v}={4\pi^2J_\perp\over 2{\cal V}}
   \sum_{q,\mu} V_{q\mu}[n_{q\mu}-h_{q\mu}]
   [n_{-q\mu}-h_{-q\mu}]\enspace,
\label{eLG7}
\end{equation}
where $V_{qx}=V_{qy}\equiv V_{q\perp}$, and,
\begin{equation}
    V_{q\perp}={\lambda_\perp^2\over 1+\lambda_\perp^2q_z^2
    +\lambda_z^2q_\perp^2}\enspace,\qquad
   V_{qz}={\lambda_\perp^2(1+\lambda_z^2q^2)\over
   (1+\lambda_\perp^2q^2)(1+\lambda_\perp^2q_z^2+\lambda_z^2q_\perp^2)}
   \enspace.
\label{eVdiag}
\end{equation}
In most of this paper we will be considering behavior in the
presence of a $uniform$ applied magnetic field, for which
${\bf h}_{q\ne 0}=0$ and hence ${\bf b}_{q\ne 0}^{\rm ind}
={\bf b}_{q\ne 0}$.

\section{Helicity Modulus, Magnetic Susceptibility, and 2D Bosons}

\subsection{Definition of the Helicity Modulus}
\label{shelicity}

If we define the supercurrent as,
\begin{equation}
   j_{\mu}=J_\perp\eta_\mu^{-2}v_{\mu}=J_\perp\eta_\mu^{-2}(
   \nabla_\mu\theta-A_\mu^{\rm ext}-A_\mu^{\rm ind})\enspace,
\label{ej}
\end{equation}
then from Eqs.(\ref{eLG4}) and (\ref{ej}) we see that
\begin{equation}
  \langle j_{q\mu}\rangle=-{\cal V}\left\langle{\partial {\cal H}\over\partial 
   A_{-q\mu}^{\rm ext}}\right\rangle=-{\cal V}{\partial{\cal F}\over\partial 
   A_{-q\mu}^{\rm ext}},
\label{ejave}
\end{equation}
where ${\cal F}=-T\ln {\cal Z}$ is the total free energy for the partition
function ${\cal Z}=\int {\cal D}\theta{\cal D}{\bf A}^{\rm ind}e^{-{\cal H}/T}$.  

Consider now a small perturbation about a uniform applied magnetic
field $h_0\hat{\bf z}$, with ${\bf A}^{\rm ext}=2\pi h_0x\hat{\bf y}
+\delta{\bf A}^{\rm ext}$.
We define the helicity modulus $\Upsilon_{\mu\nu}({\bf q})$
as the linear response coefficient between the induced supercurrent
and the perturbation $\delta{\bf A}^{\rm ext}$,  
\begin{equation}
\langle j_{q\mu}\rangle\equiv -\Upsilon_{\mu\nu}({\bf q})\delta A_{q\nu}^{\rm 
ext}
\label{ehm1}
\end{equation}
From Eqs.(\ref{ejave}) and (\ref{ehm1}) we then have
\begin{eqnarray}
   \Upsilon_{\mu\nu}({\bf q}) & = &
   -\left.{\partial\langle j_{q\mu}\rangle\over
   \partial A_{q\nu}^{\rm ext}}\right|_0
   ={\cal V}\left.{\partial^2{\cal F}
   \over\partial A_{q\nu}^{\rm ext}\partial A_{-q\mu}^{\rm ext}}
   \right|_0
\nonumber\\   
   &=&
   {\cal V}\left\langle{\partial^2{\cal H}\over\partial A_{q\nu}^{\rm ext}
   \partial A_{-q\mu}^{\rm ext}}\right\rangle_0
   -\,{{\cal V}\over T}\left\{\left\langle{\partial
   {\cal H}\over\partial A_{q\nu}^{\rm ext}}\,{\partial{\cal H}\over\partial 
   A_{-q\mu}^{\rm ext}}\right\rangle_0-\left\langle{\partial
   {\cal H}\over\partial A_{q\nu}^{\rm ext}}\right\rangle_0\left\langle
   {\partial{\cal H}\over\partial A_{-q\mu}^{\rm ext}}\right\rangle_0\right\},
\label{ehm2}
\end{eqnarray}
where the subscript ``$0$'' denotes the unperturbed 
system with $\delta {\bf A}_q^{\rm ext}=0$.  For a uniform system,
the third term on the right hand side of Eq.(\ref{ehm2})
may be ignored as $\langle j_{q\mu}
\rangle_0=-{\cal V}\langle\partial{\cal H}/
\partial A_{-q\mu}^{\rm ext}\rangle_0=0$ (for
the mixed state, we are assuming that $q$ is smaller than
any of the reciprocal lattice vectors of the vortex lattice).

Applying Eq.(\ref{ehm2}) to the Hamiltonian (\ref{eLG4}), and
using the definition of Eq.(\ref{ej}), then results in,
\begin{equation}
   \Upsilon_{\mu\nu}({\bf q})=J_\perp\eta_\mu^{-2}\left[\delta_{\mu\nu}
   -{J_\perp\eta_\mu\over {\cal V}T\eta_\nu}\left\langle v_{q\mu}
   v_{-q\nu}\right\rangle_0\right],
\label{ehm3}
\end{equation}
The form of Eq.(\ref{ehm3}), expressing $\Upsilon_{\mu\nu}({\bf q})$
in terms of a velocity correlation, is familiar as defining
the superfluid density of a neutral superfluid, or equivalently
the helicity modulus of an XY model.\cite{R7}

Alternatively, we could apply the results of Eqs.(\ref{ejave}) and
(\ref{ehm2}) to the form of ${\cal H}$ in Eq.(\ref{eLG1}) to get,
\begin{equation}
   \langle j_{q\mu}\rangle=-J_\perp\lambda_\perp^2\Big\langle [{\bf q}
   \times({\bf q}\times{\bf A}_q^{\rm ind})]_\mu\Big\rangle
   =-2\pi iJ_\perp\lambda_\perp^2\Big\langle[{\bf q}\times
   {\bf b}^{\rm ind}_q]_\mu\Big\rangle\enspace,
\label{eAmp}
\end{equation}
and,
\begin{equation}
  \Upsilon_{\mu\nu}({\bf q})=J_\perp\lambda_\perp^2\left[q^2
  \delta_{\mu\nu}-q_\mu q_\nu -{4\pi^2J_\perp\lambda_\perp^2\over
  {\cal V}T}\Big\langle [{\bf q}\times {\bf b}_q]_\mu[{\bf q}\times
  {\bf b}_{-q}]_\nu\Big\rangle_0\right],
\label{ehm4}
\end{equation}
where we have used ${\bf b}_q={\bf b}_q^{\rm ind}$ for the unperturbed
system, and $\langle {\bf b}_q\rangle_0=0$ for finite 
small ${\bf q}$.
Eq.(\ref{eAmp}) is just a statement of Amp\`{e}re's law, relating the induced
magnetic field to the flowing supercurrents.  Eq.(\ref{ehm4}) 
expresses $\Upsilon_{\mu\nu}({\bf q})$ 
in an explicitly gauge invariant form, in terms of correlations of the
fluctuating internal magnetic field ${\bf b}$.  

Finally, we can also express $\Upsilon_{\mu\nu}({\bf q})$ in terms
of vortex line correlations.  Using the form of ${\cal H}$ in 
Eq.(\ref{eLG6}), substituting in $2\pi{\bf h}_q=-i{\bf q}
\times{\bf A}_q^{\rm ext}$, and taking the appropriate derivatives
as in Eq.(\ref{ehm2}), results in,
\begin{equation}
\Upsilon_{\mu\nu}({\bf q})=J_\perp\Bigg[({\bf q}\times\hat{\bf\nu})
\cdot{\bf V}_q\cdot({\bf q}\times\hat{\bf\mu})-{4\pi^2 J_\perp
\over {\cal V}T}({\bf q}\times\hat{\bf\nu})\cdot{\bf V}_q\cdot
\left\langle{\bf n}_{-q}{\bf n}_q\right\rangle_0\cdot{\bf V}_q\cdot
({\bf q}\times\hat{\bf\mu})\Bigg]
\label{ehm5}
\end{equation}
where ${\bf V}_q$ is the vortex line interaction tensor of either
Eqs.(\ref{eVq}) or (\ref{eVdiag}).

Note that the helicity modulus is Hermitian, $\Upsilon_{\mu\nu}({\bf q})
=\Upsilon_{\nu\mu}^*({\bf q})=\Upsilon_{\nu\mu}(-{\bf q})$.  Also note
that any longitudinal component of $\delta {\bf A}^{\rm ext}_q$ produces no
response in $\langle {\bf j}_q\rangle$, since ${\bf \Upsilon}({\bf q})
\cdot{\bf q}=0$.  This is as expected since a longitudinal component
of $\delta{\bf A}^{\rm ext}_q$ produces no magnetic field, and can be eliminated 
by a gauge transformation.  Henceforth it will be simplest to work in
the London gauge in which ${\bf q}\cdot{\bf A}^{ext}_q=0$.

The tensor products in Eq.(\ref{ehm5}) can be simplified greatly if we restrict
our interest to wavevectors lying along the symmetry directions, i.e.
${\bf q}=q{\bf\hat x}$, $q{\bf\hat y}$, and $q{\bf\hat z}$.  
Changing notation for the sake of clarity, from ${\bf n}_q$ to 
${\bf n}({\bf q})$, and using Eq.(\ref{eVdiag}) for ${\bf V}_q$,
we find for the diagonal elements
\begin{equation}
   \Upsilon_\mu(q{\bf\hat\nu})\equiv \Upsilon_{\mu\mu}(q{\bf\hat\nu})=
   {J_\perp\lambda_\perp^2q^2\over 1+\lambda_\mu^2q^2}
   \left[1-{4\pi^2J_\perp\lambda_\perp^2
   \over {\cal V}T}{\langle n_\sigma(q{\bf\hat\nu})n_\sigma(-q{\bf\hat\nu})
   \rangle_0\over 1+\lambda_\mu^2 q^2}\right],
\label{ehm6}
\end{equation}
where $\mu$, $\nu$, $\sigma$ are any cyclic permutation of $x$, $y$, $z$, and
$\lambda_\mu$ is either $\lambda_z$ or $\lambda_\perp$ depending on
whether $\mu=z$ or $\mu=x$, $y$.  Note that 
$\Upsilon_\mu(q\hat{\bf\nu})\sim q^2$ as $q\to 0$.

The off diagonal elements are
\begin{equation}
   \Upsilon_{\mu\nu}(q{\bf\hat\sigma})={J_\perp\lambda_\perp^2q^2\over 
   1+\lambda_\mu^2q^2}\left[
   {4\pi^2J_\perp\lambda_\perp^2\over {\cal V}T}{\langle 
   n_\nu(q{\bf\hat\sigma})n_\mu(-q{\bf\hat\sigma})\rangle_0\over
   1+\lambda_\nu^2 q^2}\right].
\label{ehm7}
\end{equation}
However for ${\bf q}=q{\bf\hat\sigma}$, ${\bf q}\cdot{\bf 
n}_q=0$ implies that $n_\mu(q{\bf\hat\sigma})$ and 
$n_\nu(q{\bf\hat\sigma})$ 
fluctuate without 
constraint, and since the free energy of Eq.(\ref{eLG7}) is symmetric
separately in $n_\mu\to -n_\mu$ and in $n_\nu\to -n_\nu$, we
expect that the  off diagonal terms will vanish.

Henceforth we will restrict ourselves to the simple cases
given by Eq.(\ref{ehm6}). For a uniform applied magnetic
field along the $\hat{\bf z}$ direction, and taking 
here and henceforth $\mu,\nu,\sigma$ 
to be a cyclic permutation of $x,y,z$, we have the
three distinct cases, ($a$) $\Upsilon_y(q{\bf\hat z})$, ($b$)
$\Upsilon_x(q{\bf\hat y})$, and ($c$) $\Upsilon_z(q{\bf\hat x})$.
In Fig.\,\ref{f1} we show a schematic of the magnetic field lines
corresponding to these three different perturbations.  
As suggested by these diagrams, we will refer to ($a$) as the
$tilt$ perturbation, ($b$) as the $compression$ perturbation,
and ($c$) as the $shear$ perturbation.  We will find that the
first two cases are determined by the transverse and
longitudinal magnetic susceptibilities respectively.  We will find
that the presence of a perfect Meissner effect with respect to the
shear perturbation is a convenient measure of superconducting
coherence for the mixed state.  Because the screening currents
involved in the shear perturbation run parallel to the applied
magnetic field, a perfect Meissner effect for the shear perturbation
has also been termed {\it longitudinal superconductivity}.\cite{R8}

\subsection{Magnetic Susceptibilities and Renormalized Penetration 
Lengths}
\label{ssucs}

As indicated above, the helicity modulus $\Upsilon_{\mu\nu}({\bf q})$
is closely related to the magnetic susceptibility.  Combining
Amp\`{e}re's Law (\ref{eAmp}) with the definition
of $\Upsilon_{\mu\nu}({\bf q})$ in Eq.(\ref{ehm1}), we have
\begin{equation}
  \left\langle\delta{\bf A}_q^{\rm ind}\right\rangle =
  -{{\bf\Upsilon}({\bf q})\over J_\perp\lambda_\perp^2 q^2}\cdot
  \delta{\bf A}_q^{\rm ext}\enspace.
\label{eAtot}
\end{equation}
For the three cases of Eq.(\ref{ehm6}), corresponding to perturbations
$\delta{\bf A}_\mu^{\rm ext}(q\hat{\bf\nu})$ where $\mu,\nu,\sigma$ is
a cyclic permutation of $x,y,z$, ${\bf\Upsilon}({\bf q})$
is diagonal and so we can substitute into the above $2\pi \delta
b^{\rm ind}_\sigma(q\hat{\bf \nu})=-iq\delta A^{\rm ind}_\mu(q\hat{\bf\nu})$ and
$2\pi \delta h_\sigma(q\hat{\bf\nu})=-iq\delta A_\mu^{\rm ext}
(q\hat{\bf\nu})$ to get,
\begin{equation}
  -{\Upsilon_\mu(q\hat{\bf\nu})\over J_\perp\lambda_\perp^2 q^2}=
  \left.{\partial\langle b_\sigma^{\rm ind}(q\hat{\bf\nu})\rangle\over
  \partial h_\sigma(q\hat{\bf\nu})}\right|_0\equiv 4\pi\chi_\sigma
  (q\hat{\bf\nu})\enspace,
\label{echi}
\end{equation}
where ${\bf b}^{\rm ind}_q/4\pi$ is the induced
magnetization, and $\chi_\sigma(q\hat{\bf\nu})$ is the
magnetic susceptibility at wavevector $q\hat{\bf\nu}$ for a 
perturbation in applied magnetic field in direction $\hat{\bf\sigma}$.

To get a feel for the information contained in the helicity modulus,
or equivalently the magnetic susceptibility, consider first the
case of zero field, in the absence of vortex line fluctuations.
When ${\bf n}_q=0$, Eq.(\ref{ehm6}) yields 
$\Upsilon_\mu=J_\perp\lambda_\perp^2q^2/
(1+\lambda_\mu^2q^2)$.  Substituting into Eq.(\ref{echi}) 
gives,
\begin{equation}
  4\pi\chi_\sigma(q\hat{\bf\nu})={-1\over 1+\lambda_\mu^2q^2}
\label{eAtot2}\enspace.
\end{equation}
This describes a perfect Meissner effect.
As $q\to 0$, $\langle\delta b_\sigma^{\rm ind}(q\hat{\bf\nu})\rangle
=4\pi\chi_\sigma(q\hat{\bf\nu})\delta 
h_\sigma(q\hat{\bf\nu})=-\delta h_\sigma(q\hat{\bf\nu})$, and so the
total field inside the superconductor, $\langle \delta b_\sigma\rangle=
\delta h_\sigma+\langle \delta b_\sigma^{\rm ind}\rangle$ vanishes.
The perturbation $\delta h_\sigma$ is completely
expelled from the system.  The length scale on which this
expulsion takes place is $\lambda_\mu$.

In the presence of vortex line fluctuations, we can write
a phenomenological form for the helicity modulus at small $q$,
\begin{equation}
   \Upsilon_\mu(q{\bf\hat\nu})=\gamma_\mu\,{J_\perp\lambda_\perp^2q^2
   \over 1+\lambda_{\mu{\rm R}}^2q^2}\enspace.
\label{ehm8}
\end{equation}
In this case, substituting into Eq.(\ref{echi}) one gets,
\begin{equation}
  4\pi\chi_\sigma(q\hat{\bf\nu})={-\gamma_\mu\over 1+\lambda_{\mu{\rm R}}^2
  q^2}\enspace.
\label{eAtot3}
\end{equation}
Now only a fraction $\gamma_\mu$ of the applied perturbation is
expelled from the system; this expulsion takes place on the
length scale $\lambda_{\mu{\rm R}}$.  We thus see that $-\gamma_\mu$
gives the long wavelength magnetic susceptibility, while
$\lambda_{\mu{\rm R}}$ is the magnetic penetration length, as renormalized
by vortex fluctuations.  $\gamma_\mu$ and $\lambda_{\mu{\rm R}}$ are
the important physical parameters to be extracted from the
helicity modulus.

Formally, we can define $\gamma_\mu$ and $\lambda_{\mu{\rm R}}$ in terms
of the small $q$ expansion of the vortex line correlation that appears
in Eq.(\ref{ehm6}).  If we define,
\begin{equation}
   \langle n_\sigma(q{\bf\hat\nu})n_\sigma(-q{\bf\hat\nu})
   \rangle_0\equiv n_{\mu 0} + n_{\mu 1}q^2+n_{\mu 2}q^4+...\enspace,
\label{enn}
\end{equation}
then we have,
\begin{equation}
   \gamma_\mu \equiv -\lim_{q\to 0}4\pi\chi_\sigma(q\hat{\bf\nu})=
   1-{4\pi^2J_\perp\lambda_\perp^2\over {\cal V} T}n_{\mu 0}\enspace,
\label{eRg}
\end{equation}
and,
\begin{equation}
   \\ {\lambda_{\mu{\rm R}}^2\over \lambda_\mu^2} \equiv -\lim_{q\to 0}
   \left[{1\over\lambda_\mu^2 \chi_\sigma(q\hat{\bf\nu})}
   \,{d\chi_\sigma(q\hat{\bf\nu})\over dq^2} 
   \right]=1-{4\pi^2J_\perp
   \lambda_\perp^2\over {\cal V}T}\,{(n_{\mu 0}-n_{\mu 1}\lambda_\mu^{-2})
   \over\gamma_\mu}\enspace.
\label{eRl}
\end{equation}
Thus $\gamma_\mu=1$, or equivalently $n_{\mu 0}=0$,
signals a perfect Meissner screening of the perturbation
$\delta A_\mu^{\rm ext}(q\hat{\bf\nu})$.  For zero applied
magnetic field, this has a simple physical interpretation: one is
in the Meissner state if there are no infinitely large vortex rings.

Although the helicity modulus should have the form of
Eq.(\ref{ehm8}) both below and above the superconducting transition, due
to the presence of ordinary fluctuation diamagnetism above the
transition, we expect that a phase transition will be indicated by
singular behavior in the parameters $\gamma_\mu$ and $\lambda_{\mu{\rm R}}$.
In particular, a transition from a state with perfect a Meissner screening
of the perturbation $\delta A_\mu^{\rm ext}(q\hat{\bf\nu})$
will be signaled by a singular decrease of $\gamma_\mu$ from unity,
as well, presumably, by a divergence in $\lambda^2_{\mu{\rm R}}$.
For such a case, it is reasonable to interpret $n_{s}\equiv m_\mu c^2/4\pi e^2
\lambda_{\mu{\rm R}}^2$ as the density of superconducting electrons.

We stress at this point that 
$\gamma_\mu$ and $\lambda_{\mu{\rm R}}$ are describing the response of
the system to a small spatially varying perturbation about a
uniform applied field and not the response to this uniform field itself.  

\subsection{2D Boson Analogy}
\label{sanalogy}

Much work on vortex line fluctuations has been 
done utilizing an analogy between the magnetic field induced
vortex lines in the mixed phase of a three dimensional
superconductor and the imaginary time world lines of two dimensional
bosons within a Feynman path integral description of 2D quantum 
mechanics.  Here we will show the explicit connection between 
the superfluid density of these analog 2D bosons and the helicity
modulus $\Upsilon_z(q{\bf\hat x})$ giving the response to the
$shear$ perturbation of Fig.\,\ref{f1}$c$.  

In this analogy, as introduced by Nelson,\cite{R11} 
the energy of vortex line fluctuations is modeled by two pieces:
($i$) a line tension representing the vortex core energy and self
interaction, and ($ii$) a pairwise interaction between all vortex
line segments which lie in the same $xy$ plane at equal heights $z$.
This simplified vortex interaction is expected to be a reasonable
approximation when the vortex lines remain, over the length scale
$\lambda$, approximately parallel to the applied field.
This simplified vortex interaction is then mapped into
a 2D boson mass and an instantaneous pairwise boson 
interaction. The mapping results in the 
following correspondences (quantities on the left refer to the 2D 
bosons, those on the right refer to the 3D superconductor):
imaginary time, $\tau\leftrightarrow z$, distance in direction of
applied field ${\bf h}$; $\hbar_{\rm boson}/T_{\rm boson}\leftrightarrow L_z$,
length of system parallel to ${\bf h}$; 
$\hbar_{\rm boson}\leftrightarrow T$, temperature of 3D superconductor; 
boson mass, $m_{\rm boson}\leftrightarrow \tilde\epsilon_1\sim\pi J_z$,
where $\tilde\epsilon_1=\eta^{-2}\epsilon_1$, and $\epsilon_1$
is the single vortex line tension.

In Appendix \ref{s2Db}, starting from the standard definition of the
superfluid density as the long wavelength limit of the transverse
momentum susceptibility,\cite{R19,R19.1} we derive an expression Eq.(\ref{ec8}) for the 
number density of superfluid bosons, $\rho_{s\,{\rm boson}}$, for a system
of 2D interacting bosons, expressed in the form of a path integral over
boson world lines.  In Eqs.(\ref{ec9}) and (\ref{ec10}) we show
that $\rho_{s\,{\rm boson}}$ is related to the 
helicity modulus of the 2D bosons, $\Upsilon_{\rm boson}(q)$, 
by $\lim_{q\to 0}\Upsilon_{\rm boson}(q)=(\hbar^2_{\rm boson}
/m_{\rm boson})\rho_{s\,{\rm boson}}$.
We now recast the results of Appendix \ref{s2Db} 
into the language of vortex lines.

For a magnetic field induced 
vortex line $i$ parameterized by its transverse deflection ${\bf 
r}_{i\perp}(z)$ in the $xy$ plane at height $z$, the vortex line 
density is given by,
\begin{equation}
{\bf n}({\bf r}_\perp,z)=\sum_i\delta^{(2)}({\bf r}_\perp-{\bf r}_
{i\perp}(z))\left[\hat{\bf z}+{d{\bf r}_{i\perp}(z)\over dz}\right]
\enspace.
\label{enu}
\end{equation}
Using the above correspondences between the analog bosons and
the superconductor, we then have for the term that appears in the
boson path integral of Eq.(\ref{ec9}),
\begin{equation}
\int_0^{\hbar_{\rm boson}/T_{\rm boson}}d\tau\sum_i
{dr_{iy}\over d\tau}e^{iqx_i}
=n_y(q\hat{\bf x})\enspace.
\label{enr2}
\end{equation}
Eq.(\ref{ec10}) for the 2D boson helicity modulus can then be written as,
\begin{equation}
   {\Upsilon_{\rm boson}(q)\over T_{\rm boson}}=
  {1\over L_\perp^2}\langle n_y(q{\bf\hat x})n_y(-q{\bf\hat x})\rangle_0
  \enspace,
\label{e2D1}
\end{equation}
where $L_\perp$ is the length of the system in the $xy$ plane.
The vortex correlation that appears in Eq.(\ref{e2D1}) above is
precisely the same correlation that enters  Eq.(\ref{ehm6}) 
for $\Upsilon_z(q\hat{\bf x})$, which gives the response to the
shear perturbation of Fig.\,\ref{f1}$c$.
Taking the limit $q\to 0$ in Eq.(\ref{e2D1}) and combining with 
Eqs.(\ref{enn}), (\ref{eRg}) and (\ref{ec10}) then gives,
\begin{equation}
   \gamma_z=1-{4\pi^2 J_\perp\lambda_\perp^2\over L_z T}
   \left[{\Upsilon_{\rm boson}(q\to 0)\over T_{\rm boson}}\right]
   =1-{4\pi^2 J_\perp\lambda_\perp^2\over L_z T}
   \left[{\hbar^2_{\rm boson}\rho_{s\,{\rm boson}}\over m_{\rm boson}
   T_{\rm boson}}\right]
   \enspace.
\label{e2D2}
\end{equation}

This leads to the
following identifications, originally pointed out
by Feigelman and co-workers:\cite{R20} the 2D boson superfluid phase with
$\rho_{s\,{\rm boson}}>0$ corresponds to a 3D vortex line 
normal diamagnetic phase with $\gamma_z<1$; the 2D boson normal fluid phase
with $\rho_{s\,{\rm boson}}=0$ corresponds 
to a 3D vortex line phase with $\gamma_z=1$, 
and hence with longitudinal superconductivity characterized by
a perfect Meissner effect for shear perturbations.

Having made the above observation, there now exists the possibility,
as first suggested by Nelson,\cite{R11} that a Kosterlitz--Thouless (KT)
transition from superfluid to normal fluid in the analog 2D boson system, 
could appear in the
3D superconductor as a transition from a normal vortex line
liquid state to a vortex line liquid with longitudinal 
superconductivity.  
Fisher and Lee,\cite{R21} and more recently T\"{a}uber and 
Nelson,\cite{R22}
have argued that if one relaxes the periodic boundary conditions
along $\hat{\bf z}$ that is assumed in the boson analogy, and uses
instead the free boundary conditions which are more realistic
for a bulk 3D superconductor, 
the sharp KT transition no longer exists.  Nevertheless, one might
expect that a clear cross-over remnant of this KT transition should
still be observable in the superconductor.  We will return to
discuss this KT cross-over in Sec.\,\ref{sboson}.

\subsection{$\lambda\to\infty$ Approximation}
\label{slaminf}

Many numerical simulations,\cite{R23,R24,R25,R26,R27,R28,R28.1}
as well as other theoretical
approaches such as the ``lowest Landau level'' approximation,\cite{R29}
have been based upon the approximation of taking 
$\lambda_\perp\to\infty$, while keeping $J_\perp$ finite.
This approximation corresponds to taking a spatially uniform internal
magnetic field ${\bf b}$ which is equal to the applied field ${\bf 
h}$.  Such an approximation can be shown to be exact for modeling
the analog system of a 3D neutral (uncharged particles) superfluid in a 
rotating bucket.\cite{R4,R19}  It is interesting to see how the 
helicity modulus and the 2D boson analogy look within this 
$\lambda\to\infty$ limit.

In this case, the interaction between vortex lines is given by
the ``bare'' interaction tensor ${\bf V}_q^0$ of 
Eq.(\ref{eVq0}).  One can show that the correct helicity modulus 
$\Upsilon_\mu(q\hat{\bf\nu})$ is obtained by taking
the limit $\lambda_\mu\to\infty$ in Eq.(\ref{ehm6}), keeping
$J_\perp$ and $\lambda_\perp/\lambda_\mu$ constant,
\begin{equation}
  \Upsilon_\mu(q\hat{\bf\nu})=J_\mu\left[1-{4\pi^2J_\mu\over{\cal V}T}\,
  {\langle n_\sigma(q\hat{\bf\nu})n_\sigma(-q\hat{\bf\nu})\rangle_0
  \over q^2}\right]\enspace,
\label{eli1}
\end{equation}
where $J_\mu\equiv J_\perp(\lambda_\perp/\lambda_\mu)^2$ is the
coupling in direction $\hat{\bf\mu}$.

Noting that ${\cal H}_{\rm v}=(4\pi^2J_\perp/2{\cal V})\sum_q {\bf n}_q
\cdot{\bf V}_q^0\cdot{\bf n}_{-q}$ must have a finite thermal
average, and since ${\bf V}_q^0\sim 1/q^2$,
it must therefore be true that as $q\to 0$, 
\begin{equation}
  \langle n_\sigma(q\hat{\bf\nu})
  n_\sigma(-q\hat{\bf\nu})\rangle_0\sim q^2\enspace.
\label{eli2}
\end{equation}
Substituting Eq.(\ref{eli2}) into Eq.(\ref{eli1}) we see that,
in contrast to the finite $\lambda_\mu$ case where we found
$\Upsilon_\mu(q\hat{\bf\nu})\sim q^2$ as $q\to 0$, here we find
that $\lim_{q\to 0}\Upsilon_\mu(q\hat{\bf\nu})$ is in general a
finite number.  This differing dependence of the helicity modulus
on $q$, in the small $q$
limit, is one of the characteristic differences between a charged
superfluid (with a finite $\lambda$ giving a coupling to a fluctuating
vector potential) and a neutral superfluid (with $\lambda\to
\infty$ leading to a frozen vector potential).  For the $\lambda\to\infty$
case of a neutral superfluid, $\Upsilon_\mu(0)\equiv\lim_{q\to 
0}\Upsilon_\mu(q\hat{\bf\nu})$ is just proportional to the number
density of superfluid particles, as discussed in Appendix \ref{s2Db}
for the two dimensional case, and used in the preceding
section.  We therefore expect to find
$\Upsilon_z(0)>0$ for $T<T_{\rm c}$ in an ordered phase with longitudinal
superconductivity, and $\Upsilon_z(0)=0$ for $T>T_{\rm c}$ in the normal
phase.

We now consider the 2D boson analogy for this $\lambda\to\infty$
approximation.  Combining
Eq.(\ref{e2D1}) for the helicity modulus $\Upsilon_{\rm boson}(q)$ of the
analog bosons with Eq.(\ref{eli2}), we see that
\begin{equation}
\Upsilon_{\rm boson}(q)\sim q^2\qquad {\rm for\ small\ } q\enspace.  
\label{eli2.1}
\end{equation}
Thus the analog 2D bosons
have a helicity modulus characteristic of a 2D $charged$
superfluid!  This is in agreement with the results of Feigelman 
and co-workers,\cite{R20}
who show that the system of vortex lines interacting with the
true London interaction of Eqs.(\ref{eLG7}) and (\ref{eVdiag})
(as opposed to the more simplified
interaction of Nelson's model) can be viewed as a system of
analog 2D bosons whose interaction is mediated by a massive
vector potential.  As $\lambda\to\infty$, the mass associated
with this vector potential vanishes, and one has a system of
2D charged bosons interacting with 2D electrodynamics.  

We can develop the analogy further. Combining Eqs.(\ref{e2D1}) with
(\ref{eli1}) we have,
\begin{equation}
  \Upsilon_z(q\hat{\bf x})=J_z\left[1-{4\pi^2J_z\over L_zTq^2}\,
  {\Upsilon_{\rm boson}(q)\over T_{\rm boson}}\right]
\label{eli3}
\end{equation}
One can then define the proportionality coefficient
$\gamma_{\rm boson}$ of Eq.(\ref{eli2.1}) by,
\begin{equation}
  \Upsilon_{\rm boson}(q)=\gamma_{\rm boson}T_{\rm boson}{L_zTq^2\over 4\pi^2J_z}
  =\gamma_{\rm boson}\left[{\hbar^2_{\rm boson}\over 4\pi^2J_z}\right]q^2\enspace,
\label{eli4}
\end{equation}
where we have used the correspondences between superconducting
variables and analog boson variables to arrive at the last equality.
Note that when $\gamma_{\rm boson}=1$ we have $\Upsilon_z(0)=0$, and when
$\gamma_{\rm boson}<1$ we have $\Upsilon_z(0)>0$.

One can now show,\cite{R30} at least in the isotropic case, that the
term $[\hbar^2_{\rm boson}/4\pi^2 J]=[\hbar^2_{\rm 
boson}4\pi\lambda^2/\phi_0^2]$ which appears on the right
hand side of Eq.(\ref{eli4}) is just twice the magnetic energy
coupling of the analog magnetic field of the 2D electrodynamics.
We can rename this coupling $[J\lambda^2]_{\rm boson}$ in analogy with
the magnetic energy coupling of our original 3D superconductor
of Eq.(\ref{eLG1}).  Eq.(\ref{eli4}) then becomes,
\begin{equation}
  \Upsilon_{\rm boson}(q)=\gamma_{\rm boson}[J\lambda^2]_{\rm boson}q^2
\label{eli5}
\end{equation}
in complete agreement with the small $q$ limit of the form
of the helicity modulus for a charged superfluid, given in Eq.(\ref{ehm8})
(as derived for our original 3D superconductor at finite $\lambda$).
$-\gamma_{\rm boson}$ is therefore the magnetic susceptibility
of the analog 2D charged bosons.
To next order in $q^2$ we expect, in analogy with Eq.(\ref{ehm8}),
that $\Upsilon_{\rm boson}(q)$ has the form,
\begin{equation}
\Upsilon_{\rm boson}(q)=\gamma_{\rm boson}{[J\lambda^2]_{\rm boson}q^2\over 1+
\lambda_{\rm R\, boson}^2q^2}\enspace,
\label{eli6}
\end{equation}
where $\lambda_{\rm R\,boson}$ is the magnetic penetration length of the
analog 2D charged bosons.  If we take $\lambda_{\rm R\, boson}^2=1/(4\pi 
n_{s\,{\rm boson}})$, with $n_{s\,{\rm boson}}$ the number density of superfluid
charged bosons, then combining Eqs.(\ref{eli3}--\ref{eli6})
one can recover all the results found in Sec.\,V\,B.3 of 
Blatter {\it et al}.,\cite{R31}
which are therefore seen to apply in a strict sense only 
to the $\lambda\to\infty$
approximation, rather than to the finite $\lambda$ case.

We thus have the following amusing duality.  
For finite $\lambda$ we have seen in the preceding section that
the 3D superconductor, which is a charged superfluid problem, maps onto
analog 2D bosons, which is a neutral superfluid problem.  
The 3D longitudinal superconductivity transition maps onto the
2D superfluid transition.
A perfect 
Meissner effect for shear perturbations in the 3D superconductor, with
$\gamma_z=1$, represents the normal fluid state of the 2D bosons
with $\Upsilon_{\rm boson}(q\to 0)=0$; the loss 
of this perfect Meissner effect, with $\gamma_z<1$, corresponds to the 
superfluid state of the 2D bosons with $\Upsilon_{\rm boson}(q\to 0)>0$.
For $\lambda\to\infty$ however, the 3D superconductor, which now behaves like
a neutral superfluid problem, maps onto 2D bosons interacting
with 2D electrodynamics, which is a charged superfluid problem.
The 3D longitudinal superconductivity transition now maps onto a Meissner 
transition of a 2D superconductor.
The normal state of the 3D superconductor, with $\Upsilon_z(q\to 
0)=0$, corresponds to a perfect Meissner state of the charged
2D bosons, with $\gamma_{\rm boson}=1$; the 3D superconducting state,
with $\Upsilon_z(q\to 0)>0$, corresponds to the loss of this
perfect Meissner effect for the 2D charged bosons, with $\gamma_{\rm 
boson}<1$.

Note that for the analog 2D charged bosons of the $\lambda\to\infty$
approximation, vortices in the 2D
condensate wavefunction will interact with a potential that
decays exponentially on length scales greater than $\lambda_{\rm
R\, boson}$, due to the screening by the
2D analog magnetic field.  A vortex anti-vortex pair
will therefore have a finite energy barrier for unbinding, and
so at any finite $T_{\rm boson}$ there must be free vortices which
will destroy the 2D Meissner state.  Only at $T_{\rm boson}=0$ ($L_z\to
\infty$), does there remain the possibility of a sharp Meissner
transition in this 2D analog boson system, as $\hbar_{\rm boson}$ 
varies.  Such a
transition, if it exists, is driven by quantum and not thermal fluctuations
and so it is not in the Kosterlitz--Thouless universality class.
We believe that it is this transition at $T_{\rm boson}=0$, in
the $\lambda\to\infty$ model, that the work of Feigelman and 
co-workers\cite{R20} pertains to.   Recently, Te\v{s}anovi\'c
has argued\cite{R32} that such a transition must be driven by the proliferation
of closed vortex rings (boson anti-boson virtual pairs), which are left
out of the naive 2D boson mapping, and that the transition will be in the
universality class of the ordinary 3D XY model.

The above considerations suggest that taking the  
$\lambda\to\infty$ limit in our model is rather subtle and
possibly leads to discontinuous changes in the phase diagram, although
any such discontinuities will likely be obscured in a finite size
system by very strong cross-over effects.

\section{The Vortex Line Lattice: Elastic Approximation}
\label{selastic}

We consider now the mixed state of a type II superconductor. 
At low temperatures, such a state is described by the familiar
Abrikosov vortex line lattice. 
In this case, we can evaluate the vortex line correlations
that appear in the expression for the helicity modulus by using
the well known elastic approximation.\cite{R12,R12.1} 
It is now convenient to work in the Helmholtz ensemble, with a fixed
uniform density $b_0$ of magnetic field induced vortex lines,
\begin{equation}
  b_0={B\over\phi_0}\enspace, \qquad a_{\rm v}
  =\left({\textstyle{4\over3}}\right)^{1/4}{1\over\sqrt{b_0}}\enspace
\label{ef}
\end{equation}
where $a_{\rm v}$ is the lattice spacing between lines in their ground state
triangular lattice.
We will denote thermal averages in this ensemble by $\langle ...\rangle$,
dropping the subscript $``0"$ that we used earlier.

In the elastic approximation, one assumes that vortex line excitations
consist only of fluctuations of the magnetic field induced vortex lines,
transverse to the direction of the uniform applied field.  Such
fluctuations are described by the displacement field
${\bf u}_i(z)$, which gives the transverse displacement in the
$xy$ plane at height $z$, of the vortex line away from its 
position ${\bf R}_i$ in the ground state vortex lattice.  
The vortex line density is thus given by Eq.(\ref{enu}),
making the substitution ${\bf r}_{i\perp}(z)={\bf R}_i+{\bf u}_i(z)$.

If we define the Fourier transforms
\begin{equation}
   {\bf u}_q={1\over b_0}\int dz\sum_i e^{i(q_z z+{\bf q}_\perp\cdot 
   {\bf R}_i)}
   {\bf u}_i(z),\qquad {\bf u}_i(z)={1\over{\cal V}}\sum_q e^{-i(q_z z+
   {\bf q}_\perp\cdot {\bf R}_i)}{\bf u}_q\enspace,
\label{euFT}
\end{equation}
where ${\bf q}_\perp=(q_x,q_y)$ and the sum over ${\bf q}_\perp$ is 
restricted to the first Brillouin
Zone of the Abrikosov lattice, then to lowest order in ${\bf u}$ the 
vortex line density at small finite ${\bf q}$ may be written as
\begin{equation}
   {\bf n}_q=ib_0[{\bf q}\cdot{\bf u}_q
   {\bf\hat z}-q_z{\bf u}_q]\enspace.
\label{enuq}
\end{equation}
Substituting the expansion for ${\bf n}_q$ in terms of the 
${\bf u}_q$ into Eq.(\ref{eLG7}), summing over reciprocal 
lattice vectors,
and  keeping only terms up to order $u^2_q$, results in the
free energy functional for elastic vortex line displacements,
\begin{eqnarray}
   {\cal H}_{\rm el}[{\bf u}] & = & {1\over 2{\cal V}}\sum_{q\alpha\beta} 
   u_{q\alpha}\Phi_{\alpha\beta}({\bf q})u_{-q\beta}
   \nonumber\\
   & = & {1\over 2{\cal V}}\sum_q\Bigg\{[c_{44}({\bf q})q_z^2
   +c_{11}({\bf q})q_\perp^2]u_{qL}u_{-qL} + [c_{44}({\bf q})q_z^2
   +c_{66}({\bf q})q_\perp^2]u_{qT}u_{-qT}\Bigg\}\enspace.
\label{eHel}
\end{eqnarray}
Here $u_{qL}={\bf\hat q}\cdot{\bf u}_q$ 
is the longitudinal part of the
displacement, and $u_{qT}=|{\bf u}_q
-{\bf\hat q}u_{qL}|$ is the transverse
part.  $c_{44}({\bf q})$, $c_{66}({\bf q})$, and $c_{11}({\bf q})$ are the
wavevector dependent tilt, shear, and compression elastic moduli respectively.
We can now use Eqs. (\ref{enuq}) and ({\ref{eHel}) to evaluate the vortex
correlations of the helicity modulus of Eq.(\ref{ehm6}), for the 
three simple cases of Fig.\,\ref{f1}.  

\subsection{Tilt Perturbation}
\label{sstilt}

We first consider $\Upsilon_y(q{\bf\hat z})$ which
gives the response to the tilt perturbation of Fig.\,\ref{f1}$a$.
Using Eq.(\ref{enuq}), the relevant vortex correlation, to lowest
order in ${\bf u}_q$, is
\begin{equation}
  \langle n_x(q{\bf\hat z})n_x(-q{\bf\hat z})\rangle = q^2b_0^2
  \langle u_T(q{\bf\hat z})u_T(-q{\bf\hat z})\rangle = 
  {b_0^2{\cal V}T\over c_{44}(q{\bf\hat z})}\enspace,
\label{entilt}
\end{equation}
where we have used ${\cal H}_{\rm el}$ of Eq.(\ref{eHel}) to evaluate
the displacement correlation.
Expanding in $q^2$ we have,
\begin{equation}
   n_{y0}={b_0^2{\cal V}T\over c_{44}(0)},
   \qquad n_{y1}=-\,{b_0^2{\cal V}T\over c_{44}^2(0)}
   \left. {dc_{44}\over dq^2_z}\right|_{q=0}\enspace.
\label{enx01}
\end{equation}

Combining Eqs.(\ref{eJperp}), (\ref{eRg}), (\ref{eRl}), 
and (\ref{ef}) with (\ref{enx01})
above then gives for the helicity modulus parameters,
\begin{equation}
   \gamma_y=1-{B^2\over 4\pi c_{44}(0)}\enspace,
\label{ehmtiltg}
\end{equation}
and,
\begin{equation}
   {\lambda_{y{\rm R}}^2\over\lambda_\perp^2}=1-{B^2\over 4\pi 
   c_{44}(0)\gamma_y}\left[1+{1\over c_{44}(0)\lambda_\perp^2}
   \left.{dc_{44}\over dq_z^2}\right|_{q=0}\right]\enspace.
\label{ehmtiltl}
\end{equation}
Note that
from general thermodynamic arguments\cite{R33} one has,
\begin{equation}
  c_{44}(0)={B^2\over 4\pi}{dH_\perp\over dB_\perp}\enspace,
\label{ec44}
\end{equation}
where the $dH_\perp/dB_\perp$ is evaluated at the average
magnetic field $B_0{\bf\hat z}$.  Hence $\gamma_y$
is determined by the transverse magnetic susceptibility,
\begin{equation}
   \gamma_y=1-{dB_\perp\over dH_\perp}\enspace,
\label{egtilt3}
\end{equation}
as expected from our discussion in Sec.\,\ref{ssucs}.

Using our explicit results for $c_{44}$ from Appendix \ref{sselmod}, 
we have,

\begin{equation}
  \gamma_y\simeq\left\{
  \begin{array}{llll}
  {\displaystyle
  {\phi_0\over 8\pi\lambda_\perp^2B}\left[\eta^{-2}\left(\ln{H_{\rm c2}\over B}
  -1\right)+{\phi_0\over 4\pi\lambda_\perp^2B}\right]} & \ll 1 \,&{\rm for\ 
  large\ B,}&\,\lambda_\perp\gg a_{\rm v}\\  
  \qquad & \qquad & \qquad &\qquad\\
  {\displaystyle
  1-{8\pi\lambda_\perp^2B\over\phi_0}{1\over \eta^{-2}\left[\ln(H_{\rm c2}/ B)-1
  \right]+1}}&  \approx 1\,&{\rm for\ intermediate\ B,}&\, 
  \lambda_\perp\ll a_{\rm v}\ll\lambda_z\\ 
  \qquad & \qquad & \qquad &\qquad\\
  {\displaystyle
  1-{8\pi\lambda_\perp^2B\over\phi_0}{1\over \eta^{-2}\left[2\ln(\eta\kappa)-1
  \right]+1}}& \approx 1\,&{\rm for\ small\ B,}&\, \lambda_z\ll a_{\rm v}\end{array}
  \right.
\label{egtilt2}
\end{equation}

For $\lambda_{y{\rm R}}$, 
using Eq.(\ref{ehmtiltl}) and our results for $c_{44}$ from 
Appendix \ref{sselmod},
we have for large magnetic fields, $\lambda_\perp\gg a_{\rm v}$, 

\begin{equation}
   {\lambda_{y{\rm R}}^2\over\lambda_\perp^2}\simeq\left\{ \begin{array}{lll}
      {\displaystyle
      {1\over 2\lambda_\perp^2k_0^2}}\> &{\rm for\ strong\ anisotropy,}\>& 
      {\displaystyle
      {1\over 2\lambda_\perp^2k_0^2}\gg \eta^{-2}\ln\left({H_{\rm c2}
      \over B}\right)}\\ 
  \qquad & \qquad & \qquad \\
      {\displaystyle
      {1\over 2\lambda_\perp^2k_0^2}\eta^{-2}\ln\left({H_{\rm c2}\over 
      B}\right)}\>
      &{\rm for\ weak\ anisotropy,}\>& {\displaystyle 
     {1\over 2\lambda_\perp^2k_0^2}\ll \eta^{-2}\ln\left({H_{\rm c2}
      \over B}\right)}\end{array}\right.
\label{eRtilt}
\end{equation}
For intermediate magnetic fields, $\lambda_\perp\ll a_{\rm v}\ll \lambda_z$,
(where strong anisotropy is by definition implied) we have

\begin{equation}
   {\lambda_{y{\rm R}}^2\over\lambda_\perp^2}\simeq 1-\lambda_\perp^2k_0^2
   \enspace ,
\label{eRtilt2}
\end{equation}
and for weak magnetic fields, $\lambda_z\ll a_{\rm v}$, we have

\begin{equation}
   {\lambda_{y{\rm R}}^2\over\lambda_\perp^2}\simeq\left\{ \begin{array}{lll}
      {\displaystyle 
      1- 2\lambda_\perp^2k_0^2}\> &{\rm for\ strong\ anisotropy,}\>& 
       {1\over 2}\gg \eta^{-2}\ln\eta\kappa\\ 
  \qquad & \qquad & \qquad \\
      {\displaystyle
      1-{\lambda_\perp^2k_0^2\over \eta^{-2}\ln\eta\kappa}}\>
      &{\rm for\ weak\ anisotropy,}\>& {1\over 2}\ll \eta^{-2}\ln\eta\kappa
      \end{array}\right.
\label{eRtilt3}
\end{equation}
where $k_0^2=4\pi B/\phi_0\sim 1/a_{\rm v}^2$.

For strong magnetic fields, $\lambda_\perp\gg a_{\rm v}$,
Eq.(\ref{eRtilt}) gives  $\lambda_{y{\rm R}}\simeq 1/\sqrt 2 k_0
\mathrel{\hbox{\hbox to 
0pt{\lower.5ex\hbox{$\sim$}\hss}\raise.4ex\hbox{$<$}}}a_{\rm v}$,
independent of the bare $\lambda_\perp$.  
Since our definition of 
$\lambda_{y{\rm R}}$ in Eq.(\ref{ehm8}) was based on an expansion in small 
$q$, it is doubtful that we should take 
such a small $\lambda_{y{\rm R}}$ too seriously as a screening length, 
without considering higher
terms in an expansion in $q$, as well as considering the response ${\bf j}_{q+K}$
to the perturbation ${\bf A}^{\rm ext}_q$ (where ${\bf K}$ is a reciprocal lattice 
vector of the vortex lattice).  

For weak magnetic fields, $\lambda_\perp\ll a_{\rm v}$, Eqs.(\ref{eRtilt2}) and 
(\ref{eRtilt3}) give, $\lambda_{y{\rm R}}\approx \lambda_\perp$.

\subsection{Compression Perturbation}
\label{sscomp}

We next consider $\Upsilon_x(q{\bf\hat y})$ which
gives the response to the compression perturbation of Fig.\,\ref{f1}$b$.
Using Eq.(\ref{enuq}), the relevant vortex correlation, to lowest
order in ${\bf u}_q$, is

\begin{equation}
  \langle n_z(q{\bf\hat y})n_z(-q{\bf\hat y})\rangle = q^2b_0^2
  \langle u_L(q{\bf\hat y})u_L(-q{\bf\hat y})\rangle = 
  {b_0^2{\cal V}T\over c_{11}(q{\bf\hat y})}\enspace,
\label{encomp}
\end{equation}
where we have used Eq.(\ref{eHel}) to evaluate
the displacement correlation.

We therefore have,
\begin{equation}
\gamma_x=1-{B^2\over 4\pi c_{11}(0)}\enspace.
\label{ecompg}
\end{equation}
The compression modulus in the vortex line lattice
can be written as $c_{11}=c_L+c_{66}$,
where $c_L$ is the ``bulk modulus'' for an isotropic compression.
General thermodynamic arguments give,\cite{R33}
\begin{equation}
   c_L(0)={B^2\over 4\pi}{dH_z\over dB_z}\enspace.
\label{ecL0}
\end{equation}
Noting from Eqs.(\ref{eAp7a}) and (\ref{eAp10b}) that, for
large $\lambda_\perp \gg a_{\rm v}$, $c_{66}(0)\ll c_{11}$,
we have $c_{11}(0)\simeq c_L(0)$, and so,
\begin{equation}
  \gamma_x\simeq 1-{dB_z\over dH_z}
\label{ecL}
\end{equation}
is determined by the longitudinal magnetic susceptibility.
For $\lambda_\perp\ll a_{\rm v}$, Eqs.(\ref{eAp10a})
and (\ref{eAp10b}) give $c_{11}={3\over 2}c_L$, and so
$\gamma_x=1-{2\over 3}(dB_z/dH_z)$.

Using our explicit results for $c_{11}$ from Appendix \ref{sselmod}, 
we find,
\begin{equation}
  \gamma_x\simeq\left\{\begin{array}{lll}{\displaystyle
  -{\phi_0\over 16\pi\lambda_\perp^2B}}&{\rm 
   for\>large\>B,}&\lambda_\perp\gg a_{\rm v}\\
  \qquad & \qquad & \qquad \\
  {\displaystyle
  -{16\sqrt{2\pi}\lambda_\perp^2B\over 9\phi_0}\left({\lambda_\perp
  \over a_{\rm v}}\right)^{3/2}e^{a_{\rm v}/\lambda_\perp}}&{\rm
  for\>small\>B,}&\lambda_\perp\ll a_{\rm v}
  \end{array}\right.
\label{egcomp}
\end{equation}
$\gamma_x<0$ implies that the magnetic field induced in the material
is $larger$ than the applied perturbation, so there is $negative$
screening.  This may be understood from Eq.(\ref{ecL}) by noting that 
in the mixed phase one always has $dB_z/dH_z>1$.  

For the screening length, we find

\begin{equation}
   {\lambda_{x{\rm R}}^2\over\lambda_\perp^2}\simeq\left\{  \begin{array}{lll}
   {\displaystyle -\,{1\over 4\lambda_\perp^2k_0^2}} & \qquad{\rm for\>large\>B,}
   & \lambda_\perp\gg a_{\rm v}\\ \\
   {\displaystyle -\,{5\,a_{\rm v}^2\over 72\lambda_\perp^2}}&\qquad{\rm for\>
   small\>B,}&\lambda_\perp\ll a_{\rm v}
   \end{array}\right.
\label{eRcomp}
\end{equation}
Since $k_0^2=4\pi B/\phi_0\sim 1/a_{\rm v}^2$,
both cases give $\lambda_{x{\rm R}}\sim ia_{\rm v}$.  It is tempting to
interpret this imaginary $\lambda_{x{\rm R}}$ as indicating the rearrangement
of vortex lines on the length scale $a_{\rm v}$ due to the penetration of
the applied field, with no ``healing'' length at all at the surface of the
sample.  However our cautionary remarks following Eq.(\ref{eRtilt3}),
concerning the applicability of our results on the length scale 
$a_{\rm v}$, should again be noted.

\subsection{Shear Perturbation}
\label{ssshear}

The preceding two cases of the tilt and the compression perturbations
gave information about the transverse and longitudinal magnetic 
susceptibilities, via $\gamma_y$ and $\gamma_x$.  However, since
for large $B$ $\lambda_{y{\rm R}}$, $\lambda_{x{\rm R}}\sim a_{\rm v}$ is 
independent of the bare $\lambda_\perp$, it
is unclear whether they give any information about the density of
superconducting electrons, or whether they can be expected to diverge
at the superconducting to normal transition.
A more interesting case is therefore given by the third possibility,
$\Upsilon_z(q{\bf\hat x})$, which
gives the response to the shear perturbation of Fig.\,\ref{f1}$c$. 

The relevant vortex correlation we need to compute is
$\langle n_y(q{\bf\hat x})n_y(-q{\bf\hat x})\rangle$, but to lowest
order in ${\bf u}_q$, Eq.(\ref{enuq}) shows this to vanish identically.
This is merely an artifact of our Fourier transform of the displacement
field ${\bf u}$, which prohibits vortex lines from having a net tilt
away from the ${\bf\hat z}$ axis. 
To avoid this difficulty we can evaluate the correlation at
${\bf q}=q_x{\bf\hat x}+q_z{\bf\hat z}$, with finite $q_z$, and
then take the limit as $q_z\to 0$.  From Eqs.(\ref{enuq}) and
(\ref{eHel}) we get

\begin{equation}
   \lim_{q_z\to 0}\langle n_{y}({\bf q})n_y(-{\bf q})\rangle=
   \lim_{q_z\to 0}{q_z^2b_0^2{\cal V}T\over c_{66}({\bf q})q_x^2+c_{44}({\bf q})
   q_z^2}\enspace.
\label{enshear}
\end{equation}
For the case of a vortex line lattice with $c_{66}>0$, taking
$q_z\to 0$ results in a vanishing of the vortex
correlation.  
From Eqs.(\ref{eRg}) and (\ref{eRl}) we then have $\lambda_{z{\rm R}}=\lambda_z$, 
and $\gamma_z=1$.  We will see in the next section that higher order
elastic corrections lead to a temperature dependent increase 
in $\lambda_{z{\rm R}}$, but do not change $\gamma_z=1$.
The vortex line lattice thus exhibits longitudinal
superconductivity, with a perfect Meissner effect for
shear perturbations.  

If one assumes $c_{66}\equiv 0$, as might describe the case of
a vortex line liquid, Eq.(\ref{enshear}) results in
$\langle n_y(q{\bf\hat x})n_y(-q{\bf\hat 
x})\rangle=b_0^2{\cal V}T/c_{44}(q{\bf\hat x})$.  In this case we have
$\gamma_z=1-B^2/4\pi c_{44}(0)=1-dB_\perp/dH_\perp<1$,
exactly as in the case of the tilt perturbation, Eqs.(\ref{ehmtiltg})
and (\ref{egtilt3}).  To summarize,
\begin{eqnarray}
  \begin{array}{lll}
  \langle n_y(q\hat{\bf x})n_y(-q\hat{\bf x})\rangle =0\enspace,
  &\gamma_z=1\qquad &{\rm if\ }c_{66}>0\label{ens1}\\
  \langle n_y(q\hat{\bf x})n_y(-q\hat{\bf x})\rangle ={\displaystyle{
  {b_0^2{\cal V}T\over c_{44}(q\hat{\bf x})}}}
  \enspace, &\gamma_z=1-{\displaystyle{{dB_\perp\over dH_\perp}}}<1
  \qquad &{\rm if\ }c_{66}\equiv 0\label{ens2}
  \end{array}
\end{eqnarray}

The above arguments suggest that a singular decrease of $\gamma_z$
from unity (or equivalently the singular increase of $n_{z0}$ from
zero), marking the loss of longitudinal superconductivity, serves
as a convenient criterion for the superconducting to normal transition
in the mixed state of a type II superconductor.  
This is one of the main results of our paper.
If this transition
is second order, we expect that $\lambda_{z{\rm R}}$ will diverge at the
transition with $n_s\sim 1/\lambda_{z{\rm R}}^2$ the density of 
superconducting electrons.  

In considering the vortex correlation at ${\bf q}=
q_x\hat{\bf x}+q_z\hat{\bf z}$, it is the relative order in
which one takes $q_x$ and $q_z$ to zero that distinguishes between
the shear and the tilt perturbation.  It is the order corresponding
to the shear perturbation that is related to the superfluid density
of the 2D analog bosons, $\rho_{s\,{\rm boson}}$.  Note that if we had defined 
$\rho_{s\,{\rm boson}}$ in terms of the ${\bf q}=0$ winding number of Pollock
and Ceperley,\cite{R34} rather than in terms of the ${\bf q}\to 0$ transverse 
momentum correlation function as we have done in Appendix
\ref{s2Db}, the connection of $\rho_{s\,{\rm boson}}$ to the shear, as opposed
to the tilt, perturbation would become ambiguous.

The preceding discussion has been based upon elastic fluctuations
about a perfect dislocation free vortex line lattice.  Recently, Frey 
{\it et al}.\cite{R35}
have argued that, at sufficiently high magnetic field in a layered
superconductor, the proliferation of dislocations can result in the
loss of longitudinal superconductivity even in the vortex line lattice
state.

\subsection{Second Order Corrections}
\label{sssecond}

Our analysis in the preceding sections is based on Eq.(\ref{enuq}),
which gives an expansion of the vortex line density ${\bf n}$ to $linear$
order in the displacement field ${\bf u}$. In this section, we
consider the effect of higher orders, by continuing the expansion in
${\bf u}$,

\begin{eqnarray}
  {\bf n}_q&=&ib_0[{\bf q}\cdot{\bf u}_q{\bf\hat z}
   -q_z{\bf u}_q]\nonumber\\
   &-&{b_0\over{\cal V}}\sum_{q^\prime}\left\{{\textstyle {1\over 2}}
    [{\bf q}\cdot{\bf u}_{q^\prime}]\,[{\bf q}\cdot{\bf u}
    _{q-q^\prime}] {\bf\hat z} - 
    (q_z-q_z^\prime)[{\bf q}\cdot{\bf u}_{q^\prime}]\,
    {\bf u}_{q-q^\prime}\right\}\nonumber\\
    &-&{ib_0\over 2{\cal V}^2}\sum_{q^\prime,\, q^{\prime\prime}}\left\{
    {\textstyle {1\over 3}}[{\bf q}\cdot{\bf u}_{q^\prime}]\,
    [{\bf q}\cdot{\bf u}_{q^{\prime\prime}}]\,
    [{\bf q}\cdot{\bf u}_{q-q^\prime-q^{\prime\prime}}]
    {\bf\hat z}\right.\nonumber\\&\,&\qquad\qquad - 
    \left.(q_z-q_z^\prime-q_z^{\prime\prime})
    [{\bf q}\cdot{\bf u}_{q^\prime}]\,
    [{\bf q}\cdot{\bf u}_{q^{\prime\prime}}]\,
    {\bf u}_{q-q^\prime-q^{\prime\prime}}\right\}+...
\label{enu2}
\end{eqnarray}
for small but finite ${\bf q}$.

To systematically
evaluate vortex correlations using this higher order expansion,
one should also in principle extend the elastic energy of 
Eq.(\ref{eHel}) to higher order in ${\bf u}$
by taking the expansion above, and substituting
into the vortex-vortex interaction Hamiltonian
of Eq.(\ref{eLG7}).  The resulting expression is  rather complex.
For simplicity, we will instead continue to use the quadratic elastic energy
of Eq.(\ref{eHel}), however we now view the elastic moduli as
appropriately redefined temperature dependent parameters, in the 
spirit of a self-consistent phonon approximation. 

We consider here only the case of $\Upsilon_z(q\hat{\bf x})$,
corresponding to the shear perturbation, for which we need the 
correlation, $\lim_{q_z\to 0} \langle n_y({\bf q})n_y(-{\bf q})
\rangle$, with ${\bf q}=q{\bf\hat x}+q_z{\bf\hat z}$.  As we have
already seen in the preceding section, for $c_{66}>0$, the contribution to
$O(u^2)$ vanishes.  By symmetry, the next leading term is $O(u^4)$.
Using the expansion of Eq.(\ref{enu2}), and factorizing the average of
the product of the four $u$'s into all possible pairs, we find,
\begin{eqnarray}
     \langle n_y({\bf q})n_y(-{\bf q})\rangle&=&b_0^2T^2 q_\alpha q_\beta
     \sum_{q^\prime}\left\{(q_z-q_z^\prime)^2\Phi^{-1}_{\alpha\beta}
     ({\bf q}^\prime)\Phi^{-1}_{yy}({\bf q}-{\bf q}^\prime)
     \nonumber\right.\\ \qquad &+&\left.
     (q_z-q_z^\prime)q_z^\prime\Phi^{-1}_{\alpha y}({\bf q}^\prime)
     \Phi^{-1}_{\beta y}({\bf q}-{\bf q}^\prime)-q_z^2
     \Phi^{-1}_{yy}({\bf q})\Phi^{-1}_{\alpha\beta}
     ({\bf q}^\prime)\right\}\enspace,
\label{enu4}
\end{eqnarray}
where summation over $\alpha,\beta=x,y$, is implied, and
$\Phi$ is the elasticity tensor of Eq.(\ref{eHel}).  Taking
$q_z\to 0$, keeping only terms of $O(q^2)$,
and using the fact that $\Phi$ is symmetric
in ${\bf q}$ as well as its indices, we get,
\begin{equation}
    \langle n_y(q\hat{\bf x})n_y(-q\hat{\bf x})\rangle=b_0^2T^2{\cal V}q^2I
    \enspace,
\label{enu42}
\end{equation}
where I is the integral,
\begin{equation}
   I\equiv {1\over {\cal V}}\sum_{k}
    {k_z^2\over {\rm det}\Phi_{k}}
    ={1\over {\cal V}}\sum_{k}
    {k_z^2\over (c_{44}k_z^2+c_{11}k_\perp^2)
    (c_{44}k_z^2+c_{66}k_\perp^2)}\enspace.
\label{enu43}
\end{equation}
The correlation of Eq.(\ref{enu42}) vanishes as $q^2$
for $q\to 0$.  We therefore continue to find, as in the preceding
section, that $n_{z0}=0$ and $\gamma_z=1$ giving a perfect
Meissner screening of the shear perturbation.  It is straightforward
to see that this result persists to all orders in $u$.  

However, in contrast to the preceding section, we now find a finite
renomalization of the penetration length.  Comparing the expansion of
Eq.(\ref{enn}) with the result of Eq.(\ref{enu42}), we get
$n_{z1}=b_0^2T^2{\cal V}I$.  Using this in Eq.(\ref{eRl}) then gives,
\begin{equation}
   {\lambda_{z{\rm R}}^2\over\lambda_z^2}=1+{B^2\over 4\pi\lambda_z^2}
   IT\enspace.
\label{enu44}
\end{equation}
Thus the $o(u^4)$ term generates an $O(T)$ correction to 
$\lambda_{z{\rm R}}^2$.  Continuing the elastic expansion in powers of $u$
will generate corrections to $\lambda_{z{\rm R}}^2$ in the form
of a power series in $T$.

To estimate the magnitude of the correction to $\lambda_{z{\rm R}}$ of
Eq.(\ref{enu44}) we can evaluate the
integral $I$ using a crude approximation.  The elastic moduli
which appear in $I$ are functions of wavevector ${\bf k}$.  
However the dominant contributions to the integral will come at
wavevectors $k_z\simeq\sqrt{(c_{66}/c_{44})}\,k_\perp$ and
$k_z\simeq\sqrt{(c_{11}/c_{44})}\,k_\perp$, both giving
$k_z\simeq\eta k_\perp$.  The dominant $k_\perp$ 
will be $k_\perp\simeq k_0=\sqrt{4\pi B/\phi_0}$, at
the edge of the Brillouin zone.  We will therefore approximate the 
elastic moduli by their values at this dominant wavevector, denoting
these values as $\tilde c_{44}$, $\tilde c_{11}$, and $\tilde c_{66}$.
Within this approximation one can explicitly calculate the integral
to get,
\begin{equation}
  I={k_0\over 4\pi\, \tilde  c_{44}\>\sqrt{\tilde c_{44}}\>\left(
   \sqrt{\tilde c_{11}}+\sqrt{\tilde c_{66}}\right)}\enspace.
\label{eI}
\end{equation}

Within the same crude approximation we can estimate the vortex
lattice melting temperature $T_{\rm m}$.  Using the Lindemann 
criterion\cite{R2,R6,R36} that
melting occurs when $\langle u^2\rangle\simeq c_{\rm L}^2a_{\rm v}^2$
($c_{\rm L}\sim 0.15$ is the Lindemann parameter), and
keeping only the transverse fluctuations as the dominant soft mode,
we get,
\begin{equation}
  T_{\rm m}={4\pi c_{\rm L}^2a_{\rm v}^2\sqrt{\tilde c_{66}
  \tilde  c_{44}}\over k_0}
  \enspace.
\label{etmelt}
\end{equation}

Combining Eqs. (\ref{enu44}), (\ref{eI}), and (\ref{etmelt}), 
and using the estimate that $\tilde c_{11}\sim \tilde c_{66}$
at the Brillouin zone boundary, we find,
\begin{equation}
   {\lambda_{z{\rm R}}^2\over\lambda_z^2}\simeq 1+{\textstyle{1\over 2}}c_L^2
   \left({a_{\rm v}\over\lambda_z}\right)^2
   {B^2\over 4\pi\tilde c_{44}}\,{T\over T_{\rm m}}\enspace.
\label{elamR2}
\end{equation}
Using the result of Fisher\cite{R37} for $c_{44}$ at the zone boundary, for
large magnetic fields $\lambda_\perp\gg a_{\rm v}$,
\begin{equation} 
   \tilde c_{44}\simeq {\phi_0 B\over 32\pi^2\lambda_z^2}\left[
   1+\ln\left[{H_{\rm c2}\over B(1+\eta^{-2})}\right]\right]
   \enspace,
\label{ec44fish}
\end{equation}
we get,
\begin{equation}
   {\lambda_{z{\rm R}}^2\over\lambda_z^2}\simeq 1+
   {4\pi c_L^2\sqrt{4/ 3}\over 1+{\displaystyle 
   \ln\left[{H_{\rm c2}\over B(1+\eta^{-2})}\right]}}\enspace.
\label{elamR3}
\end{equation}
Taking for example $B=0.2H_{\rm c2}$, and $\eta^{-2}\ll 1$, we
estimate a $12\%$ increase in $\lambda_{z{\rm R}}^2$ over
$\lambda_z^2$ near melting, due to lowest order elastic fluctuations.  
Since the elastic moduli $c_{\alpha\alpha}
({\bf k})$ are in general larger than the values 
$\tilde c_{\alpha\alpha}$, this is an overestimate.
The above estimate does not of course include the effects of
critical fluctuations near a phase transition, which for a second
order transition should result in a divergence of $\lambda_{z{\rm R}}$
at $T_{\rm c}$.

\section{The Vortex Line Liquid}
\label{ssliquid}

In the preceding section we considered the vortex line
lattice at $T<T_{\rm m}$.  In particular we showed how the response to the
shear perturbation, given by $\Upsilon_z(q{\bf\hat x})$,
gives a useful criterion for superconducting phase coherence:
$\gamma_z\equiv\lim_{q\to 0}[\Upsilon_z(q{\bf\hat 
x})/J_\perp\lambda_\perp^2q^2]=1$, or equivalently 
$n_{z0}\equiv \lim_{q\to 0}\langle n_y(q{\bf\hat x})
n_y(-q{\bf\hat x})\rangle =0$, indicates the presence of
longitudinal superconductivity.  In this section we consider behavior
in the vortex line liquid at $T>T_{\rm m}$.  As a measure of 
superconductivity we focus on the behavior of $n_{z0}$.

\subsection{Hydrodynamic Approximation}

The simplest approximation one can make at high $T$ is to
take the Hamiltonian (\ref{eLG7}) and
regard the Fourier components of the vortex line density ${\bf n}_q$
as continuous, independently fluctuating variables, subject to
the constraint ${\bf q}\cdot{\bf n}_q=0$.  We refer to this as
a hydrodynamic approximation.\cite{R38}
This yields $\langle n_y(q{\bf\hat x})n_y(-q{\bf\hat x})\rangle
=T{\cal V}/[4\pi^2J_\perp V_\perp (q{\bf\hat x})]$.  Substituting into
Eq.(\ref{ehm6}), we find that $\Upsilon_z(q{\bf\hat x})=0$ for
all values of $q$, reflecting the fact that  well above the 
superconducting transition, an applied magnetic field will 
induce no supercurrents at all.

For temperatures closer to, but still above $T_{\rm m}$, we expect 
the system to show a finite fluctuation diamagnetism.
A better approximation can be obtained
by coarse graining the Hamiltonian (\ref{eLG7}) over
a length scale of order the inter-vortex spacing $a_{\rm v}$, and then
applying the hydrodynamic approximation to average over the
resulting coarse grained vortex density.  This coarse grained
free energy has been given by Marchetti\cite{R13} as,
\begin{equation}
  {\cal H}[{\bf n}]={1\over 2b_0^2{\cal V}}\sum_q\left\{c_L({\bf q})\delta n_{qz}
  \delta n_{-qz} + c_{44}({\bf q}){\bf n}_{q\perp}
  \cdot{\bf n}_{-q\perp}\right\}\enspace,
\label{eHA1}
\end{equation}
where $\delta n_z=n_z-b_0$, $c_L$ is the bulk modulus,
and $c_{44}$ is a tilt modulus
of the same form as for the vortex lattice.  Using this form we find,

\begin{equation}
  n_{z0}=\lim_{q\to 0}\langle n_y(q{\bf\hat x})n_y(-q{\bf\hat x})\rangle 
   ={b_0^2{\cal V}T\over c_{44}(0)}>0\enspace,
\label{eHA2}
\end{equation}
and from Eqs.(\ref{eRg}) and (\ref{ec44}),
\begin{equation}
  \gamma_z=1-{B^2\over 4\pi c_{44}(0)}=1-{dB_\perp\over dH_\perp}<1
  \enspace.
\label{eHA3}
\end{equation}
Thus, within this hydrodynamic approximation, the longitudinal 
superconductivity found in the vortex line lattice is lost for the
vortex line liquid.  Note that since Eq.(\ref{eHA3}) gives
$\gamma_z$ strictly less than unity in the vortex line liquid,
while $\gamma_z=1$ in the vortex line lattice, $\gamma_z$
presumably takes a discontinuous jump at the transition where
longitudinal superconductivity is lost.

A more detailed calculation of vortex correlations, averaging
over unbounded dislocation loops within a continuum elastic model, has
been carried out by Marchetti and Nelson\cite{R39} as a model for a hexatic
vortex line liquid.  In the limit $q\to 0$, the
result of Eq.({\ref{eHA2}) is again obtained.  

Eqs.(\ref{eHA2}) and (\ref{eHA3}) are identical to the result we 
found in Eq.(\ref{ens2}) by simply taking $c_{66}\equiv 0$ 
in the elastic approximation for the vortex line lattice.  It is 
interesting to speculate about the behavior of a ``soft'' vortex 
line lattice in which the long wavelength shear modulus vanishes,
$c_{66}({\bf q}=0)=0$, but in which a finite shear stiffness remains on 
shorter length scales, $c_{66}({\bf q}_\perp,q_z=0)>0$ for
${\bf q}_\perp>0$.  In this case, taking
the limit ${\bf q}\to 0$ as in Eq.(\ref{enshear}), we find that
$n_{z0}=0$ and longitudinal superconductivity remains.\cite{R39.1} 
As Marchetti and Nelson\cite{R39} show however, it is not possible to
describe an entangled vortex line liquid with such a simple
elastic description.\cite{R40}

\subsection{Kosterliz-Thouless Transition}
\label{sboson}

In Sec.\,\ref{sanalogy} we discussed how the KT
superfluid transition of the analog 2D bosons could appear 
in the 3D superconductor as a strong
cross-over to a vortex line liquid state with longitudinal superconductivity.
In his original model, Nelson\cite{R11} interpreted this KT transition in terms 
of a transition from an ``entangled'' to a ``disentangled'' vortex
line liquid, for sufficiently thin samples.  In this section, we
estimate the temperature $T_{\rm c}$ for this KT transition as
a function of sample thickness $L_z$ and magnetic field $B$, and 
compare this estimate with Nelson's entanglement criterion.

The 2D KT superfluid transition is characterized\cite{R14} by the fact that,
exactly at the transition, the boson helicity modulus $\Upsilon_{\rm 
boson}(q\to 0)$ takes a
discontinuous jump to zero from the universal finite value
$\Upsilon_{\rm boson}/T_{\rm boson}=2/\pi$.  $\Upsilon_{\rm boson}$ is given
by the vortex correlation of Eq.(\ref{e2D1}), which for a vortex
line liquid can be related to the tilt modulus $c_{44}$ by 
Eq.(\ref{eHA2}). Using Eq.(\ref{ec44}) for $c_{44}$ and applying
the universal jump criterion then gives for the KT transition,
\begin{equation}
  T_{\rm c}={2\over\pi}{c_{44}(0)\over b_0^2L_z}=
  {\phi_0^2\over 2\pi^2L_z}{dH_\perp\over dB_\perp}\enspace.
\label{e2D4}
\end{equation}
Thus as the thickness $L_z$ increases, $T_{\rm c}$ decreases.
In order to observe a vortex line liquid with longitudinal 
superconductivity we need the system to be thin enough that
$T_{\rm c}>T_{\rm m}$.  If we define the length,
\begin{equation}
  \Lambda\equiv {\phi_0^2\over 2\pi^2 T_{\rm m}}\enspace,
\label{e2D5}
\end{equation}
then we can rewrite Eq.(\ref{e2D4}) as,
\begin{equation}
  {T_{\rm c}\over T_{\rm m}}= {dH_\perp\over dB_\perp}
  {\Lambda\over L_z}\equiv{L_{z\,{\rm max}}\over L_z}\enspace.
\label{e2D6}
\end{equation}
We thus will have $T_{\rm c}>T_{\rm m}$ provided,
\begin{equation}
  L_z<L_{z\,{\rm max}}={dH_\perp\over dB_\perp}\Lambda = 
  \left({4\pi c_{44}(0)\over B^2}\right)\Lambda\enspace.
\label{e2D7}
\end{equation}

Assuming that $c_{44}(0)$ in the line liquid is not too different
from $c_{44}(0)$ in the line lattice, we can use our results from
Appendix \ref{sselmod} to evaluate the length $L_{z\,{\rm max}}$.
For large applied magnetic fields, such that $a_{\rm v}\ll\lambda_\perp$
or equivalently $H_{\rm c1}\ll B$, we have $dH_\perp/dB_\perp\simeq 1$
and so to leading order,
\begin{equation}
  L_{z\,{\rm max}}=\Lambda\enspace.
\label{e2D8}
\end{equation}
For small magnetic fields, such that $\lambda_z\ll a_{\rm v}$, 
we have,

\begin{equation}
  L_{z\,{\rm max}}={\phi_0\over 8\pi\lambda_\perp^2 B}
  \left\{\eta^{-2}\left[ 2\ln(\eta\kappa)-1\right]+1\right\}
  \Lambda.
\label{e2D9}
\end{equation}
For an anisotropic material in intermediate magnetic fields, 
such that $\lambda_\perp\ll a_{\rm v}\ll\lambda_z$, we have to leading
order,
\begin{equation}
  L_{z\,{\rm max}}={\phi_0\over 8\pi\lambda_\perp^2 B}
  \left\{\eta^{-2}\left[ \ln(H_{\rm c2}/B)-1\right]+1\right\}
  \Lambda\enspace.
\label{e2D10}
\end{equation}
Note that for a melting temperature of $T_{\rm m}\sim 90^\circ$K, as in YBCO,
one has for large $B$, $L_{z\,{\rm max}}=\Lambda\simeq 1400\mu{\rm m}$.
This is much thicker than typical experimental samples, which
are generally of the order $50\mu{\rm m}$.  As $B$ decreases,
$L_{z\,{\rm max}}$ only gets {\it larger}.

The above results may be compared with the original criterion 
for 2D boson superfluidity given by Nelson\cite{R11}
in terms of the ``entanglement length,''  
\begin{equation}
  \xi_z={\tilde\epsilon_1\phi_0\over 2 TB}\enspace.
\label{e2D10.1}
\end{equation}
$\xi_z$ is the length required for a vortex line to have a transverse
deflection equal to the average spacing between vortex lines, $a_{\rm 
v}$.  Only when $\xi_z<L_z$ can the vortex lines have enough transverse
wandering so that they may become geometrically entangled.
The cross-over $T_\times$ between a disentangled and an entangled vortex
line liquid is thus given by,
\begin{equation}
   T_\times={\tilde\epsilon_1\phi_0\over 2BL_z}={\phi_0^2\over2\pi^2L_z}\,
   {\phi_0\eta^{-2}\ln\kappa\over 16\lambda_\perp^2 B}\enspace,
\label{e2D10.2}
\end{equation}
where we have used $\tilde\epsilon_1=\epsilon_1\eta^{-2}$
with $\epsilon_1=(\phi_0^2/4\pi\lambda_\perp)^2\ln\kappa$
for small $B$.  One will have $T_\times>T_{\rm m}$ only for 
$L_z<L^\prime_{z\,{\rm max}}$, where,
\begin{equation}
  L^\prime_{z\,{\rm max}}={4\over\pi}\left[{\pi^2\tilde\epsilon_1
  \over\phi_0 B}\right]\Lambda=
  {\phi_0\eta^{-2}\ln(\kappa)\over 16\lambda_\perp^2 B}\,
  \Lambda\enspace.
\label{e2D11}
\end{equation}
Except for some numerical factors, $L^\prime_{z\,{\rm max}}$ 
agrees with $L_{z\,{\rm max}}$ of 
Eqs.(\ref{e2D9}) and (\ref{e2D10}), decreasing as $1/B$, for
increasing magnetic field.  In the large field limit however,
our result in Eq.(\ref{e2D8}) saturates to the finite value
$\Lambda$ instead of continuing to decrease.  The
difference between the results of Eqs.(\ref{e2D8}) and (\ref{e2D11})
arises because the later is based on the
the line tension $\tilde\epsilon_1$ for the
energy of a single vortex line tilting, while the former
is based on the
tilt modulus $c_{44}(0)$ for the collective tilting of all
lines.  This points out an important distinction: geometric
vortex line entanglement, i.e. the local wrapping of lines
around each other, is not necessarily equivalent to 
the global vortex line winding that characterizes the analog 2D 
boson superfluid phase.\cite{R34}
Nelson's entanglement length of Eq.(\ref{e2D11})
nevertheless remains the important length scale for local
geometric entanglement, which still can have a significant effect
on the dynamic behavior of the vortex line liquid if the
barriers for vortex line cutting are high.\cite{R41}

The above discussion has been based on the familiar KT transition of
an ordinary 2D superfluid, and predicts that as $L_z\to\infty$ 
at fixed $T>T_{\rm m}$ ($T_{\rm boson}\to 0$ at fixed $\hbar_{\rm boson}$) 
$T_{\rm c}\to 0$ and so one is always in the boson superfluid state,
corresponding to a normal vortex line liquid.  Feigelman\cite{R42} and 
co-workers\cite{R20} however have argued that for $\lambda\to\infty$, the long
range nature of the effective 2D boson interaction can lead to 
a normal boson fluid, and hence to a vortex line liquid with
longitudinal superconductivity, even in the $L_z\to\infty$ 
($T_{\rm boson}\to 0$) limit for $T<T_{\rm c}^\infty$, where
$T_{\rm c}^\infty$ (i.e. the critical $\hbar_{\rm boson}$ in the
boson variables) gives the 2D Meissner transition of the
analog 2D $charged$ bosons of the $\lambda\to\infty$ approximation, 
as discussed in Sec.\,\ref{slaminf}.
Such a $\lambda\to\infty$ transition would probably
lead to strong cross-over effects in the finite $\lambda$ case,
which would obscure the KT transition when
$L_z\ll L_{z\,{\rm max}}$, where the $T_{\rm c}$ of Eq.(\ref{e2D4}) 
can be very much larger than $T_{\rm c}^\infty$.

Searching for longitudinal 
superconductivity within the vortex line liquid will be one of the
main objectives of our numerical investigations, to be discussed in 
the following section.

\section{Numerical Simulations}

In this section we report the results of numerical Monte
Carlo simulations we have carried out in order to study the
behavior of the system of fluctuating vortex lines.

\subsection{Lattice Superconductor}

To carry out numerical simulations, our first step will be to
discretize the continuum to a cubic grid of $N=N_\perp^2N_z$
sites $i$.  The grid
spacing in direction ${\hat\mu}$ is taken to be
\begin{equation}
    a_\mu = \left\{
      \begin{array}{ll}
        a_\perp = \xi_\perp,\qquad &\mu=x,y\\
        a_z=d,\qquad &\mu=z
      \end{array}\right.
\label{eNS1}
\end{equation}
The grid spacing $\xi_\perp$ in the $xy$ plane is meant to approximate the core 
radius of
a vortex, while the spacing $d$ along $\hat {\bf z}$ is meant to simulate the
spacing between CuO planes of a layered high $T_c$ superconductor.
If one wants to model an anisotropic continuum superconductor, such 
as in Sec.\,\ref{slondon}, one should take $d\equiv\xi_z$,
where the anisotropic Ginzburg Landau free
energy functional gives\cite{R2} $\xi_z=\eta^{-1}\xi_\perp$.
Discretization of Eq.(\ref{eLG1}) then leads to the {\it lattice 
superconductor} model,\cite{R15,R16}
\begin{equation}
  {\cal H}[\theta_i,A_{i\mu}]=\sum_{i,\mu}\left[
   {\cal U}_\mu(\theta_{i+\hat\mu}-\theta_i-A_{i\mu})+2\pi^2 C_\mu
   (b_{i\mu}-h_{i\mu})^2\right]\enspace,
\label{eNS2}
\end{equation}
where $\theta_i$ is the phase angle on grid site $i$, 
\begin{equation}
   A_{i\mu}=\int_i^{i+\hat\mu} {\bf A}\cdot{\bf d\ell}
\label{eNS2.1}
\end{equation}
is the integral
of the total magnetic vector potential across the bond at site $i$ in
direction ${\bf\hat\mu}$, and if $\mu,\nu,\sigma$ is a cyclic
permutation of $x,y,z$, then,
\begin{equation}
   2\pi b_{i\mu}=A_{i+\hat{\bf\nu},\sigma}-
   A_{i\sigma}-A_{i+\hat{\bf\sigma},\nu}+A_{i\nu}
\label{eNS2.2}
\end{equation}
is the sum of the $A_{j\nu}$ going counterclockwise around the plaquette
at site $i$ in direction ${\bf\hat\mu}$, and
gives $2\pi$ times the flux of total magnetic field through the plaquette
(see Fig.\,\ref{fgrid}); a similar relation defines $h_{i\mu}$
in terms of $A_{i\mu}^{\rm ext}$.
The kinetic energy piece is expressed in terms of the Villain
function,\cite{R43}
\begin{equation}
  e^{-{\cal U}_\mu(\phi)/T}=\sum_{m=-\infty}^\infty
  e^{-\bar{J}_\mu\eta_\mu^{-2}(\phi-2\pi m)^2/2T}\enspace,
\label{eNS3}
\end{equation}
with couplings,
\begin{equation}
    \bar{J}_\mu = J_\perp {a_\nu a_\sigma\over a_\mu}=\left\{
      \begin{array}{ll}
        \bar{J}_\perp = J_\perp d,\qquad &\mu=x,y\\
        {\displaystyle
        \bar{J}_z=J_\perp {\xi_\perp^2\over d}},\qquad &\mu=z
       \end{array}\right.
\label{eNS4}
\end{equation}
The couplings of the magnetic energy piece are,
\begin{equation}
    C_\mu = J_\perp\lambda_\perp^2 {a_\mu\over a_\nu a_\sigma}=\left\{
      \begin{array}{ll}
        {\displaystyle
        C_\perp = \bar{J}_z \left({\lambda_\perp\over\xi_\perp}
        \right)^2},\qquad &\mu=x,y\\
        {\displaystyle
        C_z=\bar{J}_\perp \left({\lambda_\perp\over\xi_\perp}
        \right)^2},\qquad &\mu=z
       \end{array}\right.
\label{eNS5}
\end{equation}

To express the Hamiltonian in terms of 
vortex line variables, we first perform a standard duality
transformation\cite{R44} of the kinetic energy piece, and then, following
Carneiro\cite{R16} (in complete analogy with Eqs.(\ref{eLG6})
$-$(\ref{eVdiag})), complete the square in $b^{\rm ind}_{q\mu}
=b_{q\mu}-h_{q\mu}$ subject to 
the constraint that ${\bf b}^{\rm ind}_q$ is divergenceless.
Our lattice Fourier transforms are defined by,
\begin{equation}
   b_{q\mu}=\sum_i e^{i{\bf q}\cdot{\bf r}_i}b_{i\mu}\enspace,\qquad
   b_{i\mu}={1\over N}\sum_q e^{-i{\bf q}\cdot{\bf 
   r}_i}b_{q\mu}\enspace,
\label{eNS6}
\end{equation}
and the constraint that $b_{i\mu}^{\rm ind}$ is divergenceless
can be written as, ${\bf Q}^*\cdot {\bf b}_i^{\rm ind}=0$, where,
\begin{equation}
   Q_\mu\equiv 1-e^{iq_\mu a_\mu}\enspace.
\label{eNS9}
\end{equation}
The vortex part of the resulting Hamiltonian is,
\begin{equation}
  {\cal H}_{\rm v}={4\pi^2\bar{J}_\perp\over 2N}\sum_{q,\alpha}
  V_{q\alpha}[n_{q\alpha}-h_{q\alpha}][n_{-q\alpha}-h_{-q\alpha}]
\enspace,
\label{eNS10}
\end{equation}
where $V_{qx}=V_{qy}\equiv V_{q\perp}$, and,
\begin{equation}
    V_{q\perp} = {\left({\lambda_\perp\over d}\right)^2\over 
    1+\left({\lambda_\perp\over d}\right)^2|Q_z|^2
    +\left({\lambda_z\over \xi_\perp}\right)^2|{\bf Q}_\perp|^2}
\label{eNS11}
\end{equation}
\begin{equation}
    V_{qz} = {\left({\lambda_\perp\over\xi_\perp}\right)^2\left[ 1+
    \left({\lambda_z\over d}\right)^2|Q_z|^2 + 
    \left({\lambda_z\over\xi_\perp}\right)^2|{\bf Q}_\perp|^2
    \right]\over\left[1+
    \left({\lambda_\perp\over d}\right)^2|Q_z|^2 + 
    \left({\lambda_\perp\over\xi_\perp}\right)^2|{\bf Q}_\perp|^2
    \right]\left[ 1+
    \left({\lambda_\perp\over d}\right)^2|Q_z|^2 + 
    \left({\lambda_z\over\xi_\perp}\right)^2|{\bf Q}_\perp|^2
    \right]}\enspace.
\label{eNS12}
\end{equation}
$n_{q\mu}$ is the Fourier transform of the vorticity $n_{i\mu}$
piercing plaquette $i\mu$.
Eqs.(\ref{eNS10}--\ref{eNS12}) are the lattice equivalents of
the continuum Eqs.(\ref{eLG7}) and (\ref{eVdiag}).

Note that ${\cal H}_{\rm v}/T$ depends on $four$ dimensionless
parameters, which may be taken to be $\bar J_\perp/T$,
$\eta=\lambda_z/\lambda_\perp$,
$\kappa=\lambda_\perp/\xi_\perp$, and $\zeta=\xi_\perp/d$.  
The ratio the of couplings that appear in the Villain
kinetic energy terms of Eq.(\ref{eNS3}) is then $\bar J_z\eta^{-2}/\bar
J_\perp=(\lambda_\perp\xi_\perp/\lambda_z d)^2=(\zeta/\eta)^2$.
If one wants to model an anisotropic continuum, with $d=\xi_z=\eta^{-1}
\xi_\perp$, then one has $\zeta=\eta$ and there are only $three$ 
dimensionless parameters,
$\bar J_\perp/T$, $\eta$, and $\kappa$, with $\bar J_z\eta^{-2}/\bar
J_\perp=1$.  Both cases are in general different from an earlier derivation 
of the London lattice vortex line interaction\cite{R16} which 
assumed equal grid spacings in all directions, $a_\mu=a_0$
for all $\mu$, and so with $\zeta=1$ involves only the three 
dimensionless parameters,
$\bar J_\perp/T$, $\eta$, and $\lambda_\perp/a_0$, but with
$\bar J_z\eta^{-2}/\bar J_\perp=\eta^{-2}$.
Keeping the distinction $a_z\ne a_\perp$ (i.e. $d\ne\xi_\perp$)
is essential to correctly
model the effects of the anisotropic vortex core energy in either
a continuum or a layered anisotropic superconductor.

We can now define the helicity modulus for the lattice superconductor
in complete correspondence with the continuum 
Eqs.(\ref{ej}--\ref{ehm2}).  The only change needed is to replace the
system volume ${\cal V}$ with the number of grid sites $N$,
due to the slightly differing definitions of the Fourier transform in the
continuum, Eq.(\ref{eFT}), and on the lattice, Eq.(\ref{eNS6}).
As in the continuum we restrict ourselves to the three special
perturbations of Fig.\,\ref{f1}, $A^{\rm ext}_\mu(q\hat{\bf\nu})$,
where $\mu,\nu,\sigma$ are a cyclic permutation of $x,y,z$.
Taking the Fourier transform of Eq.(\ref{eNS2.2}), we get
$2\pi h_\sigma(q\hat{\bf\nu})=Q_\nu^*A^{\rm ext}_\mu(q\hat{\bf\nu})$.
Substituting for $h_\sigma$ in terms of $A^{\rm ext}_\mu$ in ${\cal H}_{\rm v}$
of Eq.(\ref{eNS10}), and then applying the definition
of helicity modulus in Eq.(\ref{ehm2}), we get for the diagonal
part of the helicity modulus tensor, 
\begin{equation}
  \Upsilon_\mu(q{\bf\hat\nu})={\bar{J}_\mu\left({\lambda_\perp
  \over a_\nu}\right)^2|Q_\nu|^2\over 1+\left({\lambda_\mu\over 
  a_\nu}\right)^2|Q_\nu|^2}\left[1-{4\pi^2\bar{J}_\mu\left(
  {\lambda_\perp\over a_\nu}\right)^2\over TN}{\langle n_\sigma
  (q{\bf\hat\nu})n_\sigma(-q{\bf\hat\nu})\rangle_0\over
  1+\left({\lambda_\mu\over a_\nu}\right)^2|Q_\nu|^2}\right]\enspace,
\label{eNS16}
\end{equation}
and for the off diagonal part,
\begin{equation}
  \Upsilon_{\mu\nu}(q{\bf\hat\sigma})={\bar{J}_\mu\left(
  {\lambda_\perp\over a_\sigma}\right)^2|Q_\sigma|^2\over
  1+\left({\lambda_\mu\over a_\sigma}\right)^2|Q_\sigma|^2}
  \left[ {4\pi^2\bar{J}_\nu\left({\lambda_\perp\over a_\sigma}
  \right)^2\over NT}{\langle n_\nu(q{\bf\hat\sigma})n_\mu(-q
  {\bf\hat\sigma})\rangle_0\over 1+\left({\lambda_\nu\over
  a_\sigma}\right)^2|Q_\sigma|^2}\right]\enspace.
\label{eNS17}
\end{equation}
Eqs.(\ref{eNS16}) and (\ref{eNS17}) are the lattice equivalents
of the continuum Eqs.(\ref{ehm6}) and (\ref{ehm7}).  The primary
difference between continuum and lattice expressions is the 
substitution, $q_\mu^2\to |Q_\mu/a_\mu|^2=(2-2\cos q_\mu 
a_\mu)/a_\mu^2$.

Expanding the vortex correlation at small $q$,
\begin{equation}
   \langle n_\sigma
   (q{\bf\hat\nu})n_\sigma(-q{\bf\hat\nu})\rangle_0=n_{\mu 0}
   +n_{\mu 1}|Q_\nu|^2  +n_{\mu 2}|Q_\nu|^4+\ldots
\label{eNS18}
\end{equation}
we can again write the diagonal part as,
\begin{equation}
  \Upsilon_\mu(q{\bf\hat\nu})=\gamma_\mu{\bar{J}_\mu
  \left({\lambda_\perp\over a_\nu}\right)^2|Q_\nu|^2\over
  1+\left({\lambda_{\mu{\rm R}}\over a_\nu}\right)^2|Q_\nu|^2}\enspace,
\label{eNS18.5}
\end{equation}
where, analogous to Eqs.(\ref{eRg}) and (\ref{eRl}),
\begin{equation}
   \gamma_\mu = 1-{4\pi^2\bar{J}_\mu\left({\lambda_\perp
   \over a_\nu}\right)^2\over NT}n_{\mu 0}
\label{eNS19}
\end{equation}
and,
\begin{equation}
  \left({\lambda_{\mu{\rm R}}\over\lambda_\mu}\right)^2=1-
  {4\pi^2\bar{J}_\mu\left({\lambda_\perp
   \over a_\nu}\right)^2\over NT}{\left[n_{\mu 0}-n_{\mu 1}\left(
   {\lambda_\mu\over a_\nu}\right)^{-2}\right]\over \gamma_\mu}
\enspace.
\label{eNS20}
\end{equation}

Noting that $Q_\nu \simeq -iq_\nu a_\nu$ for small $q$, that
$N={\cal V}/(\xi_\perp^2d)$, and that there is a slight distinction between
Fourier components defined on the lattice versus in the
continuum, $n_{q\mu}^{\rm lattice}=a_\mu^{-1}n_{q\mu}^{\rm continuum}$, all
the above expressions agree completely with their continuum
counterparts, in the limit $q\to 0$.

\subsection{Monte Carlo Method and Parameters}

To carry out Monte Carlo simulations of the lattice superconductor
model, we start with a fixed density $b_0=\xi_\perp^2B/\phi_0$ 
of magnetic field induced straight vortex lines, parallel 
to the ${\bf\hat z}$ axis.
Following Carneiro, Cavalcanti, and Garter,\cite{R28.1} we update the system,
heating from the ground state, by adding elementary closed vortex
rings that surround only a single bond of the discretizing grid
(i.e. rings of area $\xi_\perp^2$ in the $xy$ plane, or  area
$\xi_\perp d$ in the $xz$ or $yz$ planes).  The rings are added one
at a time, at random positions with random orientations, and then
accepted or rejected according to the standard Metropolis algorithm.  
When a side of such a ring coincides with, and is oppositely oriented
to, a segment of one of the initial magnetic field induced vortex lines, these
two segments will cancel out resulting in a net fluctuation of the
vortex line.  This procedure provides for a complete sampling of 
phase space for the vortex variables $n_{i\mu}$ which are subject to the 
constraints that vorticity is locally conserved, 
$\sum_{\mu}[n_{i\mu}-n_{i-\hat{\bf\mu},\mu}]=0$, and that the average
internal magnetic field is kept constant, 
$(1/N)\sum_i{\bf n}_i=b_0{\bf\hat z}$ (i.e. we are using the
Helmholtz ensemble).

Our simulation uses periodic boundary conditions in all directions.
The periodicity along $\hat{\bf z}$ makes our simulation map
exactly onto the 2D boson problem.
In order to compute energy changes for the Metropolis acceptance test,
it is convenient to use,\cite{R45}
\begin{equation}
   \Delta E=2\pi^2\bar{J}_\perp\sum_{i\mu}F_{i\mu}\Delta n_{i\mu}
\label{eNS21}\enspace,
\end{equation}
where $\Delta n_{i\mu}$ is the change in vorticity due to the
vortex ring excitation, and
\begin{equation}
   F_{i\mu}\equiv \sum_j V_\mu({\bf r}_j-{\bf r}_i)n_{j\mu}
\label{eNS22}
\end{equation}
represents the ``potential'' field of all other vortices. $V_\mu({\bf 
r}_i)=(1/N)\sum_q e^{-i{\bf q}\cdot{\bf r}_i}V_{q\mu}$ 
is the Fourier transform of the vortex line interaction of
Eqs.(\ref{eNS11}) and (\ref{eNS12}), where the sum is over all
${\bf q}$ satisfying periodic boundary conditions,
$q_\mu=2\pi \ell_\mu/N_\mu a_\mu$, $\ell_\mu=0,1,\ldots, N_\mu-1$.  
In this way,
the computation of $\Delta E$ is a {\it local} computation,
involving only the sites of the elementary vortex ring excitation.
Only when an excitation is accepted is it necessary to update
the potentials $F_{i\mu}$, a calculation of order $N$.  Since
acceptance rates are generally low below the transition, this
method is significantly faster than a direct computation involving
the long range vortex interactions.

For simplicity, we have only simulated the completely isotropic case with
$\lambda\equiv\lambda_\perp=\lambda_z$, $a_0\equiv\xi_\perp=d$,
and hence $\bar{J}_0\equiv \bar{J}_\perp=\bar{J}_z$. Hence forth 
all lengths will
be measured in units of the grid spacing $a_0\equiv 1$, and temperatures
in units of $\bar{J}_0$.  Our simulations are for the fixed
vortex density $b_0=1/15$ whose ground state, shown in 
Fig.\,\ref{fground}, is
a close approximation to a perfect triangular lattice with sides
of length $\sqrt{18}\times\sqrt{17}\times\sqrt{17}$.  
We choose $\kappa=\lambda/a_0=5$, comparable
to the vortex line spacing $a_{\rm v}/a_0\simeq 1/\sqrt{b_0}=2.87$.
We study system sizes $N_\perp=30$ in the $xy$ plane, and
$N_z=15$ and $30$ parallel to the applied magnetic field.

Our simulations are carried out heating from the ground state.
At each temperature we use typically
$5000$ sweeps to equilibrate, followed by another $8-16000$ sweeps to 
compute averages.  Each ``sweep'' refers to $N=N_\perp^2N_z$ attempts 
to add an elementary vortex ring.  Statistical errors are estimated
using the standard block averaging method.

\subsection{Results: Helicity Modulus}

In an earlier report\cite{R10} we presented an analysis of our data based
on Eqs.(\ref{eNS18}-\ref{eNS20}), fitting our computed correlations
$\langle n_\sigma(q{\bf\hat\nu})n_\sigma(-q{\bf\hat\nu})\rangle_0$
to an expansion in $|Q_\nu|^2$. Here we take a different
approach. Plotting 
$\bar{J}_0\lambda^2|Q_\nu|^2/\Upsilon_\mu(q{\bf\hat\nu})$ versus
$|Q_\nu|^2$, Eq.(\ref{eNS18.5}) shows that at 
small $q$ we should find a straight line
with intercept $\gamma_\mu^{-1}$ and slope 
$\gamma_\mu^{-1}\lambda_{\mu{\rm R}}^2$.

In Figs.\,\ref{figYi}$a-c$ we show such plots for the three 
types of perturbations shown in Fig.\,\ref{f1},
for $N_z=30$ and selected values of temperature.  $\mu=y,x,z$
correspond to the tilt, compression, and shear perturbations
respectively.  The straight lines through the data
result from least squares fits, using the smallest
eight values of $q>0$.  In virtually all cases, the fit is
quite reasonable.  In Figs.\,\ref{figamma}$a-c$ and 
\ref{figlambda}$a-c$ we show the values of $\gamma_\mu$ and
$(\lambda_{\mu{\rm R}}/\lambda)^2$ obtained from these fits.
In each case we show the result of fits to the smallest eight,
seven, six, and five values of $q>0$.  As is seen, our results
are virtually insensitive to the number of values of $q$ used,
except for the case of the compression perturbation in the
vicinity of $T_{\rm m}\simeq 1.2$, where our data is rather scattered
and statistical errors are large (see data for $T=1.2$ in 
Fig.\,\ref{figYi}$b$ and the corresponding dashed line fit).
We have also obtained values of $\gamma_\mu$ and
$(\lambda_{\mu{\rm R}}/\lambda)^2$ by fitting 
$\bar{J}_0\lambda^2|Q_\nu|^2/\Upsilon_\mu(q{\bf\hat\nu})$ 
to a second order polynomial in $|Q_\nu|^2$.  We
have found the results from such quadratic fits to be essentially unchanged 
from the values obtained from the linear fits.

In Figs.\,\ref{figgl}$a-b$ we show $\gamma_\mu$ and 
$(\lambda_{\mu{\rm R}}/\lambda)^2$ respectively for all three
types of perturbations, comparing the finite size effects
for $N_z=30$ and $N_z=15$.  The results shown are for
fits to the smallest eight values of $q$, except for the
case of the tilt perturbation for $N_z=15$ where we have
used only the smallest four values of $q$ (since the
allowed values of $q_z$ are spaced twice as far apart
for $N_z=15$ as compared to $N_z=30$). We see that finite
size effects are in general small, except for the case
of the shear perturbation $\mu=z$.

We now discuss our results for $\gamma_\mu$.  
From Fig.\,\ref{figgl}$a$ we see that
all three $\gamma_\mu\simeq 1$ at low temperatures.  
For $\gamma_z$, this is in agreement with
our expectation that there is a total Meissner effect for shear
perturbations in the vortex line lattice phase.  However the 
elastic theory results given by the first lines of  Eqs.(\ref{egtilt2}) 
and (\ref{egcomp}) would lead one to expect $\gamma_y,|\gamma_x|\ll 1$.
This is because for the relatively large $B$ simulated here, 
$B\simeq H$ and the 
susceptibilities $dB_\sigma/dH_\sigma$ that enter
$\gamma_y$ and $\gamma_x$ in Eqs.(\ref{egtilt3}) and (\ref{ecL})
are both close to unity. That we find $\gamma_y,\gamma_x\simeq 1$
at low $T$ is, we believe, an artifact of our discretizing grid
which acts like a periodic pinning potential for vortex lines.
At low $T$, the vortex lines are locked into a lattice structure
commensurate with this pinning potential.  Indeed, the fact that
the ground state of Fig.\,\ref{fground} is not a perfect equilateral
triangular lattice is due to this effect.  This periodic
pinning potential leads to an enhanced stiffness of the
effective elastic moduli, greatly reducing the susceptibilities
$dB_\sigma/dH_\sigma$ from their continuum values, and resulting
in the observed $\gamma_y,\gamma_x\simeq 1$ at low $T$.  
Indeed the periodic pinning potential of the discretizing grid
acts in many ways like the columnar pins of the ``Bose glass''
model\cite{R46} of a disordered superconductor, and $\gamma_y=1$ 
is similar to the ``transverse Meissner'' effect for tilting
the applied magnetic field that is found in that problem.
One can wonder whether the decrease of $\gamma_y$ from unity which
begins at $T\simeq 0.6$ is a smooth cross-over due to finite
energy barriers for discretized vortex fluctuations, or is rather a
sharp transition, being the periodic 
pinning analog of the Bose glass transition.

At higher temperatures, $\gamma_x$ and $\gamma_y$ decrease towards
zero at $T_{\rm m}\simeq 1.2$.  We will soon see that this $T_{\rm m}$ 
is the melting temperature of the vortex line lattice.
It is only when the vortex lattice melts that the vortex
lines also depin from the the periodic potential of the grid.
Assuming that
the effective tilt and compression moduli of the unpinned vortex line 
liquid are not greatly different from those of the continuum
vortex line lattice, one expects from Eq.(\ref{egtilt2}),
$0<\gamma_y\approx 1/(8\pi\lambda^2 b_0) =0.024\ll 1$,
and from Eq.(\ref{egcomp}), $\gamma_x\approx -1/(16\pi\lambda^2 b_0)
=-0.012 <0$.  Looking at Figs.\,\ref{figamma}$a-b$
we see that $\gamma_y$ is indeed small and positive for $T>T_{\rm m}$,
while $\gamma_x$ is small and negative.  That $\gamma_x$ is indeed
negative and not zero for $T>T_{\rm m}$ is more clearly seen 
in Fig.\,\ref{figYi}$b$ by noting that
the intercepts of the fitted lines are just $\gamma_x^{-1}$.  
The numerical values
we find for $\gamma_x$ and $\gamma_y$ in this region are in roughly
the same ratio as the above estimates, but approximately two or
three times larger in magnitude.  This rough agreement of
$\gamma_x$ and $\gamma_y$ with elastic theory gives
us confidence that, above $T_{\rm m}$, the artificial pinning
introduced by our discretizing grid
is no longer playing a significant role in the vortex line
fluctuations.

Returning to Fig.\,\ref{figgl}$a$ we see that, in contrast
to $\gamma_y$ and $\gamma_x$, $\gamma_z$ remains equal to
unity well into the vortex line liquid phase $T>T_{\rm m}$.  $\gamma_z$
only decreases from unity towards the small value expected 
from Eq.(\ref{eHA3}) for the vortex line liquid,
$\gamma_z=1-dB_\perp/dH_\perp =\gamma_y$, 
at $T_{\rm c}\simeq 1.8$.  Thus longitudinal superconductivity,
indicated by the shear Meissner effect with $\gamma_z=1$, 
persists well above $T_{\rm m}$ into the vortex line liquid.
This one of the main observations
of our simulations.  Comparing results for $N_z=15$ with $N_z=30$,
we see that the width
of this transition clearly sharpens as $N_z$ increases, however
the temperature $T_{\rm c}$, where $\gamma_z$  starts
to fall below unity, decreases only slightly.

We now consider our results for $(\lambda_{\mu{\rm R}}/\lambda)^2$.
For the tilt perturbation, comparison of Figs.\,\ref{figamma}$a$
and \ref{figlambda}$a$ show that to a very good approximation,
$\gamma_y\approx (\lambda_{y R}/\lambda)^2$ for the entire
range of $T$.  Such a result follows from Eqs.(\ref{ehmtiltg})
and (\ref{ehmtiltl}) if one makes the simple assumption that
$c_{44}({\bf q})\simeq (B^2/4\pi\lambda_\perp^2)V_{q\perp}+b_0
\tilde\epsilon_1$ where $V_{q\perp}$ is the vortex line interaction
of Eq.(\ref{eVdiag}), and $\tilde\epsilon_1=\eta^{-2}\epsilon_1$
where $\epsilon_1$ is the effective $q_z$ independent single
vortex line tension. It is interesting that $\lambda_{y{\rm R}}$
shows no increase as $T_{\rm m}$ is approached from below, as
is usually associated with a decay length near a transition.

Turning to the compression perturbation we see from
Fig.\,\ref{figlambda}$b$ that, in contrast to $\lambda_{y{\rm R}}$,
$(\lambda_{x{\rm R}}/\lambda)^2$ does
increase from unity as $T_{\rm m}$ is approached from below.
This increase is clearly noticeable at temperatures 
sufficiently below $T_{\rm m}$ so that our data still has good
statistical accuracy. This is in contrast to a similar
increase in $\gamma_x$ in Fig.\,\ref{figamma}$b$ just below $T_{\rm m}$,
which we do not believe is statistically meaningful, but is rather
just a reflection of the scatter in our data.  Thus, as the
lattice starts to depin from the discretizing grid, a fluctuation in
vortex line density decays over an increasing length scale
$\lambda_{x{\rm R}}$.  Above $T_{\rm m}$, our numerical values are consistent
with $(\lambda_{x{\rm R}}/\lambda)^2\approx\gamma_x$, as expected from
Eqs.(\ref{egcomp}) and (\ref{eRcomp}) for the case $\lambda >a_{\rm v}$.
That $\lambda_{x{\rm R}}^2$ is indeed negative here, and so 
$\lambda_{x{\rm R}}$
is imaginary, can be seen in Fig.\,\ref{figYi}$b$ by noting that
the slopes of the fitted lines are equal to 
$\gamma_x^{-1}\lambda_{x{\rm R}}^2$, and that for $T>T_{\rm m}$, 
$\gamma_x<0$.

Finally, we turn to the shear perturbation. Since this
perturbation experiences a total Meissner screening in the
superconducting state, we may expect, in analogy with the
Meissner effect at $H=0$, that $\lambda_{z{\rm R}}^{-2}\sim
n_s$ where $n_s$ is the density of superconducting electrons
(not to be confused with $\rho_{s\,{\rm boson}}$, the superfluid density
of the analog 2D bosons).  Since $n_s$ decreases
as $T$ increases, vanishing at the superconducting transition,
we expect that $\lambda_{z{\rm R}}^2$ should increase with increasing
$T$, reaching a maximum at $T_{\rm c}$ (diverging in the case of 
a second order transition).  Precisely such behavior is
seen in Fig.\,\ref{figlambda}$c$. Above $T_{\rm c}$, $\lambda_{z{\rm R}}^2$
decreases to roughly the same small values as $\lambda_{y{\rm R}}$, as is
expected from Eqs.(\ref{entilt}) and (\ref{eHA2}).
Comparing results
for $N_z=15$ with $N_z=30$ we see that, similar to the behavior of
$\gamma_z$, the transition width sharpens and the temperature
of the peak in
$\lambda_{z{\rm R}}^2$ slightly decreases as $N_z$ is increased.
It is interesting to note however, that the value of
$\lambda_{z{\rm R}}^2$ at its peak has also very slightly decreased
as $N_z$ increased.  

The possibility that longitudinal superconductivity 
can persist into the vortex line liquid has been suggested
by the 2D boson analogy.
We can therefore compare the $T_{\rm c}$ found here with the
predictions of Sec.\,\ref{sboson}.  Rewriting Eq.(\ref{e2D4})
in terms of the dimensionless parameters of our numerical
simulation, and taking $dH_\perp/dB_\perp\approx 1$,
gives $T_{\rm c}=8\pi\bar{J}_0\kappa^2/N_z$.  Using 
$\kappa=5$ and $N_z=30$ gives $T_{\rm c}/\bar J_0\simeq 21$,
ten times larger than the value $1.8$ found numerically.
We can also estimate the entanglement cross-over of 
Nelson.  Eq.(\ref{e2D10.2}) gives $T_\times=\pi\bar{J}_0\ln\kappa/2b_0N_z$, 
and using $b_0=1/15$, $\kappa=5$, and $N_z=30$ gives 
$T_\times=1.26\approx T_{\rm m}$.  This is somewhat lower
than the observed $T_{\rm c}$.  Moreover,
both the boson superfluid transition temperature and the entanglement
temperature $T_\times$ should scale with system thickness as $1/N_z$.
In contrast, comparing $N_z=15$ with $N_z=30$, we see no such
dramatic shift in the numerically observed $T_{\rm c}\simeq 1.8$.

\subsection{Results: Vortex Line Fluctuations}

To elucidate the nature of the transitions in our model, we have
measured other properties to characterize the vortex line fluctuations 
in the system.  In Fig.\,\ref{fsnap} we show snapshot views of the vortex
line configurations for $N_z=15$, at various temperatures 
$T<T_{\rm m}$, $T_{\rm m}<T<T_{\rm c}$, and $T_{\rm c}<T$.  We
show both a side perspective and a view looking down along the 
applied field.  We see clearly that for $T<T_{\rm m}$ there is
a vortex line lattice.  For $T_{\rm m}<T<T_{\rm c}$ the lattice
is disordered but the vortex lines remain for the most part disentangled.
For $T_{\rm c}<T$ the lines are highly entangled.

For a quantitative determination of the vortex line lattice 
melting temperature, we compute the structure function of
vortices within the same $xy$ plane,
\begin{equation}
  S({\bf q}_\perp)={1\over L_z}\sum_{i,j}e^{i{\bf q}_\perp
  \cdot({\bf r}_i-{\bf r}_j)}\langle n_{iz}n_{jz}\rangle
  \delta_{z_i,z_j}
\label{estruc}
\end{equation}
Below $T_{\rm m}$ we expect to see Bragg peaks at the
reciprocal lattice vectors ${\bf K}$ of the vortex line lattice,
while above $T_{\rm m}$ we expect to see approximately circular
rings characteristic of a liquid.  Let us denote by $\{{\bf K}_1\}$
the six smallest non-zero reciprocal lattice vectors, and by
$\{{\bf K}_1^\prime\}$ the six vectors obtained by reflecting the
$\{{\bf K}_1\}$ through the $\hat{\bf x}$ axis.  Then since the ground
state vortex lattice of Fig.\,\ref{fground} breaks this reflection symmetry,
while the vortex line liquid restores it, the quantity $\Delta 
S\equiv S({\bf K}_1)-S({\bf K}_1^\prime)$, averaged over the
six $\{{\bf K}_1\}$, serves as a convenient order parameter for
the melting transition.  We plot $\Delta S$, normalized by
$S_0\equiv S({\bf K}=0)$, in Fig.\,\ref{fstruc}.  We see that $\Delta S$
vanishes at $T_{\rm m}\simeq 1.2$.  In an earlier work\cite{R10}
we have shown intensity plots of $S({\bf q}_\perp)$ in the
entire ${\bf q}_\perp$ plane.  The circular rings seen above $T_{\rm 
m}$ verify that $T_{\rm m}$ is indeed a melting to a liquid, and not
a depinning to a floating vortex lattice, or some other vortex
lattice structural transition.

As another measure of vortex line fluctuations, we have
computed the fluctuation length of the
vortex lines in the directions transverse and parallel to the applied
magnetic field.  The total length of vortex lines in the ground state
is ${\cal L}_0=b_0N_zN_\perp^2$.  If, in any configuration, ${\cal L}_\mu$
is the total length of all vortex lines in direction $\mu$ (we count
length here as an absolute quantity; oppositely oriented segments do 
not cancel each other out), then we define the
normalized excess vortex line lengths as $\Delta\ell_\perp=
({\cal L}_x+{\cal L}_y)/(2{\cal L}_0)$ and $\Delta\ell_z=
({\cal L}_z-{\cal L}_0)/{\cal L}_0$.  
We plot $\Delta\ell_\perp$ and $\Delta\ell_z$ in Fig.\,\ref{fdl}.
If we assume that all vortex
fluctuations consist of purely transverse motion of the magnetic
field induced lines, then $\Delta\ell_\perp$ is the average
transverse distance traveled by a vortex line between two adjacent 
$xy$ planes.  If we further assume that these lines are fluctuating as
in a random walk, then the total transverse deflection of a line
in traveling down the entire length of the system $N_z$ is,
$u=\sqrt{N_z}\Delta\ell_\perp$.  Entanglement should occur when
$u\simeq a_{\rm v}$, or when $\Delta\ell_\perp\simeq a_{\rm v}/\sqrt{N_z}$.
From Fig.\,\ref{fdl}, and using $a_{\rm v}\simeq1/\sqrt{b_0}=3.87$, we
would estimate the entanglement temperatures as $T_\times
\simeq 2.1$ for $N_z=15$, and $T_\times\simeq 1.9$ for $N_z=30$.
These are both consistent with the $T_{\rm c}$ seen in Fig.\,\ref{figgl}.
However, if the transition at $T_{\rm c}$ is indeed caused by the onset of
entanglement due to transverse wandering of magnetic field induced
vortex lines, it is necessary to explain how just above $T_{\rm c}$, where
$\gamma_z\simeq 0$, one can have a $\Upsilon_{\rm boson}/T_{\rm boson}
\simeq N_zT/(4\pi^2\bar{J}_0\kappa^2)\approx 0.06$ (see Eq.(\ref{e2D2}))
so much smaller than the lower bound $2/\pi$ given by the 
Kosterlitz--Thouless theory\cite{R14} of the analog boson superfluid transition.
We further note that in previous simulations\cite{R26,R27} with $\lambda\to\infty$,
where samples up to thickness $N_z=200$ were studied, the above
criterion gives a $T_\times$ which is well below the observed $T_{\rm 
c}$.

Returning now to Fig.\,\ref{fdl},
we see that the above assumption of strictly transverse fluctuations
of the field induced lines, while reasonable near the melting
$T_{\rm m}$ where $\Delta\ell_z/\Delta\ell_\perp\simeq 0.035$, is not
at all reasonable near $T_c$, where 
$\Delta\ell_z/\Delta\ell_\perp\simeq 0.41$.  The excess vorticity
along ${\bf\hat z}$ can only come  from either field induced lines
which wander backwards, or from closed vortex ring excitations.
Both these types of excitations are absent from the usual 2D boson analogy.

Using an algorithm we have describe elsewhere,\cite{R27} we trace out the vortex
line paths in 
our configurations to compute the distribution $q(p)$ of the number
of closed rings of perimeter $p$, normalized by the ground state
vortex line length ${\cal L}_0=b_0N_zN_\perp^2$.  
In Fig.\,\ref{fdlrings} we compare the total length of all
vortex line fluctuations, $\Delta\ell_{\rm tot}\equiv 2\Delta\ell_\perp
+\Delta\ell_z$, with the total length of all vortex ring excitations,
$\Delta\ell_{\rm ring}\equiv\sum_p pq(p)$.  We see that $\Delta\ell
_{\rm ring}\ll\Delta\ell_{\rm tot}$ through the melting $T_{\rm m}$,
however at $T_{\rm c}=1.8$, $\Delta\ell_{\rm ring}$ has increased to
$27\%$ of $\Delta\ell_{\rm tot}$.  
In Fig.\,\ref{frings} we show a semi-log plot of $q(p)$ vs. $1/T$.
The straight lines found at low $T$ indicate thermal activation
with a constant energy barrier that increases with ring size.
At high $T\sim 2.8$ the $q(p)$ curves saturate.  Note that the thermal 
activation for rings persists up to temperatures above $T_{\rm c}$.
This suggests that, although the number of rings is becoming sizable
near $T_{\rm c}$, the transition at $T_{\rm c}$ is not directly 
associated with any critical behavior of the rings.
This behavior is the same that we saw in simulations of a 3D
XY model, corresponding to $\lambda\to\infty$, when we took
$anisotropic$ couplings;\cite{R27} for 
isotropic couplings\cite{R26} in the XY model, 
the saturation of the $q(p)$ curves coincided with $T_{\rm c}$.   In 
Fig.\,\ref{fspecht} we plot the specific heat $C$ vs. $T$, for 
$N_z=30$.  We see that $C$ rises smoothly through $T_{\rm c}$.  The peak occurs
near $T\sim 3.0$ (we only have enough data at high $T$ to locate
it very crudely), where  the $q(p)$ curves saturate.  
The peak in $C$ is thus associated with the proliferation
of the closed vortex rings, which we believe to be a non-singular cross-over 
phenomenon associated with the transition of the zero field $b_0=0$ model, 
which occurs\cite{R47} at $T_{\rm c 0}\approx 3$.  
The peak in $C$ is also probably associated\cite{R27}
with the onset of a strong diamagnetic response in the system, 
which occurs at the so-called ``mean field $H_{c2}(T)$'' line.
 
Finally, we consider the entanglement of the magnetic field induced
vortex lines.  Due to the periodic boundary conditions
along ${\bf\hat z}$, the set of points $\{ {\bf r}_{\perp i}(N_z)\}$
where the field induced vortex lines pierce the $xy$ plane at $z=N_z$, 
must be some permutation of the set of points 
$\{ {\bf r}_{\perp i}(0)\}$ where the lines pierce the $xy$ plane
at $z=0$.  Lines for which ${\bf r}_{\perp i}(N_z) ={\bf r}_{\perp j}
(0)$, with $i\ne j$, form part of 
an entangled braid when viewed in the periodically
repeated system.  We can thus classify each magnetic field induced line
as belonging to a given braid of order $m$, according to the number
of lines $m$ that are mutually entangled in the preceding sense.
We compute the distribution $n(m)$ giving the average number of lines in
a braid of order $m$, where $\sum_m n(m)=b_0N_\perp^2$ is just the
total number of field induced lines.

In the 2D boson analogy, such entangled vortex lines
represent particle exchanges.  A superfluid state of these
2D bosons is expected when there are many such exchanges, and
in particular when there is a finite probability to form large
exchanges involving a macroscopic fraction of the particles,\cite{R34}
that wrap entirely around the system in the transverse direction and
thus contribute to $n_{z0}\equiv \lim_{q\to 0}\langle n_y(q\hat{\bf 
x})n_y(-q\hat{\bf x})\rangle$.  $\gamma_z$
is a direct measure of $n_{z0}$ (see Eq.(\ref{eRg})), and
hence a measure of the presence of such large exchanges.  
In Fig.\,\ref{figR} we plot vs. $T$ the fraction of lines $R=n(1)/b_0N_\perp^2$  
which are not involved in {\it any} particle exchanges, i.e. the fraction of
unentangled vortex lines for which 
${\bf r}_{\perp i}(N_z) ={\bf r}_{\perp i}(0)$.
We see that $R=1$ and all lines remain unentangled up to
$T\simeq T_{\rm c}$, at which point $R$ decreases towards zero.
The width of the decrease in $R$ is roughly the same as the
width of the decrease in $\gamma_z$, for both $N_z=15$ and $30$.

In Fig.\,\ref{fignm} we plot the entanglement distribution $n(m)$ vs. 
$m$, for several values of $T$ near and above $T_{\rm c}=1.8$,
for $N_z=30$.  We see that the distribution broadens as $T$ increases,
indicating larger particle exchanges, however no sharp feature is obvious 
as $T$ increases through $T_{\rm c}$.  This is in contrast to what we
observed in simulations\cite{R26,R27} of the $\lambda\to\infty$ 3D XY model,
where $n(m)$ got dramatically flat and equal to unity over a wide 
range of intermediate $m$ as $T$ reached $T_{\rm c}$; these XY
simulations however used much larger system thicknesses,
$N_z\simeq 100-200$, and this is one possible reason for the
difference in behavior from the present case.

\section{Conclusions and Discussion}

The main conclusion of our numerical work is that longitudinal
superconductivity vanishes at a $T_{\rm c}$ which lies well 
within the vortex line liquid, at least
for the system sizes we have been able to investigate.  
We note that our system sizes $N_\perp=30$, and $N_z=15$, $30$,
are large compared to the microscopic length scales of our
model, $\lambda/\xi_\perp=5$ and $a_{\rm v}/\xi_\perp\simeq\sqrt{15}$.  
We have discussed a mechanism for this phenomenon in terms of the KT 
superfluid transition of the analog 2D bosons.  However the
$T_{\rm c}$ predicted by Eq.(\ref{e2D4}) is an order
of magnitude larger than the numerically observed value.
The entanglement temperature $T_\times$ of Nelson is of the
correct order of magnitude as the observed $T_{\rm c}$.
Figs.\,\ref{fsnap} and \ref{figR} also suggest a connection between
geometrical entanglement and $T_{\rm c}$.  However
upon comparing $N_z=15$ and $30$, we failed to see any sign of
the dramatic size dependence $T_\times\propto 1/N_z$ that is
expected from Eq.(\ref{e2D10.2}).  A similar size dependence
is also expected for the KT prediction of Eq.(\ref{e2D4}).

In earlier simulations\cite{R25,R26,R27} of a 3D XY model, corresponding to the 
$\lambda\to\infty$ approximation of Sec.\,\ref{slaminf}, we have
studied much thicker systems than reported on here, with $N_z$
as large as $200$.  We again found longitudinal superconductivity
to vanish at a $T_{\rm c}^\infty$ within the vortex line liquid, with
virtually no finite size effects in the apparent value of $T_{\rm 
c}^\infty$ as $N_z$ was varied.  An analysis\cite{R26,R27} 
of geometrical entanglement, 
as done here in connection with Fig.\,\ref{fdl}, gives a $T_\times$ well
below the observed $T_{\rm c}^\infty$ for the thicker systems, and
the dependence of $T_{\rm c}^\infty$ on the system anisotropy was
found\cite{R27} to be $T_{\rm c}^\infty\propto 1/\eta$, rather than the
$T_\times\propto 1/\eta^2$ predicted by Eq.(\ref{e2D10.2}).
New simulations\cite{R48} have further shown that
there is no apparent change in the large $N_z$
limiting value of $T_{\rm c}^\infty$ when
the periodic boundary conditions along the direction $\hat{\bf z}$ of the
applied magnetic field are replaced with the more
realistic free boundary conditions.  We believe that these 
$\lambda\to\infty$ simulations are therefore in good agreement with
the work of Feigelman and co-workers,\cite{R20,R42} who argued for just such a
superconducting to normal vortex line liquid transition, with a
$T_{\rm c}^\infty$ which remains finite as $L_z\to\infty$.
We believe that this transition of Feigelman {\it et al}.
applies strictly to the $\lambda\to\infty$ model, and represents
a $T_{\rm boson}=0$ Meissner transition, as $\hbar_{\rm boson}$
varies, for the analog 2D charged bosons.

Returning to our present simulations, we believe
that our results represent a finite $\lambda$ cross-over from the above 
$\lambda\to\infty$ transition at $T_{\rm c}^\infty$.  
Although we believe that the $\lambda\to\infty$
limit is extremely subtle, one may imagine the following scenario.
When $\lambda$ is large, although the analog 2D bosons behave
like a neutral superfluid on sufficiently long transverse length
scales, on small length scales they will have the $\lambda\to\infty$
behavior of $charged$ bosons.  We would then expect the 2D boson
helicity modulus to have, at finite transverse wavevector $q$,
a piece that looks like that of Eq.(\ref{eli6}).  We thus
expect a form like,
\begin{equation}
\Upsilon_{\rm boson}(q)=\Upsilon_{\rm boson}(0)+
\gamma_{\rm boson}{[J\lambda^2]_{\rm boson}q^2\over 1+
\lambda_{\rm R\, boson}^2q^2}\enspace.
\label{econ1}
\end{equation}
As $q\to 0$, it is $\Upsilon_{\rm boson}(0)$ that
determines if the 2D bosons are in a superfluid ($\Upsilon_{\rm boson}(0)>0$)
or a normal fluid ($\Upsilon_{\rm boson}(0)=0$) state; but at
sufficiently large $q$ it will be the second term that
dominates, giving the appearance of a charged boson system.
For $T<T_{\rm c}$, with $T_{\rm c}$ the 2D neutral boson
superfluid transition of Eq.(\ref{e2D4}), one has $\Upsilon_
{\rm boson}(0)=0$ and only the second term is present.
As $q\to 0$, this term vanishes, and  Eq.(\ref{e2D2})
then gives $\gamma_z=1$, i.e. we have the perfect shear
Meissner effect that we expect for the 2D boson normal fluid phase,
as discussed in Sec.\,\ref{sanalogy}.
However, if $L_z$ is thin enough that $T_{\rm c}^\infty\ll T_{\rm c}$, with 
$T_{\rm c}^\infty$ the Meissner transition of the $\lambda\to\infty$
2D charged boson model, then as one cools down to $T_{\rm c}^\infty$,
one expects $\lambda_{\rm R\, boson}$ will become large, and
possibly of order the finite transverse size $L_\perp$ of the system.  
In this case, for all available wavevectors, $q>2\pi/L_\perp$ yields 
$\lambda_{\rm R\,boson}^2q^2\gg 1$, and
the second term becomes approximately the constant, 
$\gamma_{\rm boson}[J\lambda^2]_{\rm boson}/\lambda_{\rm R\, 
boson}^2$.   Eq.(\ref{e2D2}) then gives $\gamma_z=1-[\gamma_{\rm boson}
\lambda_\perp^2/\lambda_{\rm R\,boson}^2]$.  It thus appears
as if the perfect shear Meissner effect has been lost at the lower
temperature $\sim T_{\rm c}^\infty$.   We note that for this scenario 
to agree with the small values of $\gamma_z$ that we find at
temperatures above our
numerically observed value of $T_{\rm c}$, it would be necessary to
have $\gamma_{\rm boson}\lambda_\perp^2/\lambda_{\rm 
R\,boson}^2\approx 1$; it is not apriori obvious why this
would be so.

Thus for a finite $\lambda$ simulation to see other than the
above $\lambda\to\infty$ cross-over behavior, it would be 
necessary to do one of the following.  One could increase
the transverse size $L_\perp$, keeping $L_z$ constant, until
one is in the limit where $\max[\lambda_{\rm R\, boson}]\ll L_\perp$
(although in the $\lambda\to\infty$ model $\lambda_{\rm R\,boson}$
might diverge at $T_{\rm c}^\infty$, in a finite $\lambda$
model any such divergence would be rounded out to a finite
maximum value).  In this limit, the second term in Eq.(\ref{econ1})
would be observed to be $\sim q^2$, and so one would find
$\gamma_z=1$.  One thus expects the apparent $T_{\rm c}$ to 
increase above $T_{\rm c}^\infty$ as $L_\perp$ increases above
$\max[\lambda_{\rm R\,boson}]$.  One might never actually reach
the true 2D boson neutral superfluid transition $T_{\rm c}$
of Eq.(\ref{e2D2}), since as temperature increases, thermally
excited closed vortex rings will start to proliferate, and
vortex lines can make long transverse wanderings between two
adjacent $xy$ planes; both such effects are left out of the
naive mapping to 2D boson statistical mechanics.  Alternatively,
one could keep $L_\perp$ constant, but increase $L_z$, so that
the $T_{\rm c}$ of Eq.(\ref{e2D2}) falls below $T_{\rm c}^\infty$.
For our parameters, Eq.(\ref{e2D2}) suggests that this would require 
a system of thickness $N_z= \phi_0^2/2\pi^2T_{\rm c}^\infty
=8\pi\bar J_0\kappa^2/T_{\rm c}^\infty\simeq 320$.  

Several other groups have done simulations similar to ours.
Most of these\cite{R28,R28.1} have been in the $\lambda\to\infty$ 
limit, but at much higher vortex line densities such as
$b_0=1/6$.  In these cases it was found that $T_{\rm c}\approx
T_{\rm m}$, and so no longitudinal superconductivity was
observed in the vortex line liquid.  We believe that this is
a consequence of the high densities $b_0$ which have been used.
Recently, we have studied\cite{R27} the phase diagram in such $\lambda\to\infty$
XY models, as a function of the system anisotropy $\eta$.  Increasing
$\eta$ at fixed $b_0$ can be argued to play a role similar to
increasing $b_0$ at fixed $\eta$.  We found that as $\eta$ increased,
$T_{\rm c}$ and $T_{\rm m}$ came closer together, and eventually
became indistinguishable from each other.  Similar results have
recently been reported in simulations by Koshelev.\cite{R49}

\v{S}\'{a}\v{s}ik and Stroud have done
simulations\cite{R50} for the $\lambda\to\infty$ limit
using the lowest Landau level approximation, which treats the $xy$ planes
as a continuum and so avoids the artificial pinning of our discretized
London model.  For all values of anisotropy studied they find
$T_{\rm c}\approx T_{\rm m}$.  However Te\v{s}anovi\'c\cite{R32} has argued
that the lowest Landau level approximation fails as the magnetic
field decreases, and so at such low magnetic fields, the London and the
lowest Landau level approaches need not be in agreement.  
Using a mean field analysis, Te\v{s}anovi\'c\cite{R32} has argued that
longitudinal superconductivity can persist into the vortex line
liquid in this low field limit.

Finite $\lambda$ simulations have been carried out, for the 
same discretized London model as considered here, by 
Carneiro.\cite{R8,R51}
For large line densities, he finds $T_{\rm c}\approx T_{\rm m}$,
consistent with the above $\lambda\to\infty$ results.  For
line densities comparable to our own, he finds $T_{\rm c}$ noticeably
above $T_{\rm m}$, when following our analysis based on the $q$
dependence of the helicity modulus within the Helmholtz ensemble of
fixed internal magnetic field $b_0$.  He has suggested\cite{R51} however that
the result may be very sensitive to the $q\to$ extrapolation
implied by fitting to the expansion of Eq.(\ref{eNS18}), with
different results obtained when truncating at different orders
of the expansion, or when using a different number of $q$ data points
in the fit.  However our fits of Figs.\,\ref{figamma} and \ref{figlambda}
show essentially no sensitivity to the number of $q$ data points used,
or when comparing a linear versus a quadratic order fit to the data
of Fig.\,\ref{figYi}.  Carneiro has also carried out
simulations\cite{R8,R51} in a Gibbs ensemble, in which the total transverse
magnetic field is allowed to fluctuate.  Here he concludes $T_{\rm c}
\approx T_{\rm m}$, even for dilute densities comparable to our own.  
However we believe\cite{R52} that in this case, his
${\bf q}=0$ calculation of the fluctuation in the transverse magnetic field
cannot distinguish between the shear perturbation of Fig.\,\ref{f1}$c$,
which is related to the 2D analog boson superfluid density, and
the tilt perturbation of Fig.\,\ref{f1}$a$, which is not.  We 
believe that his results are reflecting the softening of $c_{44}$ 
that occurs at the depinning/melting transition (as is observed
in our Fig.\,\ref{figamma}$a$), rather than
reflecting the loss of longitudinal superconductivity.

Most recently, simulations at finite $\lambda$ have been carried out
by Nguyen {\it et al}.,\cite{R53} who extend our work to consider behavior as
the anisotropy $\eta$ is varied.  For an isotropic system, they
find $T_{\rm c}$ well above $T_{\rm m}$, in good agreement with our
results.  However as $\eta$ increases, they find the very intriguing
result that $T_{\rm c}$ decreases, and eventually falls $below$
$T_{\rm m}$.  Such a possibility (not observed in similar 
$\lambda\to\infty$ simulations\cite{R27}) has been suggested by Frey {\it et
al}.\cite{R35} as a result of dislocations proliferating in the vortex line
lattice.  Glazman and Koshelev\cite{R6} have made similar predictions, based
on the effect that vortex lattice elastic fluctuations have in
reducing the interplanar Josephson coupling.  However Nguyen {\it et
al}. suggest that their result is due to the proliferation of 
vortex rings between adjacent $xy$ plans, and they find at high
anisotropy that $T_{\rm c}\sim\eta^{-2}$, rather than the 
$T_{\rm c}\sim\eta^{-1}$ predicted by Ref.[\onlinecite{R6}]
or the $T_{\rm c}\sim 1/\ln\eta$ predicted by Ref.[\onlinecite{R35}].
It should be noted however that Nguyen {\it el al}. base their
criterion for superconductivity on computing the helicity modulus
at the single smallest non-zero value of $q$ allowed by their finite size
system.  We have argued above that a more careful analysis should
be based on parameters extracted from the $q$ dependence
of the helicity modulus, as $q\to 0$.  Conclusions based on 
calculations at specific values of finite $q$ can more easily be led 
astray by subtle cross-over effects such as we have discussed above.
Clearly more systematic studies, using our $q\to 0$ analysis, and
making a more extensive study of finite size dependencies, need
to be done for both the isotropic and anisotropic cases.

Our result of Eq.(\ref{e2D7}) suggests that one should find
$T_{\rm c}>T_{\rm m}$, and hence longitudinal superconductivity
within a region of the vortex line liquid, whenever a sample
is thinner than $L_{z\,{\rm max}}\approx 1400\mu{\rm m}$,
for a $T_{\rm m}\approx 90^\circ$K.  Virtually all experimental
single crystal samples fall below this critical thickness.
One can therefore ask whether any experimental evidence favors
our conclusions.  Naively, one would expect a vortex line
liquid with longitudinal superconductivity to show a finite
linear resistivity transverse to the applied magnetic field,
but zero linear resistivity parallel to the applied field.
However, in MC simulations of a $\lambda\to\infty$ XY model,\cite{R25}
we found that in the intermediate
phase $T_{\rm m}<T<T_{\rm c}$ vortex density correlations decayed
anomalously slowly with time.  This suggests that vortex lines
may be moving more slowly than diffusion, and if so, it is
not obvious what to expect for the transverse resistivity.
Experimentally, it is very difficult to obtain accurate measurements
of the longitudinal resistivity, due to the slab geometry of
single crystal samples, and the non-uniformity of current
distributions.  Transverse resistivity measurements are intimately
related to the vortex pinning impurities in the sample, and
so are also not unambiguously characterized.

Nevertheless, the following suggestive observations have been made.
Experiments by Steel {\it et al}.\cite{R54} on artificially prepared
MoGe/Ge layered superconductors, found that the d.c. resistivity parallel to 
the applied field decreased sharply, and showed an
onset of strong nonlinear behavior, at a temperature above that where
the transverse resistivity vanished.  Experiments by Kwok {\it et 
al}.\cite{R55} on YBCO, studying the pinning of vortex lines to twin grain
boundaries in a system with a well controlled small number of twin planes, 
found evidence for a sharp lock-in pinning transition at a
temperature above vortex lattice melting (where the melting transition
was determined by the observation of a sharp drop in
transverse resistivity).  Such a lock-in transition within the
vortex line liquid may suggest a transition in the nature of vortex line
fluctuations, as at our $T_{\rm c}$.  Early experiments by Safar
{\it et al}.,\cite{R56} using a flux transformer geometry, similarly showed
evidence for the onset of coherence parallel to the applied field
at a ``$T_{th}$'' above the temperature where the transverse resistivity
vanished.  However more recent flux transformer experiments by L\'{o}pez {\it et
al}.\cite{R57} showed that these two temperatures merged when the sample was
made purer, with all twin grain boundaries eliminated.  Moore\cite{R58} has
recently proposed an interpretation which argues
that the single transition observed in these newer transformer
experiments is the result of some very rapidly increasing 
longitudinal length scale, rather than being a first order vortex
lattice melting transition, as is the usual interpretation.
If correct, such a rapidly increasing longitudinal length scale
might be associated with our $T_{\rm c}$.

The experimental evidence cited above remains, at best, inconclusive.
There are several possible reasons why observing a $T_{\rm c}>T_{\rm 
m}$ might be experimentally difficult.  Firstly, as in our simulations,
the relevant temperature is likely to be the $\lambda\to\infty$ transition
$T_{\rm c}^\infty$, rather than the much higher $T_{\rm c}$ of
Eq.(\ref{e2D4}).  In recent simulations\cite{R27} of the $\lambda\to\infty$ XY
model we found that $T_{\rm c}^\infty$ and $T_{\rm m}$ merged as the
anisotropy $\eta$ increased. How far apart the corresponding 
$T_{\rm c}^\infty$ and $T_{\rm m}$ for any particular real material are
likely to be, remains unknown.  Secondly, real layered high
temperature superconductors are likely to have an interplanar
Josephson coupling that is proportional to the cosine of the
phase angle difference across adjacent planes.  The non quadratic
nature of such a cosine interaction leads to a coupling between
spin wave and vortex fluctuations that is absent in both
our continuum model and our discretized model of Eq.(\ref{eNS2})
using the Villain interaction. As either magnetic field, temperature,
or anisotropy increases, large interplanar phase differences
can be induced by elastic vortex line fluctuations, leading to
a large decrease in the effective interplanar energy coupling
constant.  Such a
``decoupling'' cross-over, as discussed by Glazman and 
Koshelev,\cite{R6} and Daemen {\it et al}.,\cite{R59} 
might obscure any true critical behavior at
a higher $T_{\rm c}$.  Finally, the free boundary conditions of
a real superconductor, as opposed to the periodic boundary conditions
of the 2D boson mapping and of our simulations, might lead to a more
effective washing out of the 2D boson superfluid transition\cite{R21} 
than we have imagined, for samples of experimental thickness.

\section*{Acknowledgments}

In the course of this work we have benefited greatly from
discussions with C. Ciordas-Ciurdariu, M. Feigelman, 
A. E. Koshelev, M. C. Marchetti, P. Muzikar, 
D. R. Nelson, Z. Te\v{s}anovi\'c, and A. P. Young.
This work has been supported by U.S. Department of Energy
Grant DE-FG02-89ER14017.

\appendix
\section{Superfluid Density of 2D Bosons in the Path Integral 
Formulation}
\label{s2Db}

The superfluid density of a system of $N$ interacting bosons 
can be defined in terms of the response of the 
system to the presence of a heat bath moving with velocity
${\bf v}({\bf r})$ (``moving walls'').  In the following, all
position, velocity, and wave vectors are two dimensional vectors in the
$xy$ plane.

The average 
momentum density $\langle p_{q\mu}\rangle_v$ that results in
linear response to the heat bath velocity $v_{-q\nu}={1\over L^2}\int d^2r
v_\mu({\bf r})e^{-i{\bf q}\cdot{\bf r}}$ is given by
\begin{equation}
   \langle p_{q\mu}\rangle_v=\chi_{\mu\nu}({\bf q})v_{-q\nu}.
\label{ec1}
\end{equation}
For an isotropic 2D system, the momentum density susceptibility
can be written in terms of its longitudinal and transverse pieces,
\begin{equation}
   \chi_{\mu\nu}({\bf q})={\bf\hat q}_{\mu}{\bf\hat q}_{\nu}\chi_L({\bf q})
   +[\delta_{\mu\nu}-{\bf\hat q}_\mu{\bf\hat q}_\nu]\chi_T({\bf q}).
\label{ec2}
\end{equation}
The number density of superfluid bosons $\rho_s$ is then given in 
terms of the transverse susceptibility by\cite{R19,R19.1}
\begin{equation}
   m\rho_s = m\rho -\lim_{q\to 0}\chi_T({\bf q})
\label{ec3}
\end{equation}
where $m$ is the boson particle mass, and $\rho=N/L^2$ 
is the total boson density.

For a system of interacting bosons in the presence of a moving heat 
bath, the Hamiltonian in the reference frame of the heat bath
is given by,
\begin{equation}
  {\cal H}=\sum_i{1\over 2m}({\bf p}_i-m{\bf v}({\bf r}_i))^2
   +V(\{{\bf r}_i-{\bf r}_j\})
\label{ec4}
\end{equation}
where the interaction $V$ depends only on the bosons relative positions.

The 
partition function is ${\cal Z}={\rm Tr}[e^{-\beta{\cal H}}]$, and the free 
energy is ${\cal F}=-T\ln {\cal Z}$, with $T=1/\beta$.
Consider now that ${\bf v}$ points only in the ${\bf\hat y}$ direction
and varies only in the ${\bf\hat x}$ direction (so the ${\bf v}$
is purely transverse).  
If we write ${\cal H}[{\bf v}]={\cal H}[0]+\delta{\cal H}[{\bf v}]$,
then since in the $q\to 0$ limit that ${\bf v}$ becomes a uniform constant
$\delta{\cal H}$ and ${\cal H}[0]$ commute, one has

\begin{equation}
   \lim_{q\to 0}
   L^2\left.{\partial^2{\cal F}\over\partial v_y(q\hat{\bf x})
   \partial v_y(-q\hat{\bf x})}
   \right|_{v=0}= m\rho- \lim_{q\to 0} \chi_T(q\hat{\bf x}) =m\rho_s
\label{ec5}
\end{equation}

To evaluate $\rho_s$ in terms of the path integral 
formalism,\cite{R60} one now
writes the Lagrangian associated with the Hamiltonian of 
Eq.(\ref{ec4}), and transforms from real time $t$ to imaginary
time $\tau=it$.  One gets
\begin{equation}
  {\cal L}(\tau)=-\sum_i{m\over 2}\left({d{\bf r}_i\over
   d\tau}\right)^2-V(\{{\bf r}_i-{\bf r}_j\})+im\sum_i
   \left({d{\bf r}_i\over d\tau}\right)\cdot {\bf v}({\bf r}_i).
\label{ec6}
\end{equation}
The partition function is then given by
\begin{equation}
  {\cal Z}=\int{\cal D}[\{{\bf r}_i(\tau)\}]e^{\hbar^{-1}
   \int_0^{\hbar\beta} d\tau
  {\cal L}(\tau)}.
\label{ec7}
\end{equation}
where the sum is over all possible boson world lines $\{{\bf 
r}_i(\tau)\}$ subject to permuted periodic boundary conditions, 
i.e. $\{{\bf r}_i(0)\}={\cal P}\{{\bf r}_i(\beta)\}$ where
${\cal P}$ is any permutation of the $N$ bosons.

Applying Eq.(\ref{ec5}) to the above form for ${\cal Z}$ then results 
in
\begin{equation}
  m\rho_s = \lim_{q\to 0}{Tm^2\over 
   L^2\hbar^2}\left\langle
   \left(\int_0^{\hbar\beta} d\tau\sum_i{dr_{iy}\over d\tau}e^{iqx_i}
   \right)
   \left(\int_0^{\hbar\beta} d\tau^\prime
   \sum_j{dr_{jy}\over d\tau^\prime}e^{-iqx_j}
   \right)\right\rangle_0
\label{ec8}
\end{equation}
where $\langle\ldots\rangle_0$ denotes an average over world lines
weighted by the Lagrangian factor as in Eq.(\ref{ec7}),
only now taking ${\bf v}=0$ in ${\cal L}$.

Note that the heat bath velocity ${\bf v}({\bf r})$ enters the
Hamiltonian (\ref{ec4}) and the Lagrangian (\ref{ec6}) with
precisely the same form as would a 2D external magnetic vector potential 
given by ${\bf v}=(\hbar/ m){\bf A}^{\rm ext}$ (where, as in 
Sec.\,\ref{slondon}, 
the units of ${\bf A}^{\rm ext}$ are such that ${\bf\nabla}\times{\bf A}
^{\rm ext}=(2\pi/\phi_0)H\hat{\bf z}$, with $H\hat{\bf z}$ the 2D magnetic
field).  
In analogy with Eq.(\ref{ehm2}) we can thus define the helicity
modulus of the 2D bosons as,
\begin{eqnarray}
  \Upsilon_{\rm boson}(q)&=&L^2{\partial^2{\cal F}\over\partial A_y(q\hat{\bf x})
  \partial A_y(-q\hat{\bf x})}=L^2{\hbar^2\over m^2}\,
  {\partial^2{\cal F}\over\partial v_y(q\hat{\bf x})
  \partial v_y(-q\hat{\bf x})}
\nonumber\\
  &=&{T\over L^2}
  \left\langle
  \left(\int_0^{\hbar\beta} d\tau\sum_i{dr_{iy}\over d\tau}e^{iqx_i}
  \right)\left(\int_0^{\hbar\beta} d\tau^\prime
  \sum_j{dr_{jy}\over d\tau^\prime}e^{-iqx_j}
  \right)\right\rangle_0\enspace,
\label{ec9}
\end{eqnarray}
with,
\begin{equation}
   \lim_{q\to 0}\Upsilon_{\rm boson}(q)={\hbar^2\over m}\rho_s
   =T\langle W_y^2\rangle_0\enspace,
\label{ec10}
\end{equation}
where $W_y$ is the $y$ component of the ``winding number'' introduced
by Pollock and Ceperley\cite{R34} in their path integral approach to the
superfluid transition in boson systems.

Our derivation above can be modified in a straightforward way
to deal with a boson interaction mediated by a gauge field,
as is the case for the more realistic London interaction
between vortex lines.\cite{R20}  One just replaces the pair potential
$V(\{{\bf r}_i-{\bf r}_j\})$ with the necessary coupling to
the gauge field, and free field energy terms.  However the
coupling of the bosons to an external vector potential remains
unchanged.  Thus the expression for the 2D boson helicity modulus 
in terms of boson world lines remains unchanged from Eq.(\ref{ec9}).

\section{Elastic Moduli}
\label{sselmod}

In this appendix we summarize some results concerning the elastic
moduli which appear in Eq.(\ref{eHel}).  Although calculations of these
moduli have appeared elsewhere,\cite{R12,R12.1,R37} 
our explicit computation of the 
order $q^2$ dependence at small $q$ we believe is new.

As shown by Sudb{\o} and Brandt,\cite{R12.1}
the elastic tensor $\Phi_{\alpha\beta}({\bf q})$ can be expressed in
terms of the vortex line interaction tensor $V_{\alpha\beta}({\bf q})$, as

\begin{equation}
   \Phi_{\alpha\beta}({\bf q})={B^2\over 4\pi\lambda_\perp^2}\sum_K
   \left\{q_z^2V_{\alpha\beta}({\bf K-q})+({\bf K-q})_\alpha
   ({\bf K-q})_\beta V_{zz}({\bf K-q})-K_\alpha K_\beta V_{zz}({\bf K})
   \right\}
\label{eAp1}
\end{equation}
where $\{ {\bf K} \}$ are the reciprocal lattice vectors of the vortex lattice.

For ${\bf B}=B{\bf\hat z}$ the elastic moduli we are 
interested in can be expressed in terms of
$\Phi_{\alpha\beta}({\bf q})$ as

\begin{equation}
   c_{66}(q{\bf\hat y})={1\over q^2}\Phi_{xx}(q{\bf\hat y}),\qquad   
   c_{11}(q{\bf\hat x})={1\over q^2}\Phi_{xx}(q{\bf\hat x}),\qquad
   c_{44}(q{\bf\hat z})={1\over q^2}\Phi_{xx}(q{\bf\hat z}).
\label{eAp2}
\end{equation}

For the London interaction, the sum over ${\bf K}$ in Eq.(\ref{eAp1})
is divergent, and some method must be employed to
make it converge.  As shown by Brandt,\cite{R12} this can be achieved for 
$c_{66}(q{\bf\hat y})$ and $c_{11}(q{\bf\hat x})$  by 
subtracting off the self energy of a line interacting with
itself.  This then gives,

\begin{eqnarray}
  c_{66}(q{\bf\hat y}) &=& {B^2\over 4\pi\lambda_\perp^2}\left\{\sum_K
  F_{66}[{\bf K},q]-{\phi_0\over B}\int {d^2k\over (2\pi)^2}F_{66}
  [{\bf k},q]\right\}
\label{eAp4a}\\
  c_{11}(q{\bf\hat x}) &=& {B^2\over 4\pi\lambda_\perp^2}\left\{\sum_K
  F_{11}[{\bf K},q]-{\phi_0\over B}\int {d^2k\over (2\pi)^2}F_{11}
  [{\bf k},q]\right\},
\label{eAp4b}
\end{eqnarray}
where we find, after expanding $\Phi_{xx}$ to $O(q^4)$, averaging over 
the orientation of the vortex lattice in the $xy$ plane, and
substituting in for $V_{zz}$ from Eq.(\ref{eVq})

\begin{eqnarray}
   F_{66}[{\bf k},q] &=& {d\over dk^2}
   \left\{ {\textstyle{ 1\over 4}}k^4 \dot{V}_{zz}({\bf k})+
   \left[{\textstyle{ 1\over 8}}k^4
   \ddot{V}_{zz}({\bf k})+{\textstyle{ 1\over 24}}k^6
   {\buildrel ... \over V}_{zz}({\bf k})
   \right]q^2\right\}
\label{eAp5a1}\\
   &=&-{\textstyle{1\over 2}}\lambda_\perp^2\left\{{1\over 
   (1+\lambda_\perp^2k^2)^2}-{1\over (1+\lambda_\perp^2k^2)^3}\right.
   \nonumber\\
   & \> &\qquad\qquad\qquad\qquad+\left.\left[
   {1\over (1+\lambda_\perp^2k^2)^3}-{3\over (1+\lambda_\perp^2k^2)^4}
   +{2\over (1+\lambda_\perp^2k^2)^5}\right]\lambda_\perp^2q^2\right\}
\label{eAp5a2}
\end{eqnarray}
and

\begin{eqnarray}
   F_{11}[{\bf k},q]&=&{d\over dk^2} 
   \left\{k^2V_{zz}({\bf k}) + {\textstyle{ 3\over 4}}k^4
   \dot{V}_{zz}({\bf k})+\left[
   k^2\dot{V}_{zz}({\bf k})+{\textstyle{ 9\over 8}}
   k^4\ddot{V}_{zz}({\bf k})+{\textstyle{ 5\over 24}}k^6
   {\buildrel ...\over V}_{zz}({\bf k})\right] q^2\right\}
\label{eAp5b1}\\
   &=&-{\textstyle{1\over 2}}\lambda_\perp^2\left\{{1\over 
   (1+\lambda_\perp^2k^2)^2}-{3\over (1+\lambda_\perp^2k^2)^3}\right.
   \nonumber\\
   & \>  &\qquad\qquad\qquad\qquad+\left.\left[
   {1\over (1+\lambda_\perp^2k^2)^3}-{9\over (1+\lambda_\perp^2k^2)^4}
   +{10\over (1+\lambda_\perp^2k^2)^5}\right]\lambda_\perp^2q^2\right\}
\label{eAp5b2}
\end{eqnarray}
where $\dot{V}_{zz}\equiv dV_{zz}/dk^2$.  

To treat the tilt modulus $c_{44}(q{\bf\hat z})$, self interactions
of the vortex lines are important.  One therefore handles the convergence
of the sum in Eq.(\ref{eAp1}) by introducing a convergence factor into
the London interaction of Eq.(\ref{eAp1}), 
$V_{q\alpha\beta}\to  V^c_{q\alpha\beta}\equiv
g(\xi_\perp^2q_\perp^2+\xi_z^2q_z^2)V_{q\alpha\beta}$.  
Here $g(x)\to 1$ for $x<1$,
$g(x)\to 0$ for $x> 1$, and one uses an anisotropic cutoff to model 
the vortex core, $\xi_z/\xi_{\perp}=
\lambda_{\perp}/\lambda_z=1/\eta$.  Averaging over the orientation of
the vortex lattice in the $xy$ plane, and substituting in for $V_{\mu\mu}$ 
from Eq.(\ref{eVq}) we find

\begin{equation}
   c_{44}(q{\bf\hat z})={B^2\over 4\pi\lambda_\perp^2}\sum_K F_{44}[{\bf K},q]
\label{eAp8a}
\end{equation}
where

\begin{eqnarray}
   F_{44}[{\bf k},q]&=&
   \left\{V^c_{xx}({\bf k})+\textstyle{1\over 2}k^2\dot{V}^c_{zz}({\bf k})+
   \left[\dot{V}^c_{xx}({\bf k})+\textstyle{1\over 4}k^2\ddot{V}^c_{zz}({\bf k})
   \right]q^2\right\}
   \\
   &=&{1\over 2}\left\{{\lambda_\perp^2g\over 
   1+\lambda_z^2k^2}+{\lambda_\perp^2g\over (1+\lambda_\perp^2k^2)^2}+
   \dot{g}-{\dot{g}\over 1+\lambda_\perp^2k^2}\right.
   \nonumber\\
   &  - &\left.\left[
   {\lambda_\perp^2g\over (1+\lambda_z^2k^2)^2}+
   {\lambda_\perp^2g\over (1+\lambda_\perp^2k^2)^3}-
   {\dot{g}\over 1+\lambda_z^2k^2}-{\dot{g}\over (1+\lambda_\perp^2k^2)^2}
   -{\ddot{g}\over 2}+{\ddot{g}\over 2
   (1+\lambda_\perp^2k^2)}\right]\lambda_\perp^2q^2\right\}
\label{eAp8}
\end{eqnarray}
where $\dot{V}^c_{\mu\mu}\equiv dV^c_{\mu\mu}/dq_z^2$ and $\dot{g}\equiv
dg/dq_z^2$.

We consider first the limit of large magnetic fields,
$\lambda_\perp \gg a_{\rm v}$ ($a_{\rm v}$ is 
the spacing between vortex lines).
In this case one can approximate the sum over ${\bf K}$ by 

\begin{equation}
   \sum_K F[{\bf K},q]=F[0,q]+{2\pi\over (\Delta K)^2}\int^{\infty}_{k_0}
   dk k F[{\bf k},q],
\label{eAp6}
\end{equation}
where $(\Delta K)^2=4\pi^2 B/\phi_0\equiv \pi k^2_0$ is the area per reciprocal
lattice vector, and $k_0\sim 1/a_{\rm v}$ is the edge of an approximate circular 
Brillouin Zone.  Carrying out the integrations, we get

\begin{eqnarray}
   c_{66}(q{\bf\hat y})
   &=&{B^2\over 4\pi}\left\{ {\lambda_\perp^2k_0^2\over
   4(1+\lambda_\perp^2k_0^2)^2}-\left[{\lambda_\perp^2k_0^2\over
   4(1+\lambda_\perp^2k_0^2)^4}\right]\lambda_\perp^2q^2\right\}
\label{eAp6a}\\
  c_{11}(q{\bf\hat x})&=&{B^2\over 4\pi}\left\{1-
  {1\over 4(1+\lambda_\perp^2k_0^2)}-
  {3\over 4(1+\lambda_\perp^2k_0^2)^2}-\left[1+{1\over 
  4(1+\lambda_\perp^2k_0^2)^3}-{5\over 4(1+\lambda_\perp^2k_0^2)^4}
  \right]\lambda_\perp^2q^2\right\}
\label{eAp6b}\\
   c_{44}(q{\bf\hat z}) & = & {B^2\over 4\pi}\left\{1+{1\over 
   2\lambda_\perp^2k_0^2}\left[\eta^{-2}\ln\left({1+\kappa^2\eta^2\over 1+
   \lambda_z^2k_0^2}\right)+{1\over 1+\lambda_\perp^2k_0^2}-\eta^{-2}\right]
   \right. \nonumber\\ 
   &\, &\qquad\qquad\qquad -\left.\left[1+{1\over 2
   \lambda_z^2k_0^2(1+\lambda_z^2k_0^2)}+{1\over 
   4\lambda_\perp^2k_0^2(1+\lambda_\perp^2k_0^2)^2}\right]\lambda_\perp^2
   q^2\right\}
\label{eAp6c}
\end{eqnarray}
where for $c_{44}$ we have taken the cutoff $\xi_\perp\to 0$ 
in all non-divergent terms, and $\kappa\equiv\lambda_\perp/\xi_\perp$.

Expanding for large $\lambda_\perp k_0$, we get to the
lowest non trivial order

\begin{eqnarray}
   c_{66}(q{\bf\hat y})&=&{B^2\over 4\pi}\,{1\over 4\lambda_\perp^2k_0^2}
   \left\{1-{2\over\lambda_\perp^2k_0^2}-
   {\lambda_\perp^2q^2\over\lambda_\perp^4k_0^4}\right\}
\label{eAp7a}\\
  c_{11}(q{\bf\hat x})&=& {B^2\over 4\pi}\left\{ 1-{1\over 4
  \lambda_\perp^2k_0^2}
  -\left[1+{1\over 4\lambda_\perp^6k_0^6}\right]\lambda_\perp^2q^2\right\}
\label{eAp7b}\\
   c_{44}(q{\bf\hat z})&=&{B^2\over 4\pi}\left\{1+{1\over 
   2\lambda_\perp^2k_0^2}\left[\eta^{-2}\left(\ln\left({H_{c2}\over B}
   \right)-1\right)+{1\over\lambda_\perp^2k_0^2}
   \right]\nonumber\right.\\ &\>&\qquad\qquad\qquad\qquad\left.
   -\left[1+{1\over 2\lambda_z^4k_0^4}+{1\over 4\lambda_\perp^6k_0^6}
   \right]\lambda_\perp^2q^2\right\},
\label{eAp7c}
\end{eqnarray}
where $H_{c2}\equiv\phi_0/4\pi\xi_\perp^2$.

Next we consider the case of small magnetic fields, $\lambda_\perp\ll a_{\rm v}$.
Here it is convenient to use

\begin{equation}
  \sum_K F[{\bf K},q]={\phi_0\over B}\sum_R\tilde F[{\bf R},q]
  \qquad {\rm where}\qquad \tilde F[{\bf r},q]\equiv \int {d^2k\over (2\pi)^2}
  e^{-i{\bf k}\cdot {\bf r}}F[{\bf k},q]
\label{eAp9}
\end{equation}
and $\{ {\bf R}\}$ are the direct Bravais lattice vectors of the vortex 
lattice.   

For the shear and compression moduli, $c_{66}$ and $c_{11}$, 
the subtraction terms in 
Eqs.(\ref{eAp4a}) and (\ref{eAp4b}) cause the ${\bf R}=0$ term of the sum
in Eq.(\ref{eAp9}) to vanish.  Since the range of the interaction $V_{zz}$
is $\lambda_\perp\ll |{\bf R}|$, 
it will be a good approximation in the sum over ${\bf R}$ to keep only the
six smallest vectors with $|{\bf R}|=a_{\rm v}$.
The Fourier transforms of Eqs.(\ref{eAp5a2}) and 
(\ref{eAp5b2}), can now be obtained with the help of

\begin{equation}
   \int {d^2k\over (2\pi)^2}{e^{-i{\bf k}\cdot {\bf r}}\over (1+ 
   \lambda^2k^2)^n}={1\over 2^n\pi (n-1)!\lambda^2}
   \left({r\over\lambda}\right)^{n-1}K_{1-n}\left({r\over\lambda}\right)
\label{eAp9.5}
\end{equation}
where $K_\nu$ is the modified Bessel function of the second kind 
of order $\nu$, whose asymptotic form at large $x$ is

\begin{equation}
    K_\nu(x)\sim\sqrt{{\pi\over 2x}}e^{-x}
\label{eAp9.6}
\end{equation}
Keeping only the leading terms in $a_{\rm v}/\lambda_\perp$, we find

\begin{eqnarray}
   c_{66}(q{\bf\hat z})&=&{3B\phi_0\over 64\sqrt{2\pi^3}\lambda_\perp^2}
   e^{-a_{\rm v}/\lambda_\perp}\left({a_{\rm v}\over\lambda_\perp}\right)^{3/2}
   \left[ 1-\textstyle{1\over 24}a_{\rm v}^2q^2\right]
\label{eAp10a}\\
   c_{11}(q{\bf\hat x})&=&{9B\phi_0\over 64\sqrt{2\pi^3}\lambda_\perp^2}
   e^{-a_{\rm v}/\lambda_\perp}\left({a_{\rm v}\over\lambda_\perp}\right)^{3/2}
   \left[ 1-\textstyle{5\over 72}a_{\rm v}^2q^2\right].
\label{eAp10b}
\end{eqnarray}

For the tilt modulus $c_{44}$ there are two cases to consider, depending on 
the strength of the anisotropy.  For very small magnetic fields such that
$\lambda_z\ll a_{\rm v}$, all terms in Eq.(\ref{eAp8}) may be treated according
to the approximation implied by Eq.(\ref{eAp9}).  Here the ${\bf R}=0$
term dominates all others, and we find

\begin{equation}
   c_{44}(q{\bf\hat z})={B^2\over 4\pi}{1\over 2\lambda_\perp^2k_0^2}
   \left\{\eta^{-2}\left[2\ln\left(\eta\kappa\right)-1
   \right]+1-\left[\textstyle{1\over 2}+\eta^{-2}\right]\lambda_\perp^2
   q^2\right\}.
\label{eAp11}
\end{equation}
For the intermediate case $\lambda_\perp\ll a_{\rm v}\ll 
\lambda_z$, we must combine approximations, using 
Eq.(\ref{eAp6}) for terms involving $\lambda_z^2k^2$, and Eq.(\ref{eAp9})
for terms involving $\lambda_\perp^2k^2$.  We find

\begin{equation}
   c_{44}(q{\bf\hat z})={B^2\over 4\pi}\left\{ {\textstyle{1\over 2}}+
   {1\over 2\lambda_\perp^2k_0^2}\left(\eta^{-2}\left[\ln\left({H_{c2}
   \over B}\right)-1\right]+1\right)-\left[{\textstyle{1\over 2}}+
   {1\over 4\lambda_\perp^2k_0^2}\right]\lambda_\perp^2q^2\right\}.
\label{eAp12}
\end{equation}

\begin{figure}
\caption{Schematic representation of three possible
perturbations of the external magnetic field: ($a$) the
tilt perturbation, ($b$) the compression perturbation,
and ($c$) the shear perturbation.
}
\label{f1}
\end{figure}

\begin{figure}
\caption{Labeling conventions for the lattice superconductor.
$A_{i\mu}$ are directed $outwards$ from site $i$ on the bonds of the 
direct lattice. $b_{i\mu}$ are directed $inwards$ towards the dual
site $i$ on the bonds of the dual lattice, piercing the plaquettes
of the direct lattice as shown.
}
\label{fgrid}
\end{figure}

\begin{figure}
\caption{Ground state for vortex line density $b_0=1/15$ on a cubic grid.
Solid circles indicate the locations of the straight vortex lines as
they pierce the $xy$ plane.
}
\label{fground}
\end{figure}

\begin{figure}
\caption{Helicity modulus plotted as $\bar J_0\lambda^2|Q_\nu|^2/
\Upsilon_\mu(q\hat{\bf\nu})$ vs. $|Q_\nu|^2$ for various values
of $T$. The straight lines are fits to Eq.(\ref{eNS18.5}), and
determine the parameters $\gamma_\mu$ and $(\lambda_{\mu{\rm R}}/
\lambda)^2$ for ($a$) the tilt, ($b$) the compression, and ($c$) 
the shear perturbations of Fig.\,\ref{f1}.
}
\label{figYi}
\end{figure}

\begin{figure}
\caption{Plots of $\gamma_\mu$ vs. $T$ as obtained from the
straight line fits of Fig.\,\ref{figYi}, fitting to the $8$, $7$, $6$,
and $5$ smallest values of $q$, for ($a$) the tilt, 
($b$) the compression, and ($c$) the shear perturbations of Fig.\,\ref{f1}.
Little sensitivity is seen to the number of values of $q$ used
in the fit.
}
\label{figamma}
\end{figure}

\begin{figure}
\caption{Plots of $(\lambda_{\mu{\rm R}}/\lambda)^2$ vs. 
$T$ as obtained from the
straight line fits of Fig.\,\ref{figYi}, fitting to the $8$, $7$, $6$,
and $5$ smallest values of $q$, for ($a$) the tilt, 
($b$) the compression, and ($c$) the shear perturbations of Fig.\,\ref{f1}.
Little sensitivity is seen to the number of values of $q$ used
in the fit.
}
\label{figlambda}
\end{figure}

\begin{figure}
\caption{Finite size comparison of the parameters ($a$) $\gamma_\mu$
and ($b$) $(\lambda_{\mu{\rm R}}/\lambda)^2$, for the
tilt ($\triangle$), the compression ($\bigcirc$), and the
shear ($\diamondsuit$) perturbation.  Open symbols are data for 
$N_z=15$, while solid symbols are data for $N_z=30$.
}
\label{figgl}
\end{figure}

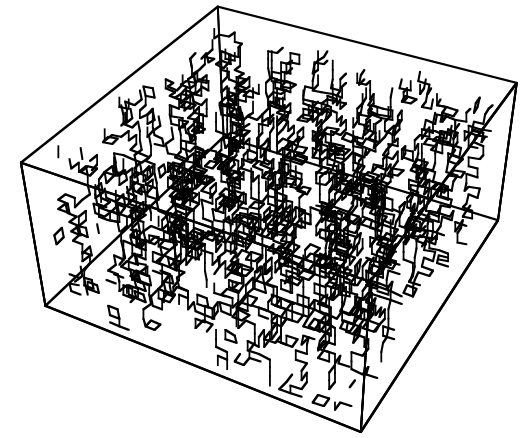
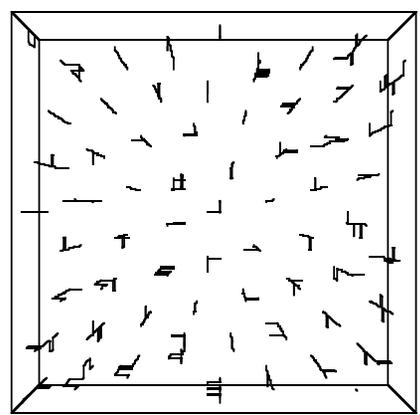
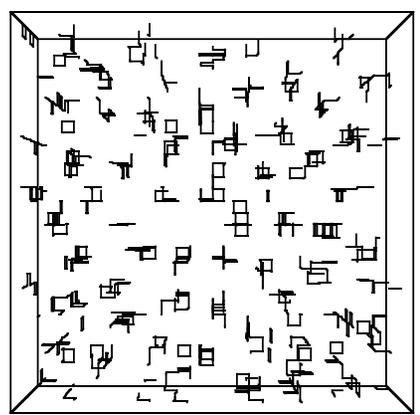
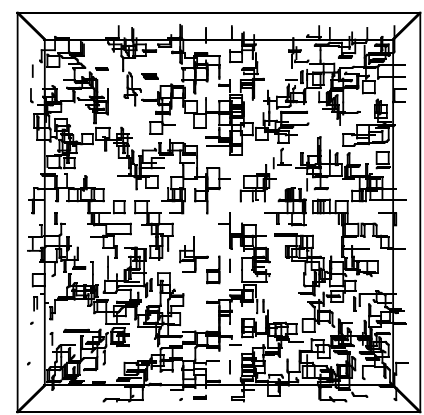
\begin{figure}
\caption{Snapshots of vortex line configurations for $N_z=15$, for
($a$) $T=1.0<T_{\rm m}$, ($b$) $T_{\rm m}<T=1.6<T_{\rm c}$, and
($c$) $T_{\rm c}<T=2.2$.  The bottom row is the view looking down
along the applied magnetic field.
}
\label{fsnap}
\end{figure}

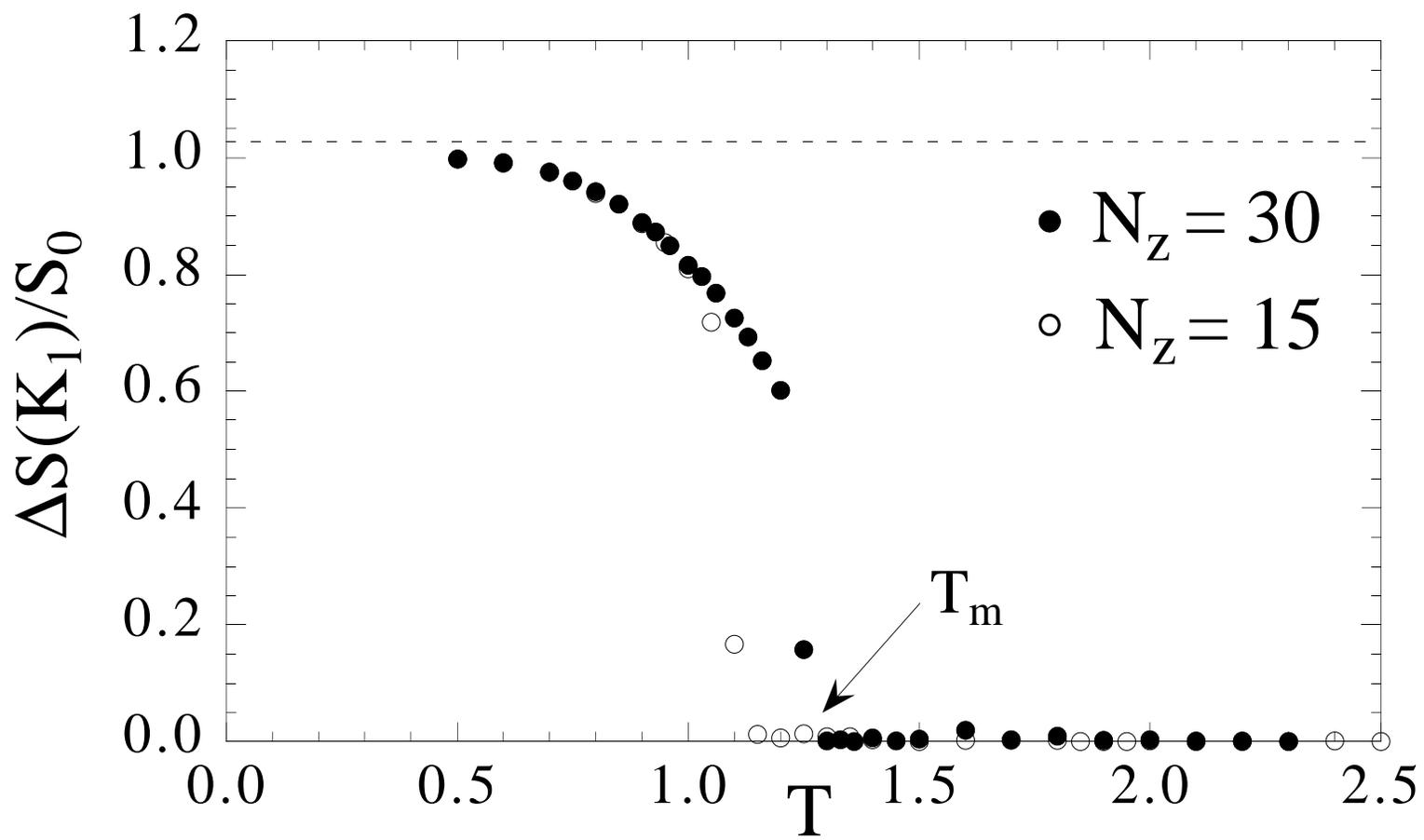
\begin{figure}
\caption{Plot of structure function peak heights, $\Delta S({\bf K}_1)
/S_0$, vs. $T$ for $N_z=15$ and $30$.
}
\label{fstruc}
\end{figure}

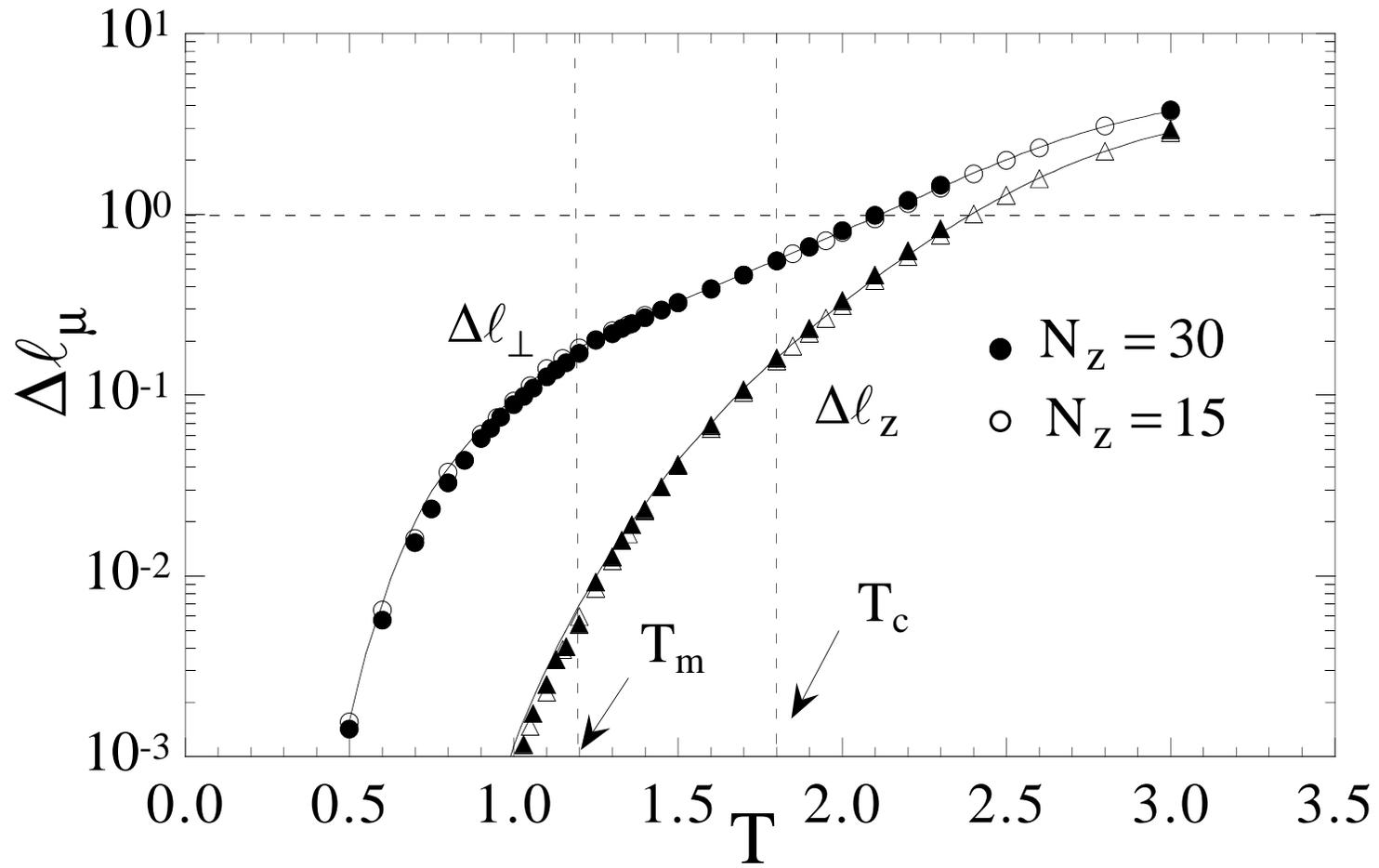
\begin{figure}
\caption{Average normalized fluctuation length of vortex lines 
$\Delta\ell_z$ and
$\Delta\ell_\perp$, parallel and transverse to the applied magnetic
field, vs. $T$.  We see that $\Delta\ell_z\ll\Delta\ell_\perp$
for $T<T_{\rm m}$, indicating that there are only transverse
fluctuations of the magnetic field induced vortex lines.  This
is no longer true near $T_{\rm c}$.  Open symbols are for $N_z=15$,
solid symbols are for $N_z=30$, and the solid lines are guides
to the eye.
}
\label{fdl}
\end{figure}

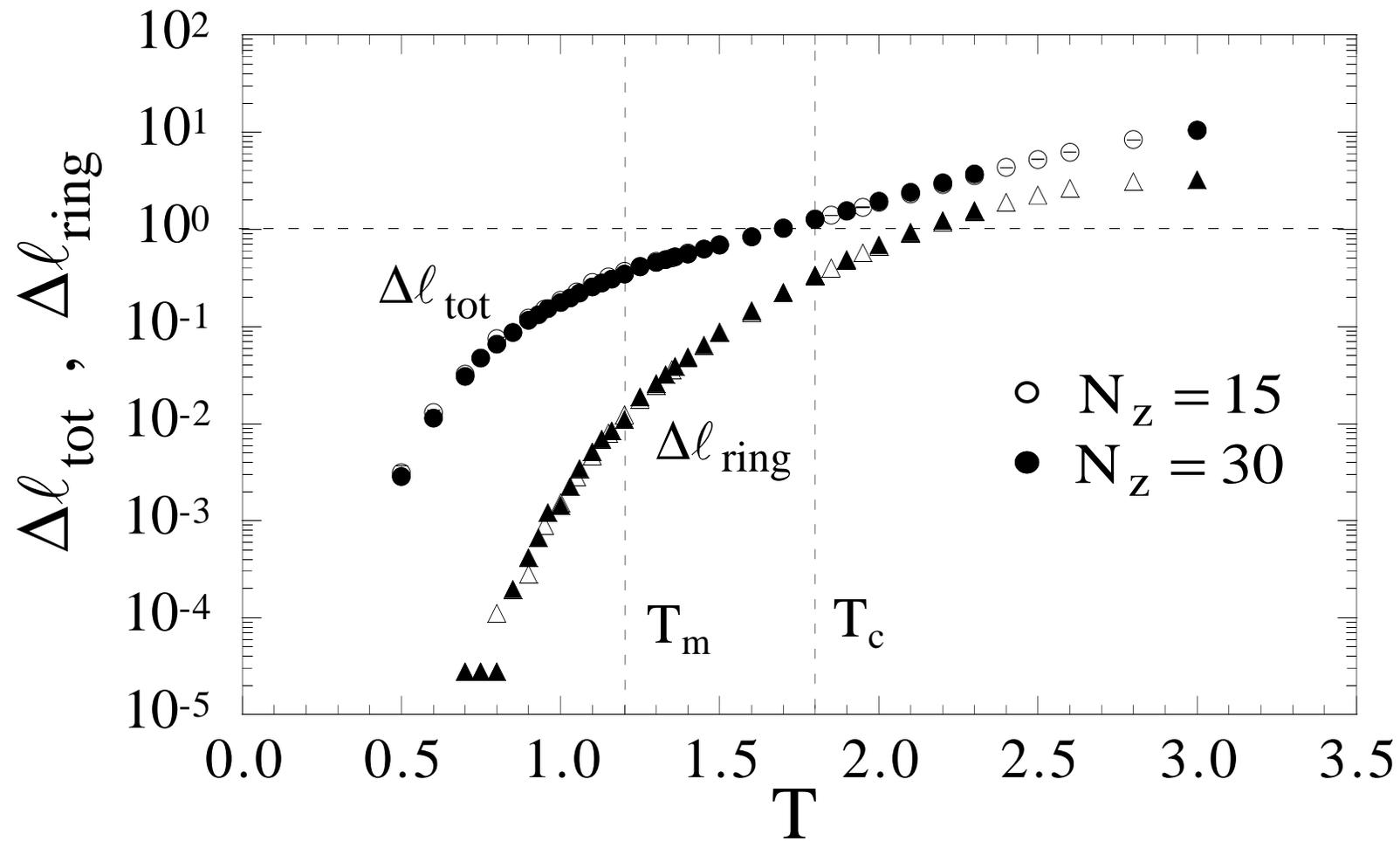
\begin{figure}
\caption{Average normalized length of all vortex line fluctuations
$\Delta\ell_{\rm tot}$, and average length in closed vortex ring
excitations $\Delta\ell_{\rm ring}$, vs. $T$.  Open symbols are for $N_z=15$,
and solid symbols are for $N_z=30$.
}
\label{fdlrings}
\end{figure}

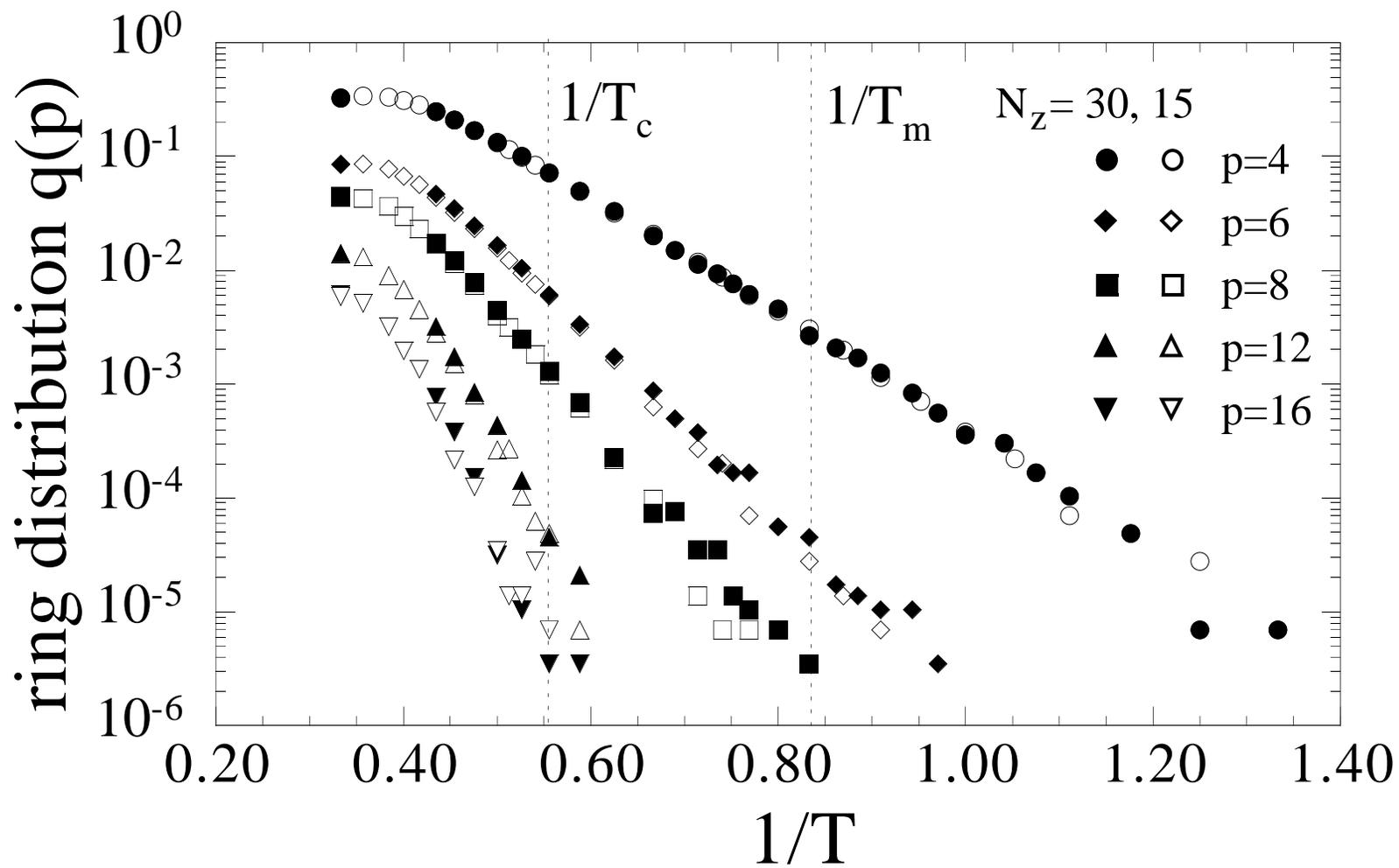
\begin{figure}
\caption{Semi-log plot of $q(p)$, the distribution of closed vortex rings 
of perimeter $p$, vs. $1/T$, for several values of $p$.  Straight 
lines at low $T$ indicate thermal activation.  $q(p)$ saturates
at a temperature above $T_{\rm c}$.  Open symbols are for $N_z=15$,
and solid symbols are for $N_z=30$.
}
\label{frings}
\end{figure}

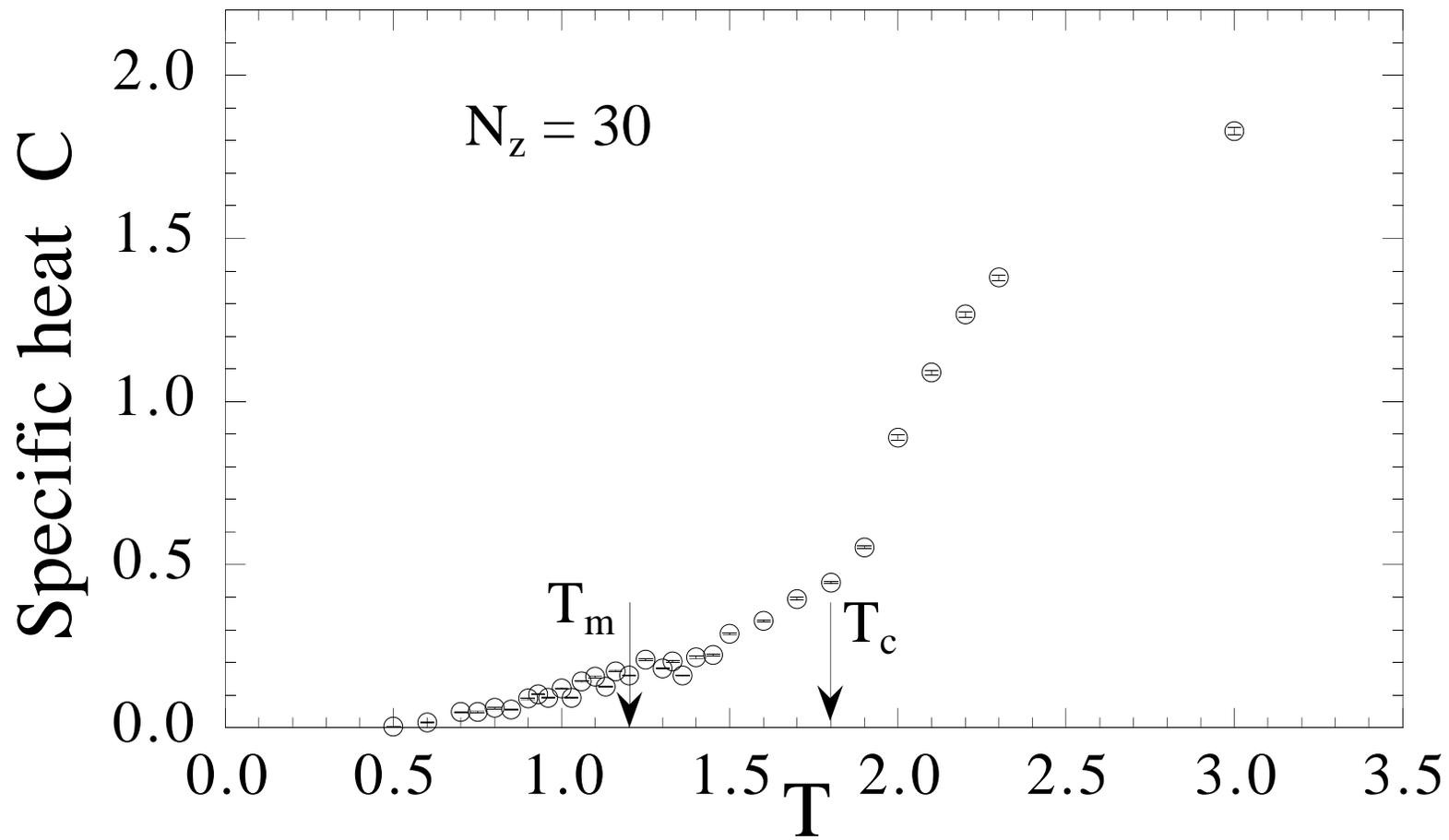
\begin{figure}
\caption{Specific heat $C$ vs. $T$ for $N_z=30$.  The peak in $C$
occurs above $T_{\rm c}$.
}
\label{fspecht}
\end{figure}

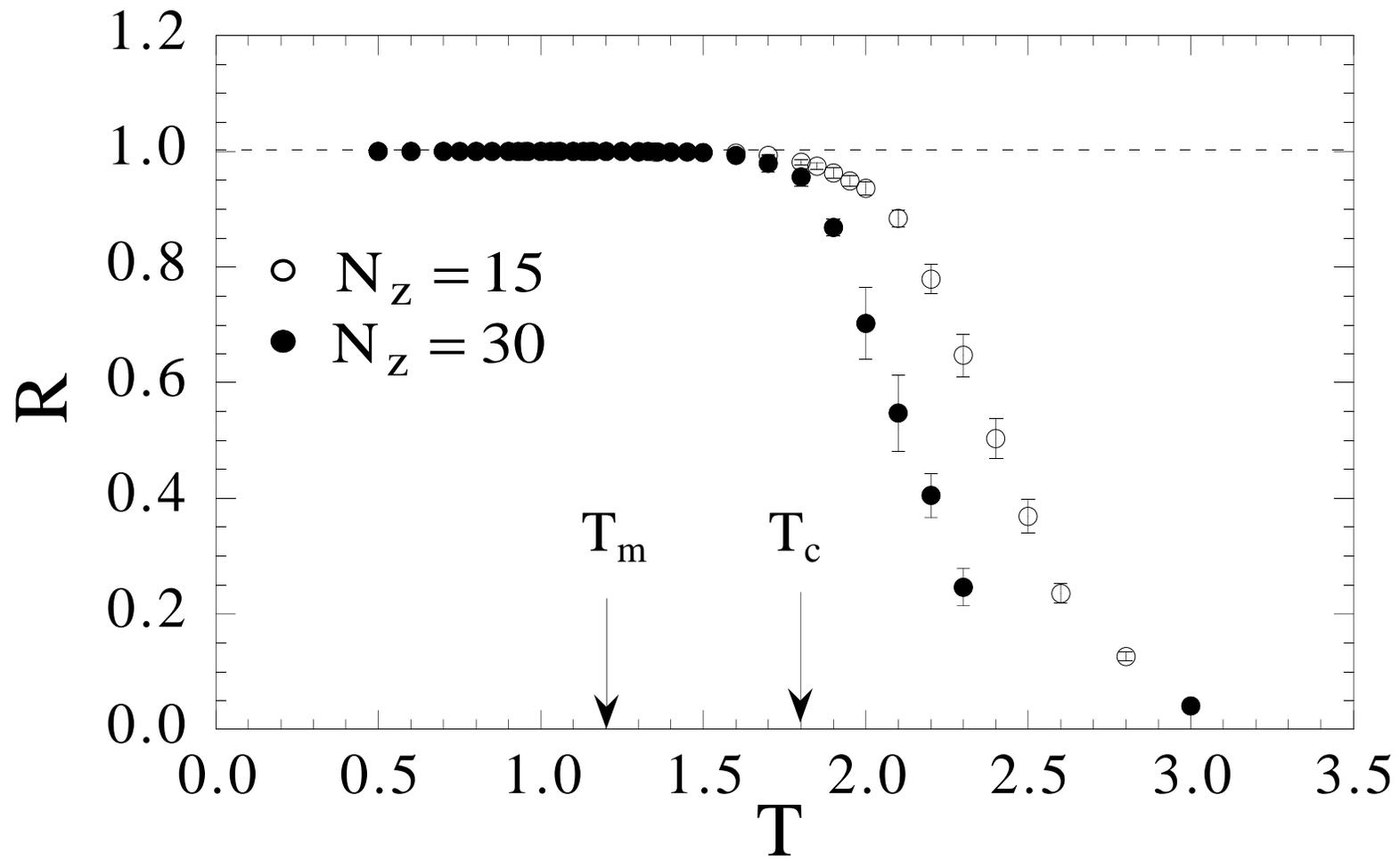
\begin{figure}
\caption{Fraction of unentangled magnetic field induced vortex lines $R$
vs. $T$, for $N_z=15$ and $30$.  Lines start to entangle at $T_{\rm c}$.
}
\label{figR}
\end{figure}

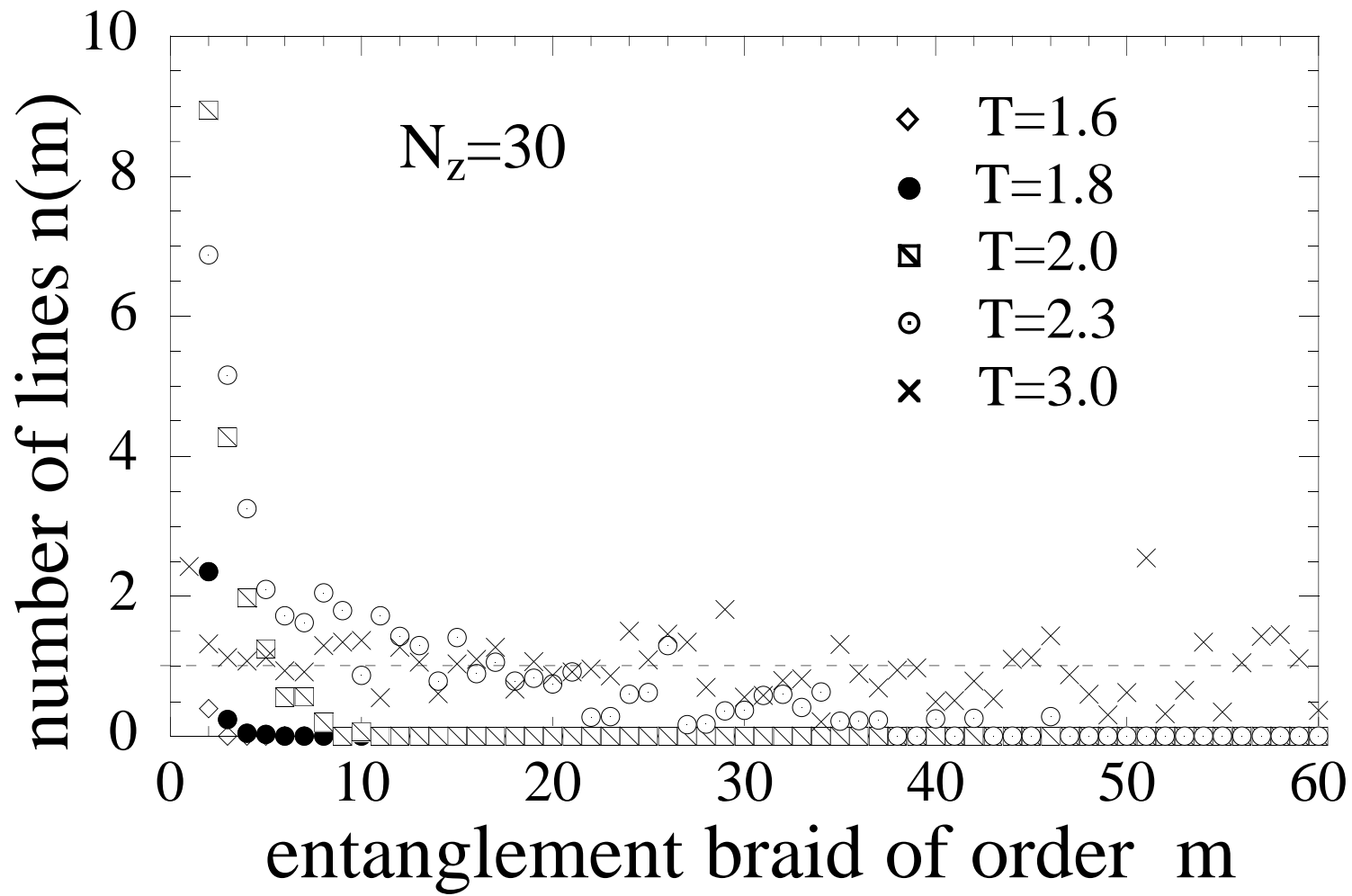
\begin{figure}
\caption{Distribution of entanglement of the magnetic field induced
vortex lines. $n(m)$ is the number of lines that participate in
entanglement braids of order $m$.  Data is shown for several 
temperatures near $T_{\rm c}=1.8$, for $N_z=30$.
}
\label{fignm}
\end{figure}

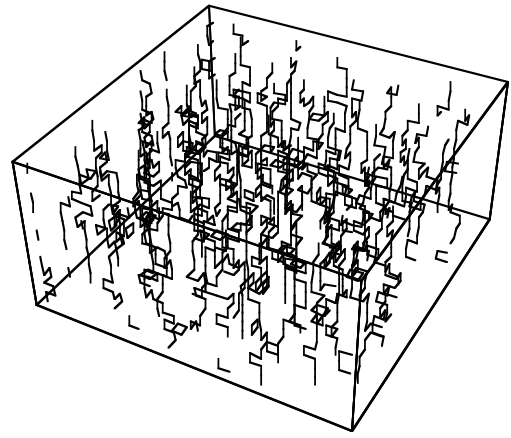
\end{document}